%% file: PNNP_Minor_nodiff.tex
\documentclass[10pt,journal,compsoc]{IEEEtran}
\usepackage{booktabs}
\usepackage{epsfig}
\usepackage{subfigure}
\usepackage{enumitem}
\usepackage{multirow}
\usepackage{amsmath}
\usepackage{amssymb}
\usepackage{algorithm}
\usepackage{algorithmicx}
\usepackage{algpseudocode}
\usepackage{makecell}
\usepackage{bbding}
\usepackage{diagbox}
\usepackage{siunitx}
\usepackage{colortbl}
\usepackage{fullwidth}
\input{LizhiPackage}
\usepackage{hyperref}
\newcommand{\HalfCheck}{\Checkmark\kern-1.2ex\raisebox{1ex}{\rotatebox[origin=c]{125}{\textbf{--}}}}
\definecolor{mygray}{gray}{.9}
\graphicspath{{figures/},{pairs/}}
\ifCLASSOPTIONcompsoc
\usepackage[nocompress]{cite}
\else
\usepackage{cite}
\fi

\ifCLASSINFOpdf
\else
\fi

\begin{document}
	%\title{Physics-guided Noise Neural Proxy\\for Practical Low-light Raw Image Denoising}
	\title{Learning Physics-informed Noise Models from Dark Frames for Low-light Raw Image Denoising}
	
	\author{Hansen~Feng, 
		Lizhi~Wang,~\IEEEmembership{Member,~IEEE,}
		Yiqi~Huang,
		Yuzhi~Wang,
		Lin~Zhu,~\IEEEmembership{Member,~IEEE,}\\
		and~Hua~Huang,~\IEEEmembership{Senior Member,~IEEE}% <-this % stops a space
		\IEEEcompsocitemizethanks{
			\IEEEcompsocthanksitem The authors gratefully acknowledge the anonymous reviewers for their valuable comments. This work is supported by National Natural Science Foundation of China (62322204, 62131003).\protect
			\IEEEcompsocthanksitem Hansen Feng, Yiqi Huang and Lin Zhu are with the School of Computer Science and Technology, Beijing Institute of Technology, Beijing, 100081, China. Email: $\{$fenghansen, huangyiqi, linzhu$\}$@bit.edu.cn\protect
			\IEEEcompsocthanksitem Lizhi Wang and Hua Huang are with the School of Artificial Intelligence, Beijing Normal University, Beijing 100875, China and also with Engineering Research Center of Intelligent Technology and Educational Application, Ministry of Education, Beijing 100875, China. Email: $\{$wanglizhi, huahuang$\}$@bnu.edu.cn\protect
			\IEEEcompsocthanksitem Yuzhi Wang is with the Megvii Technology, Beijing, 100086, China. Email: justin.w.xd@gmail.com\protect
			\IEEEcompsocthanksitem Corresponding author: Lizhi Wang\protect
			% \IEEEcompsocthanksitem Part of this work is done during the internship of Hansen Feng at Megvii Technology.
		}% <-this % stops an unwanted space
	}
	
	\markboth{Journal of \LaTeX\ Class Files,~Vol.~14, No.~8, August~2015}%
	{Shell \MakeLowercase{\textit{et al.}}: Bare Demo of IEEEtran.cls for Computer Society Journals}
	\IEEEtitleabstractindextext{%
		\begin{abstract}
			Recently, the mainstream practice for training low-light raw image denoising methods has shifted towards employing synthetic data. Noise modeling, which focuses on characterizing the noise distribution of real-world sensors, profoundly influences the effectiveness and practicality of synthetic data. Currently, physics-based noise modeling struggles to characterize the entire real noise distribution, while learning-based noise modeling impractically depends on paired real data. In this paper, we propose a novel strategy: learning the noise model from dark frames instead of paired real data, to break down the data dependency. Based on this strategy, we introduce an efficient physics-informed noise neural proxy (PNNP) to approximate the real-world sensor noise model. Specifically, we integrate physical priors into neural proxies and introduce three efficient techniques: physics-guided noise decoupling (PND), physics-aware proxy model (PPM), and differentiable distribution loss (DDL). PND decouples the dark frame into different components and handles different levels of noise flexibly, which reduces the complexity of noise modeling. PPM incorporates physical priors to constrain the synthetic noise, which promotes the accuracy of noise modeling. DDL provides explicit and reliable supervision for noise distribution, which promotes the precision of noise modeling. PNNP exhibits powerful potential in characterizing the real noise distribution. Extensive experiments on public datasets demonstrate superior performance in practical low-light raw image denoising. The source code will be publicly available at the \href{https://fenghansen.github.io/publication/PNNP}{project homepage}.
			% 两端对齐
		\end{abstract}
		}
	\maketitle

	\IEEEdisplaynontitleabstractindextext

	% \fi
	\IEEEpeerreviewmaketitle

	\IEEEraisesectionheading{\section{Introduction}\label{sec:introduction}}
	
	\IEEEPARstart{W}{ith} the increasing prevalence of cameras in mobile phones, it has become essential to deliver high-quality photography experiences, especially in low-light conditions. The inevitable noise in low-light conditions can lead to significant information loss, causing low-light raw image denoising a critical task. Learning-based denoising methods~\cite{TIP17/DnCNN,TIP18/FFDNet,CVPR19/CBDNet,NIPS19/VDN,CVPR21/MPRNet,CVPR21/NBNet,CVPR21/HINet,CVPR21/SwinIR,CVPR22/Restormer}, \ie, learning the noisy-to-clean mapping via the neural network, have become the mainstream approaches for low-light denoising, as they can learn data priors from massive data to fill the losing information. However, collecting a large-scale high-quality denoising dataset is a time-consuming and laborious process, which requires careful preparations to ensure data quality~\cite{CVPR17/DND,CVPR18/SIDD,CVPR18/SID,CVPR20/ELD,TPAMI21/ELD,IJCV23/DarkVision,TPAMI23/RViDeformer,TPAMI23/PMN}. Therefore, training with synthetic data based on a noise model is an efficient and practical alternative~\cite{CVPR19/Unprocess,ECCV20/Yuzhi,CVPR20/ELD,TPAMI21/ELD,ICCV21/SFRN}.
	
	Noise modeling aims to proxy the sensor noise for synthesizing noise and guiding denoising processes.
	Physics-based noise modeling~\cite{TPAMI94/CCD,nakamura2017image,P-G,EMVA1288,2011/CMOS,TED07/CMOS,arxiv2014/CMOS,CVPR20/ELD,TPAMI21/ELD}, as a classical and well-established approach, encompasses the modeling of noise by considering the underlying physical mechanisms and statistical properties of the sensor. For noise with known physical mechanisms, physics-based noise modeling provides accurate and interpretable representations, such as signal-dependent shot noise that can be equivalently modeled as Poisson noise~\cite{TPAMI94/CCD,EMVA1288}.
	However, in cases where the physical mechanisms underlying noise are partially known, physics-based methods generally deviate from the real-world sensor noise model. This deviation is particularly pronounced in the modeling of signal-independent noise~\cite{nakamura2017image,EMVA1288,2011/CMOS,TED07/CMOS,arxiv2014/CMOS}.
	To fill this gap, SFRN~\cite{ICCV21/SFRN} captures real dark frames as signal-independent noise, but limited sampling cannot comprehensively cover the entire spectrum of continuous camera settings and noise distribution.
	In summary, the principal challenge in physics-based noise modeling is the imprecise proxy of sensor noise, especially signal-independent noise, under low-light conditions.
	
	Learning-based noise modeling~\cite{NoiseFlow, ECCV20/DANet, ECCV20/CAGAN, NIPS21/PNGAN, CVPR22/DUS, CVPR23/LLD, ICCV23/LRD} has emerged as a potential alternative in recent years.
	Learning-based noise modeling typically involves learning the clean-to-noise mapping through neural networks trained on paired real data. 
	By leveraging large-scale high-quality datasets, learning-based noise modeling introduces a new approach to approximate the real-world sensor noise model, especially when the underlying physical mechanisms of the sensor are not fully known.
	However, due to underdeveloped data acquisition settings and inevitable environmental disturbances, acquiring high-quality paired real data remains a significant challenge~\cite{TPAMI23/PMN}. From a data perspective, the {dependence on high-quality paired real data} in learning-based noise modeling poses a paradox with the motivation for noise modeling.
	Moreover, existing methods struggle to accurately extract noise models from complex noise distributions~\cite{CVPR22/DUS,CVPR23/LLD}. The {lack of high-precision measurement techniques} for accurately approximating complex noise distributions typically results in the failure of learning-based noise modeling in real-world scenarios~\cite{ICCV21/SFRN,TPAMI21/ELD}.
	In summary, the principal challenges in learning-based noise modeling are the inherent difficulties of acquiring high-quality paired real data and approximating complex noise distributions.
	
	% \begin{enumerate}%[leftmargin=*]
		%     \item Due to underdeveloped data acquisition settings and inevitable environmental disturbances, paired real data typically suffer from spatial and intensity misalignment. As a result, neural networks may mistakenly interpret these misalignments as noise to be modeled.
		
		%     \item Due to the coupling of excessive noise models within the paired real data, the clean-to-noise mapping becomes complex and the noise distribution is difficult to measure. As a result, neural networks often converge unstably and finally reach suboptimal solutions.
		% \end{enumerate}
	
	To address the challenges above, we propose a novel strategy: \textbf{learning the noise model from dark frames instead of paired real data}.
	On one side, capturing dark frames is significantly more accessible than collecting paired real data, which {breaks the dependence on high-quality paired real data} for noise modeling. On the other side, dark frames only represent signal-independent noise, which {reduces the difficulty of noise distribution approximation}.
	Based on the strategy, we introduce a physics-informed noise neural proxy (PNNP) framework. Our framework leverages the physical priors of sensors to decouple the complex noise, constrain the optimization process, and provide reliable supervision. The comprehensive exploitation of dark frames empowers our framework to attain a superior approximation of the real-world sensor noise model.
	
	Firstly, we propose a physics-guided noise decoupling (PND) to handle different levels of noise in a flexible manner. Based on the statistical properties of noise, we decouple the dark frame into frame-wise, band-wise, and pixel-wise components. We model the frame-wise and band-wise noise via physics-based noise modeling while employing a neural network to proxy the pixel-wise noise. {Our noise decoupling strategy separates known noise components from dark frames, thereby reducing the complexity of noise modeling.}
	
	Moreover, we propose a physics-aware proxy model (PPM) incorporating physical priors to constrain the synthetic noise. Referring to the imaging process of sensors, we introduce an interpretable network structure to achieve physical constraints. {Our proxy model restricts the optimization degrees of freedom based on physical priors, thereby promoting the accuracy of noise modeling.}
	
	Finally, we propose a differentiable distribution loss (DDL) to efficiently supervise the distribution learning. By interpolating the cumulative distribution function (CDF) of the noise distribution, we introduce a differentiable loss function to directly measure pixel-wise noise distribution. Our loss function provides explicit and reliable supervision for noise modeling, thereby promoting the precision of noise modeling.
	
	Our main contributions can be summarized as follows:
	\begin{enumerate}[leftmargin=0.52cm]
		
		\item[1.] We introduce the strategy of learning the noise model from dark frames instead of paired real data, and then propose a physics-informed noise neural proxy framework to effectively integrate the physics-based and learning-based noise modeling.
		
		\item[2.] We propose a physics-guided noise decoupling to enable the neural network to approximate pixel-wise noise only, which reduces the complexity of noise modeling.
		
		\item[3.] We propose a physics-aware proxy model to ensure that the synthesized noise adheres to the physical priors, which promotes the accuracy of noise modeling.
		
		\item[4.] We introduce a differentiable distribution loss to provide explicit and reliable supervision, which promotes the precision of noise modeling.
	\end{enumerate}
	
	The efficient integration of physics-based and learning-based noise modeling is the core competitiveness of our framework. 
	% Our framework introduces statistical constraints into neural proxies based on physical priors, while also learning the continuous representation of real sensor noise through neural proxies. PNNP introduces a novel approach in comparison to existing noise modeling approaches.
	Benefiting from physics-informed designs, PNNP not only exhibits low data dependency but also facilitates easy training. These user-friendly features emphasize the practicality of PNNP.
	Extensive experiments on low-light raw image denoising datasets~\cite{CVPR18/SID,CVPR20/ELD,TPAMI21/ELD} have demonstrated the superiority of our framework compared to existing noise modeling methods.
	
	\section{Related Works}
	\subsection{Low-light Raw Image Denoising}
	Due to the inherent advantages of raw images characterized by high-bit depth and high linearity, low-light raw image denoising has garnered significant attention in the industry and has been subject to extensive research in various fields, including astronomy~\cite{astronomy}, microscopy~\cite{microscopic} and mobile photography~\cite{TOG16/HDR+}.	
	Classical denoising methods relying on image priors such as sparsity~\cite{K-SVD,elad2006image}, low rank~\cite{WNNM}, self-similarity~\cite{NLM,BM3D}, and smoothness~\cite{wavelet,TV}, can generally be applied to low-light raw image denoising. 
	However, the effectiveness of image priors is ultimately limited in low-light conditions, making it challenging to achieve satisfactory denoising results. With the development of deep learning, learning-based methods~\cite{CVPR18/SID,ICCV19/DRV,TPAMI21/LowLightSurvey} have emerged as the mainstream approach for low-light raw image denoising.
	
	Training a neural network requires large-scale high-quality data, which is a significant challenge in low-light raw image denoising. On one hand, the extreme imaging settings increase the difficulty of data collection. The combination of complex physical environments and challenging imaging settings typically results in underdeveloped datasets, characterized by either insufficient quantity or compromised quality (\eg, misalignment). On the other hand, the extreme imaging settings amplify errors in the noise model. The classical noise model~\cite{P-G} fails to generate reliable training data, highlighting the indispensability of research on noise modeling in practice.
	
	\subsection{Noise Modeling}
	\begin{table}[t!]
		\small
		\caption{Comparison between noise modeling methods. ``Approach" represents the technical pathways of these methods, among which ``DDL" is our novel noise distribution measurement method. The last two columns compare the data dependency of these methods. Calibration materials are easy to capture, while paired real data is challenging to collect.}
		\vspace{-6pt}
		\label{tab:methods}
		\setlength{\tabcolsep}{5pt}
		\begin{center}
			{%
				% \vspace{-6pt}
				\begin{tabular}{lccc}
					\toprule
					{Method} & {Approach} & \makecell{Calibration\\Materials} & \makecell{Paired\\Real Data}\\
					\midrule
					P-G~\cite{P-G} & Physics & \Checkmark & \XSolidBrush\\
					ELD~\cite{TPAMI21/ELD} & Physics & \Checkmark & \XSolidBrush\\
					SFRN~\cite{ICCV21/SFRN} & Physics & \Checkmark & \XSolidBrush\\
					\midrule
					NoiseFlow~\cite{NoiseFlow} & Flow & \XSolidBrush & \Checkmark\\
					DANet~\cite{ECCV20/DANet} & GAN & \XSolidBrush & \Checkmark\\
					CA-GAN~\cite{ECCV20/CAGAN} & GAN & \XSolidBrush & \Checkmark\\
					PNGAN~\cite{NIPS21/PNGAN} & GAN & \XSolidBrush & \Checkmark\\
					\midrule
					Starlight~\cite{CVPR22/DUS} & Physics + GAN & \Checkmark & \Checkmark\\
					LLD~\cite{CVPR23/LLD} & Physics + Flow & \Checkmark & \Checkmark\\
					LRD~\cite{ICCV23/LRD} & Physics + GAN & \Checkmark & \Checkmark\\
					%\rowcolor{gray}
					PNNP (Ours) & Physics + DDL & \Checkmark & \XSolidBrush\\
					\bottomrule
				\end{tabular}%
			}
		\end{center}
		\vspace{-10pt}
	\end{table}
	
	Physics-based noise modeling methods, inspired by the physics prior and statistical characteristics of sensor noise, are widely employed in the industry~\cite{ECCV20/Yuzhi}.
	Based on the correlation of signals, sensor noise is typically divided into signal-dependent noise and signal-independent noise. The noise model parameters are often specific to the sensors, thus requiring calibrating noise parameters from calibration materials (\eg, dark frames and flat-field frames) for specific cameras~\cite{TPAMI94/CCD,P-G,EMVA1288}. 
	In raw image denoising, the most classic noise modeling approach is the Poisson-Gaussian model or simplified heteroscedastic Gaussian model~\cite{P-G}.
	%	The calibrated noise parameters can also serve as noise priors for certain Variance-Stabilizing Transformation (VST) methods~\cite{TIP11/VST,TIP13/VST,ECCV20/Yuzhi,IJCV22/zhangyi}, thereby further enhancing denoising performance.
	However, classic noise modeling approaches tend to be limited in low-light conditions~\cite{SPIE04/imaging,TED07/CMOS,EMVA1288,arxiv2014/CMOS,TPAMI23/FPNR,nakamura2017image}.
	In response to this challenge, there are several noise modeling works specifically targeting low-light conditions in recent years.
	\cite{CVPR20/ELD,TPAMI21/ELD} propose a physics-based noise modeling method for extreme low-light photography, improving the accuracy of noise modeling in low-light conditions.
	\cite{ICCV21/SFRN} propose to directly sample from dark frames and perform high-bit recovery to synthesize noisy data, trading off noise accuracy for data diversity.
	\cite{ACMMM22/PMN,TPAMI23/PMN} introduce a linear dark shading model and a shot noise augmentation method, opening up a new perspective on the relationship between paired real data and noise modeling.
	
	To outperform the physics-based approach~\cite{NoiseFlow, ECCV20/DANet, ECCV20/CAGAN, NIPS21/PNGAN, CVPR22/DUS, CVPR23/LLD, ICCV23/LRD}, the learning-based approach has gained significant attention in recent years. Existing learning-based methods typically involve constituting a noise neural proxy through neural networks trained on paired real data. However, the utilization of imprecise distribution measurement methods often limits the effectiveness of the learning-based approach compared to the physics-based approach.
	For instance, \cite{NoiseFlow} employs normalized flow~\cite{NIPS18/Glow} to transform Gaussian noise into real noise. Nevertheless, flow-based methods struggle to approximate complex noise distributions in low-light conditions~\cite{TPAMI21/ELD,ICCV21/SFRN}. Similarly, GAN-based noise modeling methods~\cite{ACCV20/GAN, ECCV20/DANet, ECCV20/CAGAN, NIPS21/PNGAN} face challenges in effectively approximating noise distribution, which is often reported to be unstable and difficult to converge~\cite{ICCV21/SFRN,CVPR23/LLD}.
	To address the challenges, some advanced noise modeling methods~\cite{CVPR22/DUS,CVPR23/LLD,ICCV23/LRD} integrate physics-based methods into learning-based methods, ushering in a new era of learning-based noise modeling.
	
	Recently, based on the concept of learnability enhancement, PMN~\cite{ACMMM22/PMN,TPAMI23/PMN} reform paired real data according to noise modeling. Inspired by the PMN, we revisit the data dependency of existing noise modeling methods. 
	%We observe that existing learning-based methods often overlook or incorporate data defects in paired real data as part of the noise, leading to overfitting the training data and underfitting the noise model. 
	{We observe that existing learning-based methods often overlook data defects in paired real data. Consequently, they either disregard these defects or misinterpret them as noise, resulting in overfitting to the training data while failing to accurately capture the true noise characteristics.}
	The observation encourages us to propose a practical method to break through the status quo. As shown in Table~\ref{tab:methods}, our method not only facilitates easy training but also exhibits low data dependency, highlighting its superiority in practice.
	
	%	We have identified various signal-wise defects in the existing paired real data present in denoising datasets, including noise residues, spatial misalignment and brightness inconsistency. Existing learning-based noise modeling methods tend to overlook or incorporate these defects as part of the noise, resulting in overfitting to the training scene and underfitting to the noise model. Addressing the issue of data quality and enabling reliable measurements both serve as the foundation of our work.
	
	\begin{figure*}[t!]
		\begin{center}
			\includegraphics[width=\textwidth]{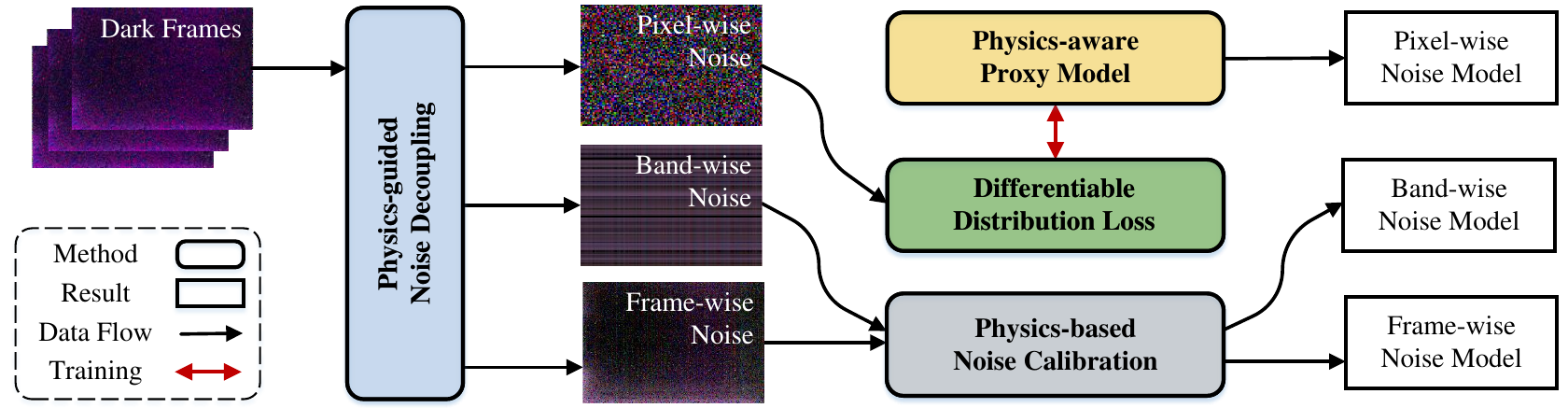}
		\end{center}
		\caption{The overview of our physics-informed noise neural proxy (PNNP) framework. We have visualized the noise as sRGB images for viewing.}
		\label{fig:framework}
	\end{figure*}
	
	\section{Method}
	\subsection{Strategy and Framework}\label{subsec:framework}
	To facilitate a clear understanding of our strategy and framework, we will begin with a brief overview of the general sensor imaging model.
	
	During the sensor imaging process, the incident scene irradiance $I$ undergoes several transformations, including conversion to charges and subsequently voltages, amplification by amplifiers, and quantization into the digital signal $D$ by an analog-to-digital converter (ADC)~\cite{EMVA1288}. For a raw image captured by a sensor, the imaging model can be expressed as
	\begin{equation}\label{eq:raw-simple}
		D = K (I + N_{p}) + N_{indep},
	\end{equation}
	where $K$ represents the overall system gain, $N_p$ is the signal-dependent photon shot noise, and $N_{indep}$ is signal-independent noise. 
	The general noise model can be extracted as
	\begin{equation}\label{eq:pg}
		N = KN_{p} + N_{indep}.
	\end{equation}	
	In the general noise model, the shot noise model based on the photoelectric effect is relatively reliable due to well-defined physical processes, which can be modeled as
	\begin{equation}\label{eq:possion}
		(I + N_p) \sim{\mathcal P(I)},
	\end{equation}
	where $\mathcal P(\cdot)$ denotes the Poisson distribution.
	In contrast, the origin of signal-independent noise is highly complex, and there is no consensus on its noise model in the imaging community~\cite{SPIE04/imaging,TED07/CMOS,EMVA1288,arxiv2014/CMOS,nakamura2017image,TPAMI23/FPNR}. %Noise modeling aims to investigate strategies for accurately and efficiently characterizing sensor noise.
	In low-light conditions, the contribution of signal-independent noise is much higher compared to that in normal-light conditions, thus further emphasizing the need for accurate noise modeling.
	
	To achieve accurate noise modeling, some advanced noise modeling methods~\cite{CVPR22/DUS,CVPR23/LLD,ICCV23/LRD} attempt to integrate physics-based methods into learning-based methods. However, they heavily depends on the high-quality paired real data, which limits their practicality.
	To release the noise modeling from the constraints of data dependency and further enhance the accuracy of noise modeling, we propose a novel strategy: learning the noise model from dark frames instead of paired real data.	
	Motivated by this new strategy, we introduce a novel noise modeling framework, physics-informed noise neural proxy~(PNNP), as shown in Fig.~\ref{fig:framework}.
	
	The first step of PNNP is physics-guided noise decoupling. Dark frames are the images captured under a lightless environment, which contains complete signal-independent noise. We decouple dark frames into three independent levels, which can be present as
	\begin{equation}\label{eq:synthtic}
		N_{indep} = N_{frame} + N_{band} + N_{pixel},
	\end{equation}
	where $N_{frame}$ represents the temporal stable frame-wise noise, $N_{band}$ represents the row- or col-related band-wise noise, and $N_{pixel}$ represents the independent and identically distributed (i.i.d.) pixel-wise noise.
	
	We leverage the frame-wise noise and band-wise noise to calibrate their respective noise models using established physics-based noise calibration methods~\cite{CVPR20/ELD,TPAMI21/ELD,ACMMM22/PMN}. Notably, our primary focus lies in accurately modeling the pixel-wise noise, a challenging task for previous physics-based noise modeling methods. Therefore, we just employ a powerful neural network as the proxy model of pixel-wise noise. The decoupled pixel-wise noise exhibits the favorable characteristic of spatial independence, which serves as a valuable physical prior. To fully exploit the physical prior, we introduce the PPM to represent the pixel-wise noise and utilize the DDL to measure the distribution distance. Through precise and efficient training, we can obtain the pixel-wise noise model. Detailed implementation specifics of the PND, PPM, and DDL will be elaborated in subsequent sections of the paper.
	
	In summary, the PNNP simplifies the complexity of noise modeling, combining the advantages of physics-based and learning-based approaches to enhance the accuracy of noise modeling.
	
	\begin{figure}[t!]
		\begin{center}
			\includegraphics[width=\linewidth]{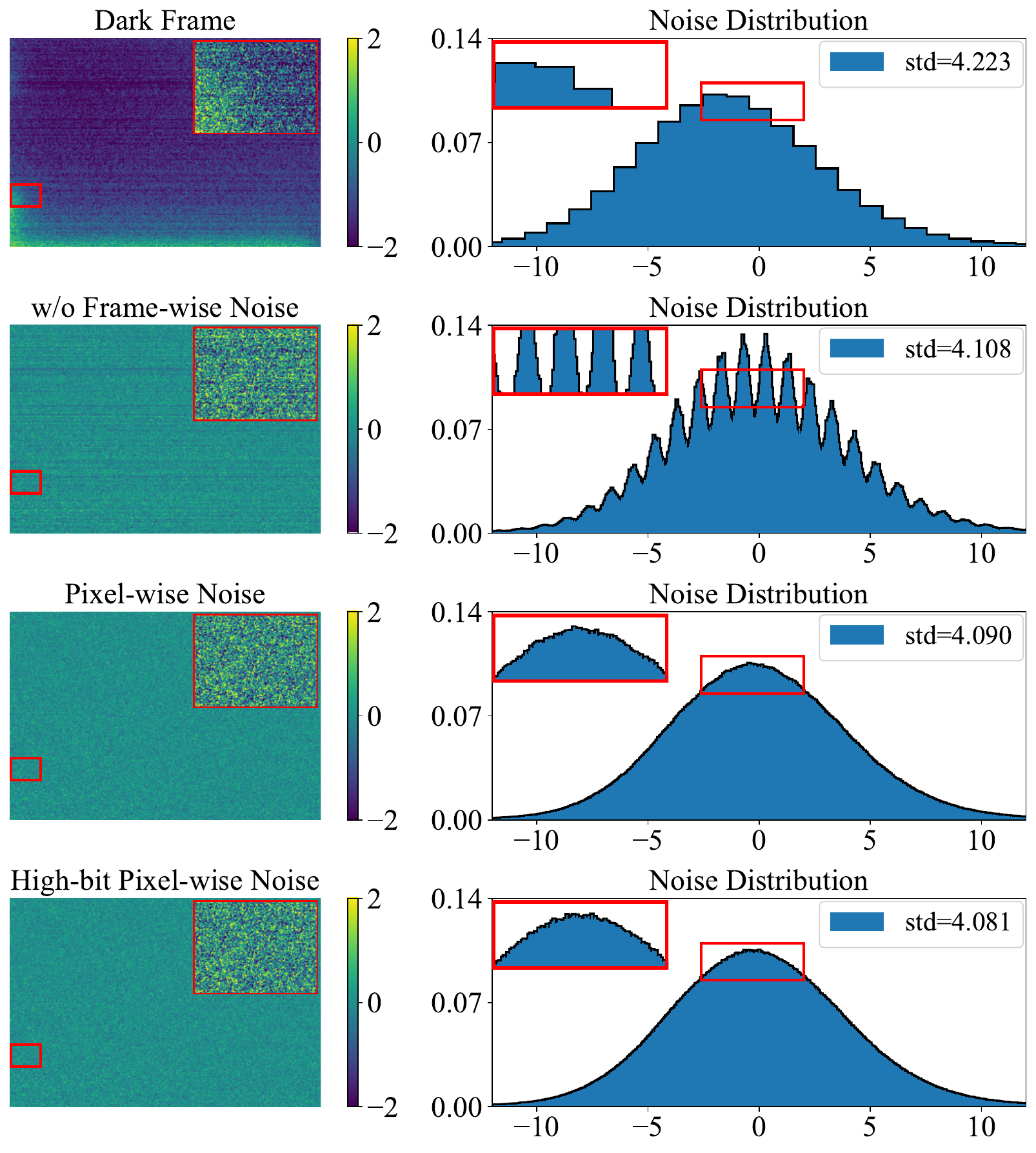}
		\end{center}
		\caption{The analysis of PND process. The first and second columns plot the noise and its corresponding noise distribution, respectively. ``std" refers to the standard deviation of the noise, which characterizes the intensity of the noise. The numerical range is adjusted for the best viewing. {\color{red}Red} boxes highlight zoomed-in regions for detailed observation.}
		\label{fig:pnd}
	\end{figure}
	
	\begin{figure*}[t!]
		\begin{center}
			\includegraphics[width=\textwidth]{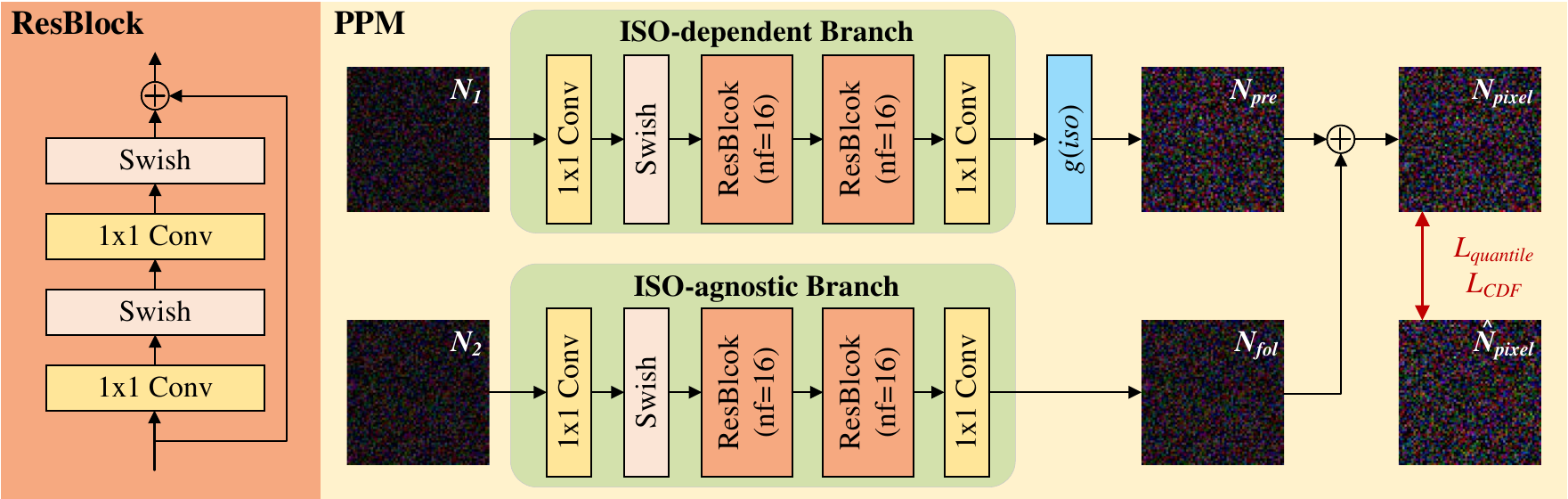}
		\end{center}
		\caption{The overview of our PPM. ``nf" represents the number of filters. All 1$\times$1 convolutions in PPM have 16 channels.}
		\label{fig:network}
	\end{figure*}
	
	\subsection{Physics-guided Noise Decoupling} \label{subsec:PND}
	
	In this section, we will introduce the principles and procedures of physics-guided noise decoupling in detail.
	
	Previous research~\cite{ACMMM22/PMN} highlights the advantages of noise decoupling in reducing data mapping complexity and improving denoising performance. We extend this notion to noise modeling, demonstrating that noise decoupling offers advantages beyond denoising. By employing different modeling approaches to address distinct levels of noise, we can effectively reduce the complexity of noise modeling. In line with the physical mechanisms and statistical characteristics of noise, we decouple dark frames into three independent noise types with significant mode differences: frame-wise noise, band-wise noise, and pixel-wise noise. The decoupling process of a dark frame is shown in Fig.~\ref{fig:pnd}.
	The notion of PND is that \textbf{noise neural proxy should only focus on the unknown part of noise}. Simplified real noise brings superiority to the learnability of data mapping.
	
	Initially, the original dark frame, captured in a lightless environment, is represented as discrete signals with quantization intervals. In the decoupling process, we first employ frame-wise decoupling~\cite{TPAMI23/PMN} to calibrate and eliminate the frame-wise noise $N_{frame}$. As a result, the dark frame devoid of frame-wise noise no longer exhibits significant spatial non-uniformity and retains various random noise components. Subsequently, we employ band-wise decoupling~\cite{TPAMI21/ELD} to calibrate and eliminate the band-wise noise $N_{band}$, resulting in a dark frame without spatial correlation, where only i.i.d. pixel-wise noise remains.
	Due to the low-bit quantization of the original dark frame, the noise distribution of the pixel-wise noise exhibits periodic fluctuations caused by quantization-induced signal misalignment, as shown in Fig.~\ref{fig:pnd}. To mitigate the impact of quantization-induced signal misalignment on the noise neural proxy, we perform high-bit reconstruction~\cite{ICCV21/SFRN} to eliminate the influence of quantization noise. The resulting high-bit pixel-wise noise $N^{HB}_{pixel}$ serves as the ground truth for subsequent noise neural proxy modeling. The observed significant variations in noise distribution and noise intensity (standard deviation) in Fig.~\ref{fig:pnd} indicate the effectiveness of our physics-guided noise decoupling.
	
	Next, we will detail the principles and procedures of frame-wise decoupling, band-wise decoupling, and high-bit reconstruction, respectively.
	
	Frame-wise noise, also known as dark shading \cite{nakamura2017image}, represents the temporal stable component of signal-independent noise and encompasses both spatial invariant Black Level Error (BLE) \cite{ICME21/RethinkNM, TPAMI21/ELD} and spatial variant Fixed Pattern Noise (FPN) \cite{2011/CMOS,EMVA1288,arxiv2014/CMOS,TIP14/FPNR,TPAMI23/FPNR}.
	Following previous works~\cite{TPAMI23/PMN}, we first compute the average of dark frames at the same ISO to estimate the ISO-specific frame-wise noise. Then, we utilize a linear dark shading model to fit the ISO-specific frame-wise noise at different ISO as the final frame-wise noise model
	\begin{equation}
		N_{frame} = N_{FPNk} \cdot iso + N_{FPNb} + BLE(iso),
	\end{equation}
	where $N_{FPNk}\in \mathbb R^{H \times W}$ and $N_{FPNb}\in \mathbb R^{H \times W}$ are the coefficient maps of the FPN that needs to be regressed, $iso$ is the ISO value, $BLE(iso)$ is the BLE at a specific ISO value $iso$. We introduce small perturbations to each parameter during training in order to simulate calibration errors.
	
	%	It is worth noting that the frame-wise noise used for denoising is reconstructed from the frame-wise noise model instead of the ISO-specific frame-wise noise. This choice offers two main advantages. Firstly, estimating the linear dark shading model leverages more dark frames across different ISO levels, thereby reducing the influence of random noise. Secondly, parameterizing the linear dark shading model allows us to introduce small perturbations to each parameter during training to simulate calibration errors.
	
	The band-wise noise can be primarily divided into row noise $N_{row}$ and column noise $N_{col}$, which are typically associated with the working frequency and readout mechanism of the sensor circuit
	\begin{equation}
		N_{band} = N_{row} + N_{col}.
	\end{equation}
	
	Following previous works~\cite{CVPR20/ELD,TPAMI21/ELD}, we model the band-wise noise as zero-mean Gaussian noise
	\begin{equation}
		N_{row}\sim \mathcal N(0, \sigma^2_{row}), N_{col}\sim \mathcal N(0, \sigma^2_{col}),
	\end{equation}
	where $N_{row}\in \mathbb R^{H \times 1}$ and $N_{col}\in \mathbb R^{1 \times W}$.
	
	The band-wise noise can be easily estimated by computing the mean of each row or column in the dark frames. The goodness of fit can be evaluated using the coefficient of determination in the probability plot, denoted as R$^2$~\cite{statistical}. In our experiments, the R$^2$ for fitting the band-wise noise with a Gaussian distribution exceeds 0.999, indicating a strong adherence of the band-wise noise to the assumption of a Gaussian distribution.
	
	After the aforementioned physics-guided noise decoupling, the remaining noise corresponds to the pixel-wise noise, which can be considered as the aggregation of all i.i.d. noise within the sensor. As introduced at the beginning of this section, the pixel-wise noise obtained from the noise decoupling exhibits quantization-induced signal misalignment. Therefore, we employ the high-bit reconstruction technique proposed in SFRN \cite{ICCV21/SFRN} to reconstruct the pixel-wise noise. It is worth noting that SFRN does not apply noise decoupling before high-bit reconstruction, resulting in non-i.i.d. signals that affect the performance. The effectiveness of high-bit reconstruction technique relies on our noise decoupling strategy, which is not considered in existing works.
	
	In summary, the physics-guided noise decoupling strategy separates known noise components from dark frames, thereby reducing the complexity of noise modeling.

	\subsection{Physics-aware Proxy Model} \label{subsec:PPM}
	
	As shown in Fig.~\ref{fig:network}, we propose the PPM to constrain the optimization process based on physical priors. Our overall network structure adopts a lightweight dual-branch design. The input of PPM is the random noise $N_{1}$ and $N_{2}$ that follows a standard normal distribution. We divide the input into two parts, which are separately fed into the ISO-dependent branch and the ISO-agnostic branch. Two branches of the network share identical structures, which include 1$\times$1 convolutions, Swish activation functions~\cite{swish}, and ResBlocks composed of these operations. Following the physical imaging process of sensors, the output $N_{pre}$ from the ISO-dependent branch undergoes amplification using a gain layer $g(iso)$, while the output $N_{fol}$ from the ISO-agnostic branch remains unchanged regardless of ISO variations. The sum of the outputs from these dual branches is pixel-wise noise $N_{pixel}$, representing the noise transformation result of our proxy model. We use the quantile loss $L_{quantile}$ and CDF loss $L_{CDF}$ to supervise the model, where the high-bit pixel-wise noise $N^{HB}_{pixel}$ is the ground truth. The process can be summarized as
	\begin{equation}
		N_{pixel} = {\rm PPM} (N_{1}, N_{2}, iso),
	\end{equation}
	where $N_{1}, N_{2}, N_{pixel}\in \mathbb R^{H \times W}$.
	
	We will provide a detailed explanation of the two physics-aware designs incorporated in our network.
	
	Firstly, we introduce the design of the pixel-wise independent module. Considering the i.i.d. nature of pixel-wise noise, all module designs utilize spatially independent operations, employing 1$\times$1 convolutions instead of 3$\times$3 convolutions. 
	
	The representation space of a $1\times1$ convolution kernel is strictly a subset of that of a $3\times3$ convolution kernel. Intuitively, replacing $1\times1$ convolutions with $3\times3$ convolutions will not degrade the performance. However, $3\times3$ convolutions can pose an optimization trap in practice. The presence of non-zero parameters in 3$\times$3 convolution kernels (except the central parameter) inevitably introduces spatial correlation, which contradicts the prior assumption of pixel-wise spatial independence. 
	
	It is worth emphasizing that most of learning-based methods~\cite{ECCV20/DANet, ECCV20/CAGAN, NIPS21/PNGAN, CVPR22/DUS}, including flow-based methods~\cite{NoiseFlow, CVPR23/LLD}, introduce spatial correlations. Moreover, the distinctive feature of PND, decoupling noise into spatially independent pixel-wise noise, is a sufficient condition where the efficacy of 1$\times$1 convolutions shines. Therefore, the usage of 1$\times$1 convolutions in the proxy model is simple yet nontrivial.
	By adhering to the structural constraint of 1$\times$1 convolutions, the proxy model significantly reduces its optimization freedom while adhering to the prior assumption of spatial independence.
	
	Next, we introduce the design of the ISO-aware dual branch. The noise pattern is significantly affected by analog amplifiers in sensor circuits. Inspired by the physical principle of noise amplification in analog amplifiers, we divide the network into ISO-dependent and ISO-agnostic branches. Typically, sensor gains exhibit linear growth with ISO. Therefore, we design an ISO-dependent gain layer $g(iso)$ based on the linear gain prior, which is equivalent to the analog gain $K$. In this paper, we define the gain layer $g(iso)$ as a learnable piecewise linear function and initialize it with parameters obtained from noise calibration. 
	By leveraging the explicit modeling of the ISO-aware dual branch network, the proxy model effectively bridges the noise model across different ISO levels while maintaining interpretability.
	
	In summary, the physics-aware proxy model restricts the optimization degrees of freedom based on physical priors, thereby promoting the accuracy of noise modeling.
	
	\subsection{Differentiable Distribution Loss} \label{subsec:DDL}
	The choice of loss function poses a major challenge in learning-based noise modeling, as it is difficult to accurately measure the distance between noise distributions based on a set of noise. Previous methods can only measure the distribution through indirect approaches to train neural networks.
	A common approach is to employ Generative Adversarial Networks (GANs)~\cite{NIPS14/GAN}. GANs employ neural networks to extract noise features and estimate the distance between these features, indirectly quantifying the distance between noise distributions. However, GANs are known to be prone to instability and overfitting, often requiring careful initialization for effective convergence in noise modeling.
	Another common approach is to employ normalizing flow~\cite{NIPS18/Glow}. The invertible neural network of the flow-based model is bijective. By learning the transformation process from unknown real noise to known Gaussian noise, the model can simultaneously learn the reverse process, which is equivalent to noise modeling. However, the strict reversibility also limits the performance of flow-based models, leading to underfitting in noise models.
	
	Our novel noise decoupling strategy introduces pixel-wise noise without spatial dependency, thereby presenting a fresh opportunity to design loss functions for noise modeling.
	We observe that the distribution function of i.i.d. pixel-wise noise can serve as a representation of the noise model itself. Motivated by this insight, we propose the DDL that leverages cumulative distribution functions (CDFs) for precise and interpretable measurements of the noise distribution. Instead of forcing complex data to fit measurement methods, we simplify the data to align with reliable measurement techniques.
	
	\begin{figure}[t!]
		\begin{center}
			\subfigure[Distribution Functions]{\label{fig:DDLa}
				\includegraphics[width=0.474\linewidth]{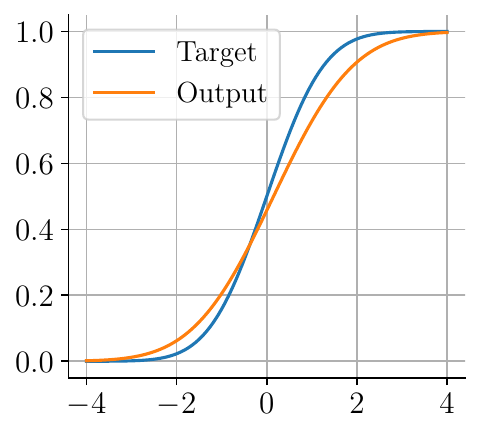}
			}
			\subfigure[Detailed View of CDF]{\label{fig:DDLb}
				\includegraphics[width=0.474\linewidth]{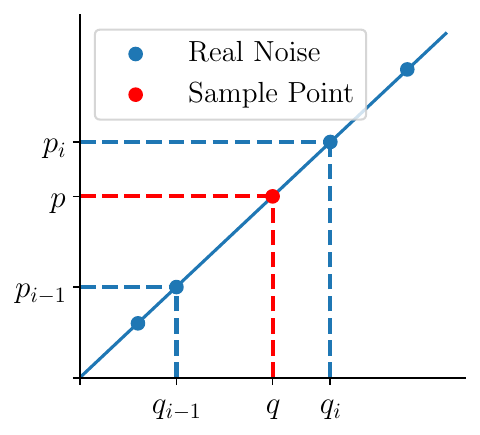}
			}
		\end{center}
		\caption{The principle of DDL. (a) Cumulative distribution function plot, used to observe and measure the distribution distance between the target and the output. (b) Detailed view of CDF, used to illustrate the linear interpolation operation during the query process.}
		\label{fig:DDL}
	\end{figure}
	
	DDL incorporates two novel distribution losses: quantile loss $L_{quantile}$  and CDF loss $L_{CDF}$. Quantiles are values that divide a probability distribution into equal parts, while the CDF characterizes the probability of a random variable taking on a value less than or equal to a given value~\cite{statistical}.
	In Fig.~\ref{fig:DDLa}, quantiles and CDF values form the horizontal and vertical axes of the cumulative distribution function plot, respectively. The CDF plot inspires us to measure the distance between distribution points using differences in their corresponding quantiles and CDF values, which forms the basis for $L_{quantile}$ and $L_{CDF}$.
	
	Notably, to train neural networks with distribution loss functions, the entire loss computation procedure must be fully differentiable, despite the discrete nature of noisy samples. We exploit the property that sorting and normalization operations inherently preserve gradients, allowing us to derive a differentiable approximation of the CDF for these discrete data points. Consequently, this ensures the differentiability of both CDF querying and the overall loss computation pipeline.
	
	For instance, when computing the CDF, given a set of $n$ noisy samples, the CDF of the $i$-th smallest value, $p_i$, is expressed as $F(q_i)=i/n$. To query the CDF, binary search locates the smallest index $i$ such that $q_i$ exceeds a given value $q$. As shown in Fig.~\ref{fig:DDLb}, linear interpolation is then used to calculate the query result $p$ between adjacent samples $q_{i-1}$ and $q_{i}$:
	\begin{equation}
		p = F(q) = (i - \frac{q_{i} - q}{q_{i} - q_{i-1}}) / n,
	\end{equation}
	where negative infinity is appended to the dataset and adjustments are made to support arbitrary queries.
	
	The above procedures involve solely differentiable operations, ensuring the differentiability of the distribution loss function. A similar but omitted approach is applied to derive the quantile loss.
	
	Subsequently, we formally define our quantile loss $L_{quantile}$ and CDF loss $L_{CDF}$. For arbitrary pixel-wise noise $N_{pixel}$, its CDF is denoted as $F$, and its quantile function is denoted as $F^{-1}$. Given pixel-wise noise $N_{pixel}$, high-bit pixel-wise noise $N^{HB}_{pixel}$, and a set of query value sampling points $Q=\{q_k|k=1,2,...,N\}$, the CDF loss is defined as
	\begin{equation}
		L_{CDF} = \sum_{k=1}^{N}|F_{out}(q_{k})-F_{real}(q_{k})|,
	\end{equation}
	where $F_{out}$ and $F_{real}$ represent the CDFs of pixel-wise noise $N_{pixel}$, high-bit pixel-wise noise $N^{HB}_{pixel}$, respectively.
	
	Similarly, the quantile loss $L_{quantile}$ is defined as
	\begin{equation}
		L_{quantile} = \sum_{k=1}^{N}|F_{out}^{-1}(p_{k})-F_{real}^{-1}(p_{k})|,
	\end{equation}
	where $F_{out}^{-1}$ and $F_{real}^{-1}$ represent the quantile functions of pixel-wise noise $N_{pixel}$, high-bit pixel-wise noise $N^{HB}{N}_{pixel}$, respectively.
	
	Each query operation typically provides supervision for only two samples around the query value. Therefore, the sampling strategy for query values plays a crucial role in the efficiency of the loss function. We design a random sampling strategy that introduces slight perturbations to fixed samples, enhancing the robustness of supervision. Initially, we use uniform sampling on the standard normal distribution as a baseline, covering the entire noise distribution in a uniform manner. Gaussian perturbations are then added around the baseline to increase the diversity of supervised points. We clip the query values to avoid numerical overflow.
	
	In summary, the distribution loss function provides explicit and reliable supervision for the noise model, thereby promoting the precision of noise modeling.

	\begin{table*}[t!]
		\small
		\caption{Quantitative results (PSNR/SSIM) of different methods on the ELD dataset and SID dataset. The {\color{red}red} color indicates the best results and the {\color{blue}blue} color indicates the second-best results.}
		\vspace{-6pt}
		\label{tab:compare}
		\setlength{\tabcolsep}{4pt}
		\begin{center}
			{%
				% \vspace{-6pt}
				\begin{tabular}{lcc >{\columncolor{mygray}}c ccc>{\columncolor{mygray}}c}
					\toprule
					{} & \multicolumn{3}{c}{ELD Dataset~\cite{TPAMI21/ELD}} & \multicolumn{4}{c}{SID Dataset~\cite{CVPR18/SID}} \\
					\cmidrule(lr){2-4}\cmidrule(lr){5-8}
					\multirow{-2.5}{*}{Method} & {$\times$100} & {$\times$200} & {Average} & {$\times$100} & {$\times$250} & {$\times$300} & Average \\ 
					\midrule
					Paired Data & {44.47} / {0.9676} & {41.97} / {0.9282} & {43.22} / {0.9479} & {42.06} / {0.9548} & {39.60} / {0.9380} & {36.85} / {\color{blue}0.9227} & {39.32} / {0.9374} \\
					P-G~\cite{P-G} & {42.05} / {0.8721} & {38.18} / {0.7827} & {40.12} / {0.8274} & {39.44} / {0.8995} & {34.32} / {0.7681} & {30.66} / {0.6569} & {34.52} / {0.7666} \\
					ELD~\cite{TPAMI21/ELD} & {45.45} / {0.9754} & {43.43} / {0.9544} & {44.44} / {0.9649} & {41.95} / {0.9530} & {39.44} / {0.9307} & {36.36} / {0.9114} & {39.05} / {0.9303} \\
					SFRN~\cite{ICCV21/SFRN} & {\color{blue}46.38} / {0.9793} & {\color{blue}44.38} / {0.9651} & {\color{blue}45.38} / {0.9722} & {42.81} / {0.9568} & {40.18} / {0.9343} & {37.09} / {0.9175} & {39.82} / {0.9349} \\
					NoiseFlow~\cite{NoiseFlow} & {43.21} / {0.9210} & {40.60} / {0.8638} & {41.90} / {0.8924} & {41.08} / {0.9394} & {37.45} / {0.8864} & {33.53} / {0.8132} & {37.09} / {0.8750} \\
					% CA-GAN~\cite{ECCV20/CAGAN} & & {41.48} / {0.9330} & {39.26} / {0.8771} & {38.66} / {0.9209} & {35.30} / {0.8463} & {32.02} / {0.7685} \\
					Starlight~\cite{CVPR22/DUS} & {43.80} / {0.9358} & {40.86} / {0.8837} & {42.33} / {0.9098} & {40.47} / {0.9261} & {36.26} / {0.8575} & {33.00} / {0.7802} & {36.33} / {0.8494} \\
					LLD~\cite{CVPR23/LLD} & {45.78} / {0.9789} & {44.08} / {0.9635} & {44.93} / {0.9712} & {42.29} / {0.9562} & {40.11} / {0.9373} & {36.89} / {0.9152} & {39.56} / {0.9348} \\
					LRD~\cite{ICCV23/LRD} & {46.16} / {\color{blue}0.9829} & {43.91} / {\color{blue}0.9677} & {45.04} / {\color{blue}0.9753} & {\color{blue}43.16} / {\color{blue}0.9581} & {\color{blue}40.69} / {\color{blue}0.9406} & {\color{blue}37.48} / {0.9190} & {\color{blue}40.24} / {\color{blue}0.9378} \\
					PNNP (Ours) & {\color{red}47.31} / {\color{red}0.9877}& {\color{red}45.47} / {\color{red}0.9791}& {\color{red}46.39} / {\color{red}0.9834}& {\color{red}43.63} / {\color{red}0.9614}& {\color{red}41.49} / {\color{red}0.9498}& {\color{red}38.01} / {\color{red}0.9353}& {\color{red}40.83} / {\color{red}0.9479} \\
					\bottomrule
				\end{tabular}%
			}
		\end{center}
	\end{table*}
	
	\section{Experiment}\label{sec:exp}
	%	In this section, we first introduce the experimental setting including the implementation details and the compared methods (Section~\ref{subsec:setting}).
	%	% Secondly, we show our discovery of dark shading.
	%	Then, we evaluate our method in comparison to other existing methods on public datasets (Section~\ref{subsec:public}). 
	%	After that, we conduct comprehensive ablation studies to analyze our methods (Section~\ref{subsec:ablation}). 
	%	Finally, we evaluate the generalizability of our method on real imaging scenarios of a mobile phone (Section~\ref{subsec:phone}). 
	
	\subsection{Experimental Setting} \label{subsec:setting}
	\subsubsection{Implementation details}
	\noindent \textbf{Noise Modeling.}
	The dark frames for noise modeling on the SID dataset~\cite{CVPR18/SID} and ELD dataset~\cite{CVPR20/ELD,TPAMI21/ELD} are captured using a SonyA7S2 camera, which shares the same sensor as the public datasets but not the same camera.
	We prepare 5 dark frames per ISO for our PNNP. After physics-guided noise decoupling, we randomly crop 1024$\times$1024 patches at each step to train the PPM.
	We train the PPM with 1000 steps per ISO using Adam~\cite{Adam}. We sample 10$^6$ query values per step according to the random sampling strategy of DDL.
	The learning rate follows a variation pattern similar to SGDR~\cite{ICLR17/SGDR}. The base learning rate is set to 1$\times$10$^{-2}$ and the minimum learning rate is set to 1$\times$10$^{-5}$.
	The training setting on the LRID dataset~\cite{TPAMI23/PMN} shares the same procedure.
	
	\begin{figure*}[!t]
		\small
		\setlength\tabcolsep{2pt}
		\renewcommand\arraystretch{0.8}
		\begin{center}
			\begin{tabular}{@{} c c @{}}
				\begin{tabular}{@{} c @{}}
					\includegraphics[width=.3955\linewidth]{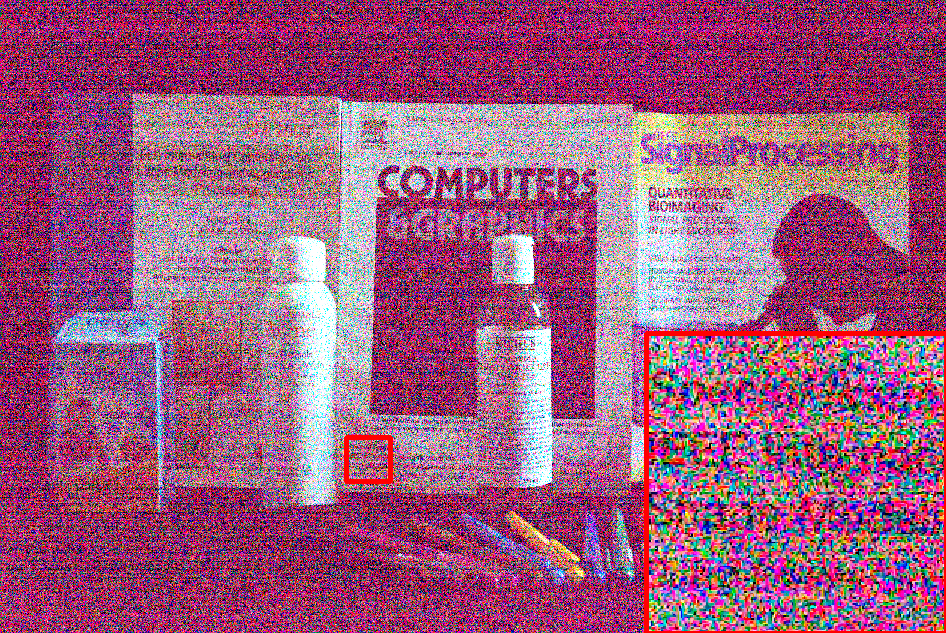} \\
					Noisy Image \\
					21.48/0.1347 \\
					\\
					\includegraphics[width=.3955\linewidth]{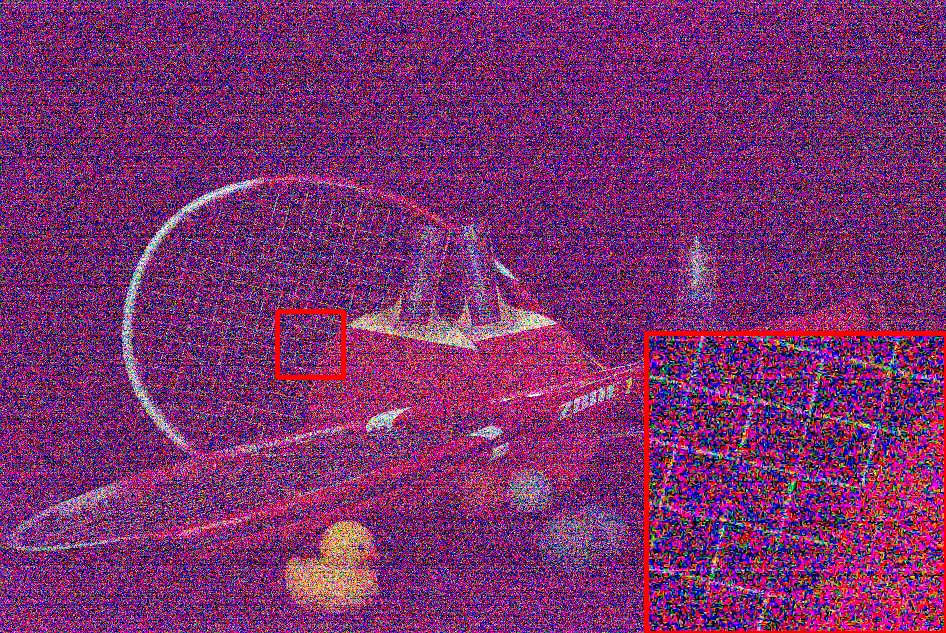} \\
					Noisy Image \\
					22.32/0.0836 \\
				\end{tabular} & 
				\begin{tabular}{@{} c c c c c @{}}
					\includegraphics[width=.1127\linewidth]{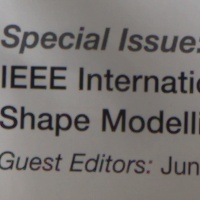} &
					\includegraphics[width=.1127\linewidth]{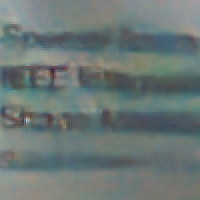} &
					\includegraphics[width=.1127\linewidth]{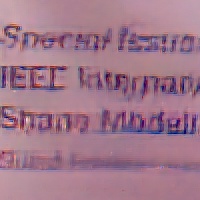} &
					\includegraphics[width=.1127\linewidth]{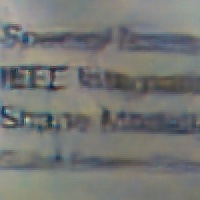} &
					\includegraphics[width=.1127\linewidth]{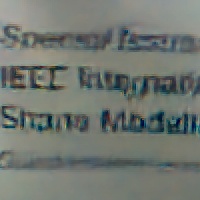} \\
					Reference & Paired Data & P-G\cite{P-G} & ELD\cite{TPAMI21/ELD} & SFRN\cite{ICCV21/SFRN}\\
					PSNR/SSIM & 36.76/0.9043 & 32.68/0.7412 & 37.63/0.9342 & {\color{blue}38.61}/0.9467 \\
					\includegraphics[width=.1127\linewidth]{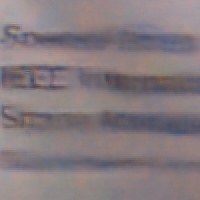} & 
					\includegraphics[width=.1127\linewidth]{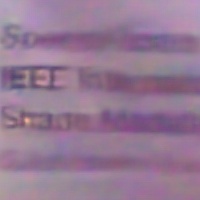} &
					\includegraphics[width=.1127\linewidth]{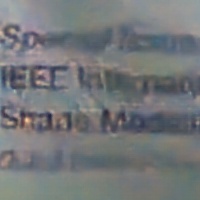} &
					\includegraphics[width=.1127\linewidth]{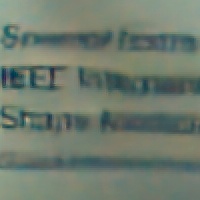} &
					\includegraphics[width=.1127\linewidth]{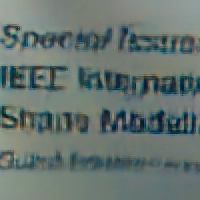} \\
					NoiseFlow\cite{NoiseFlow} & Starlight\cite{CVPR22/DUS} & LLD\cite{CVPR23/LLD} & LRD\cite{ICCV23/LRD} & PNNP (Ours) \\
					34.48/0.8126 & 34.56/0.8414 & 38.36/0.9460 & 38.40/{\color{blue}0.9512} & {\color{red}39.56}/{\color{red}0.9650} \\
					
					\\
					\includegraphics[width=.1127\linewidth]{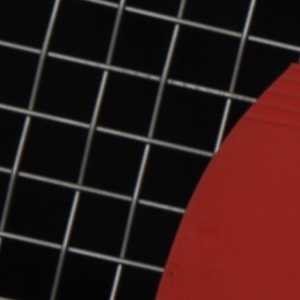} &
					\includegraphics[width=.1127\linewidth]{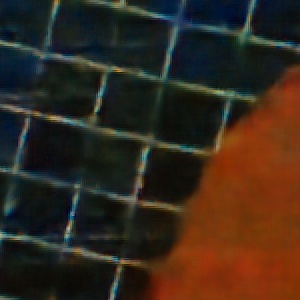} &
					\includegraphics[width=.1127\linewidth]{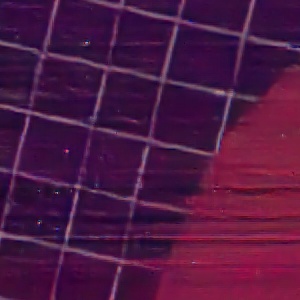} &
					\includegraphics[width=.1127\linewidth]{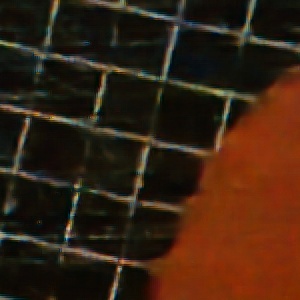} &
					\includegraphics[width=.1127\linewidth]{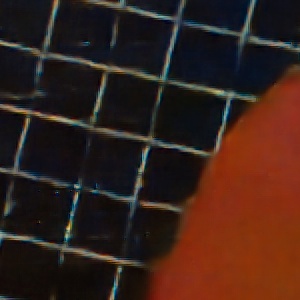} \\
					Reference & Paired Data & P-G\cite{P-G} & ELD\cite{TPAMI21/ELD} & SFRN\cite{ICCV21/SFRN} \\   
					PSNR/SSIM & {40.96}/{0.8788} & {37.41}/{0.7361} & {39.10}/{0.8164} & {43.86}/{\color{blue}0.9504} \\
					\includegraphics[width=.1127\linewidth]{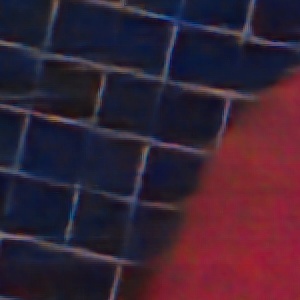} & 
					\includegraphics[width=.1127\linewidth]{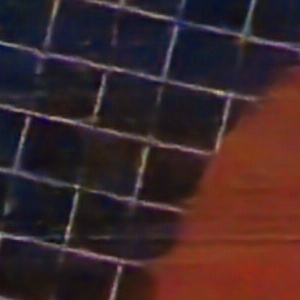} &
					\includegraphics[width=.1127\linewidth]{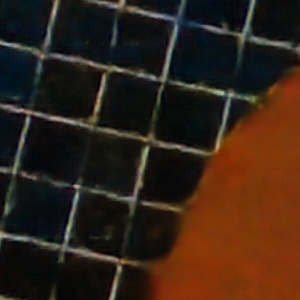} &
					\includegraphics[width=.1127\linewidth]{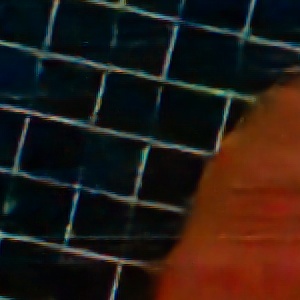} &
					\includegraphics[width=.1127\linewidth]{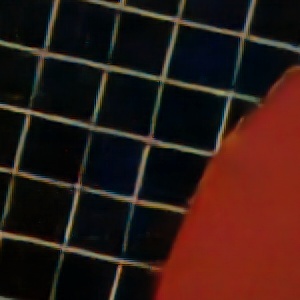}\\
					NoiseFlow\cite{NoiseFlow} & Starlight\cite{CVPR22/DUS} & LLD\cite{CVPR23/LLD} & LRD\cite{ICCV23/LRD} & PNNP (Ours) \\
					{43.07}/{0.9494} & {35.37}/{0.6497} & {43.12}/{0.9366} & {\color{blue}43.89}/{0.9454} & {\color{red}45.80}/{\color{red}0.9773}\\
					
				\end{tabular}
			\end{tabular}
		\end{center}
		\caption{Raw image denoising results on images from the ELD dataset. The {\color{red}red} color indicates the best results and the {\color{blue}blue} color indicates the second-best results. \textbf{(Best viewed with zoom-in)}}
		\label{fig:ELD-compare}
	\end{figure*}

	\setlength{\tabcolsep}{2pt}
	\begin{figure*}[!t]
		\small
		\setlength\tabcolsep{2pt}
		\renewcommand\arraystretch{0.8}
		\begin{center}
			\begin{tabular}{@{} c c @{}}
				\begin{tabular}{@{} c @{}}
					\includegraphics[width=.3955\linewidth]{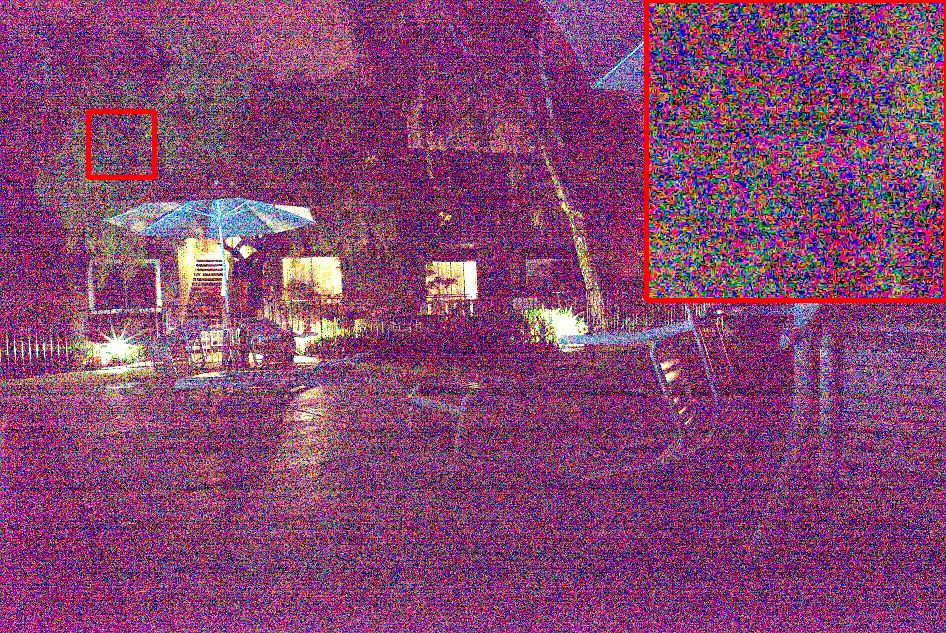} \\
					Noisy Image \\
					18.33/0.088 \\
					\\
					\includegraphics[width=.3955\linewidth]{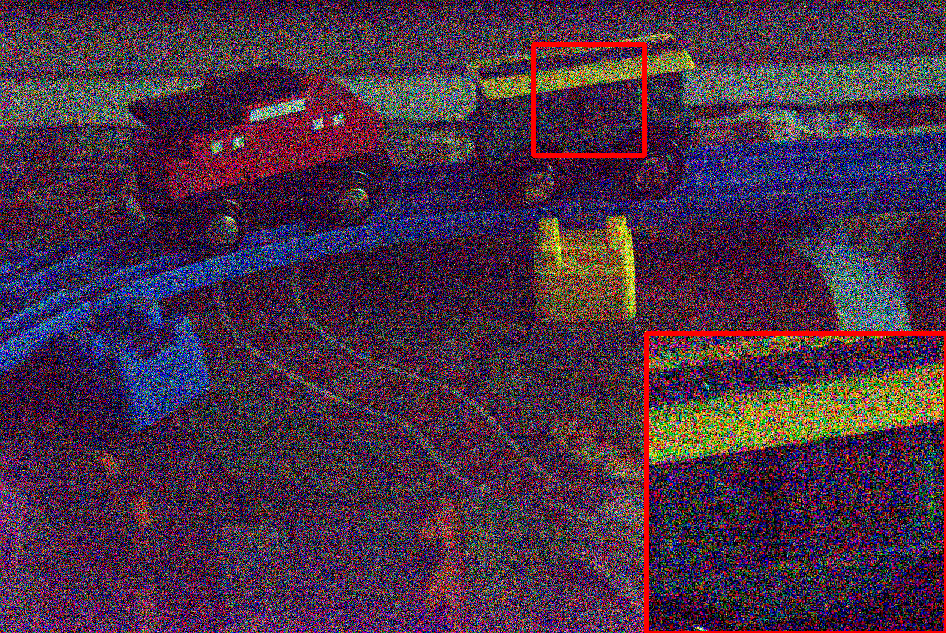} \\
					Noisy Image \\
					19.11/0.1075 \\
				\end{tabular} & 
				\begin{tabular}{@{} c c c c c @{}}
					\includegraphics[width=.1127\linewidth]{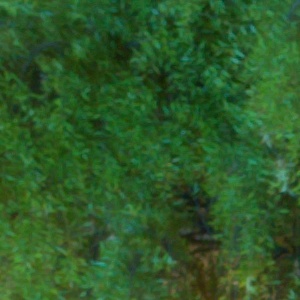} &
					\includegraphics[width=.1127\linewidth]{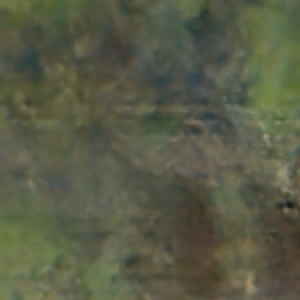} &
					\includegraphics[width=.1127\linewidth]{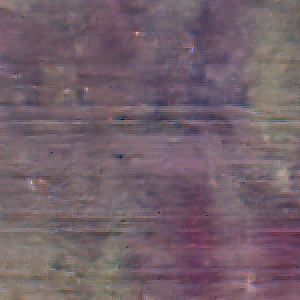} &
					\includegraphics[width=.1127\linewidth]{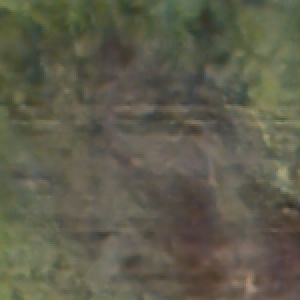} &
					\includegraphics[width=.1127\linewidth]{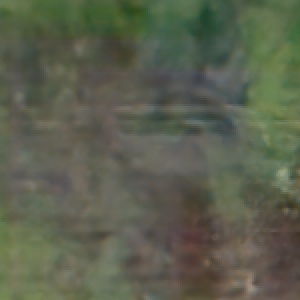} \\
					Reference & Paired Data &P-G\cite{P-G} & ELD\cite{TPAMI21/ELD} & SFRN\cite{ICCV21/SFRN}\\
					PSNR/SSIM & 32.16/{\color{blue}0.8859} & 27.27/0.7293 & 32.21/0.8769 & {\color{blue}32.54}/0.8834 \\    
					
					\includegraphics[width=.1127\linewidth]{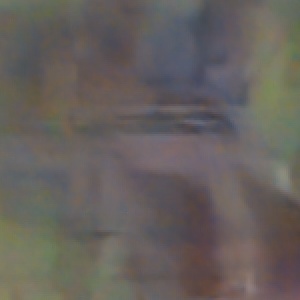} &
					\includegraphics[width=.1127\linewidth]{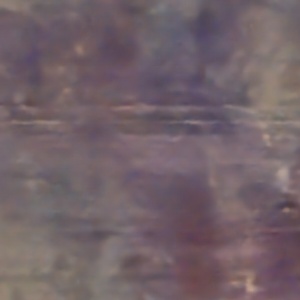} &
					\includegraphics[width=.1127\linewidth]{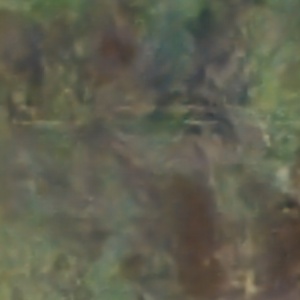} &
					\includegraphics[width=.1127\linewidth]{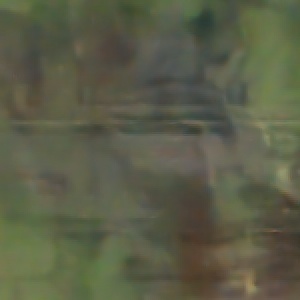} &
					\includegraphics[width=.1127\linewidth]{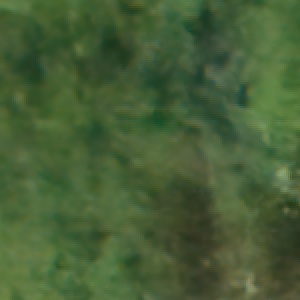} \\
					NoiseFlow\cite{NoiseFlow} & Starlight\cite{CVPR22/DUS} & LLD\cite{CVPR23/LLD} & LRD\cite{ICCV23/LRD} & PNNP (Ours) \\
					29.19/0.7943 & 28.67/0.8014 & 31.93/0.8754 & 32.46/0.8777 & {\color{red}33.22}/{\color{red}0.8970} \\
					\\
					\includegraphics[width=.1127\linewidth]{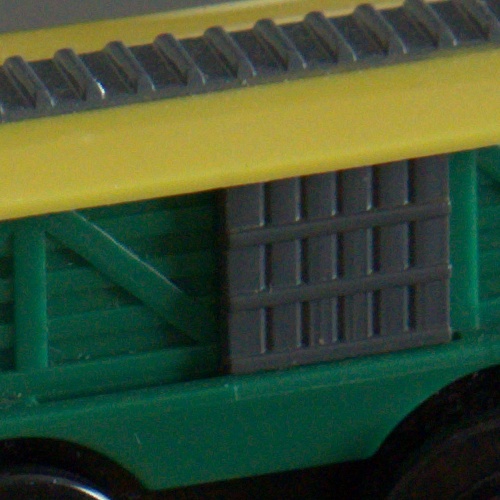} &
					\includegraphics[width=.1127\linewidth]{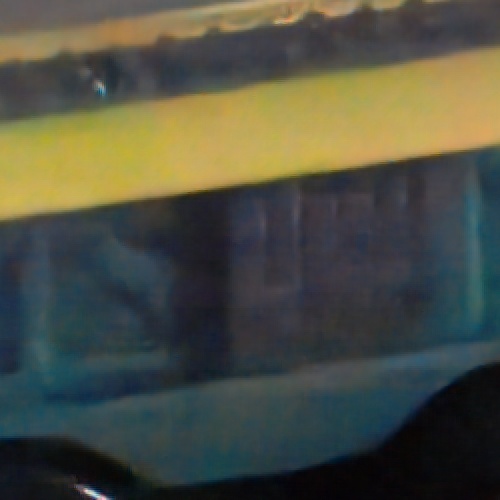} &
					\includegraphics[width=.1127\linewidth]{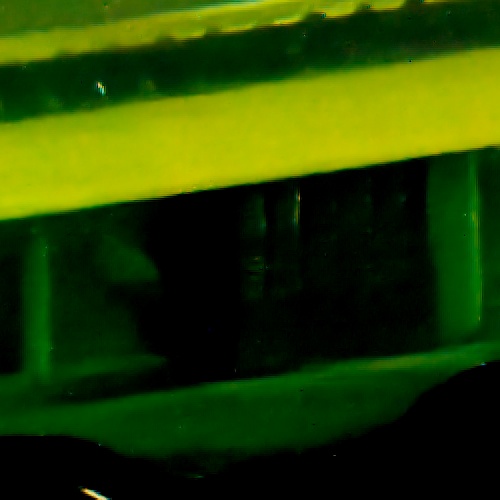} &
					\includegraphics[width=.1127\linewidth]{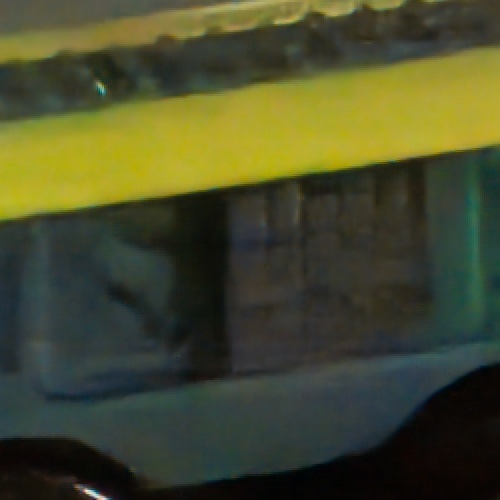} &
					\includegraphics[width=.1127\linewidth]{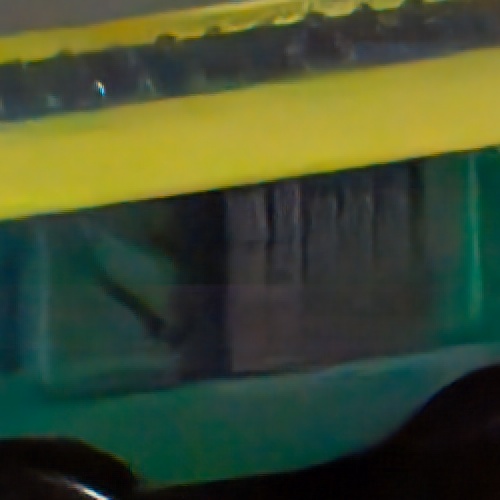} \\
					Reference & Paired Data & P-G\cite{P-G} & ELD\cite{TPAMI21/ELD} & SFRN\cite{ICCV21/SFRN}\\  
					PSNR/SSIM & 39.10/0.9447 & 29.78/0.6097 & 38.39/0.9424 & {\color{blue}39.52}/{\color{blue}0.9465} \\       
					
					\includegraphics[width=.1127\linewidth]{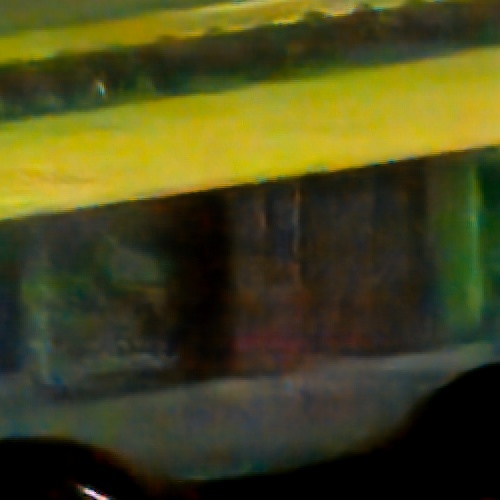} &
					\includegraphics[width=.1127\linewidth]{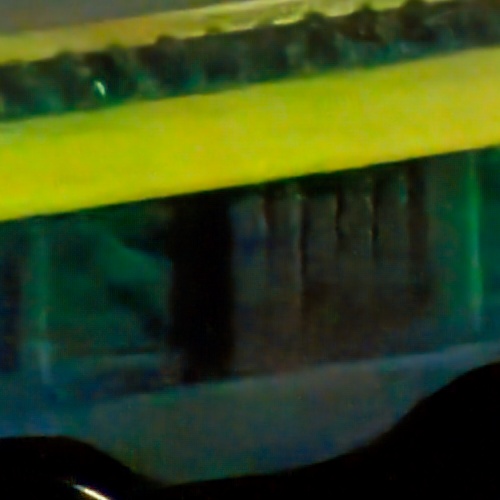} &
					\includegraphics[width=.1127\linewidth]{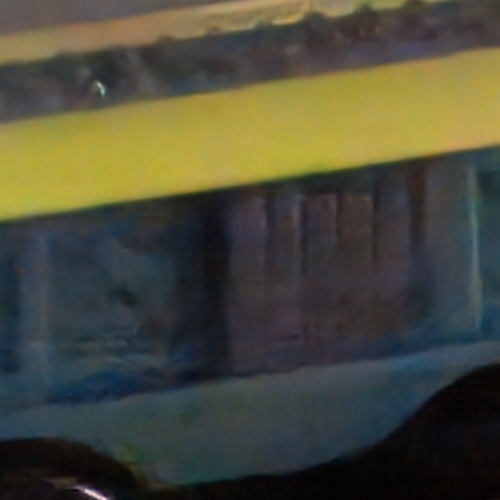} &
					\includegraphics[width=.1127\linewidth]{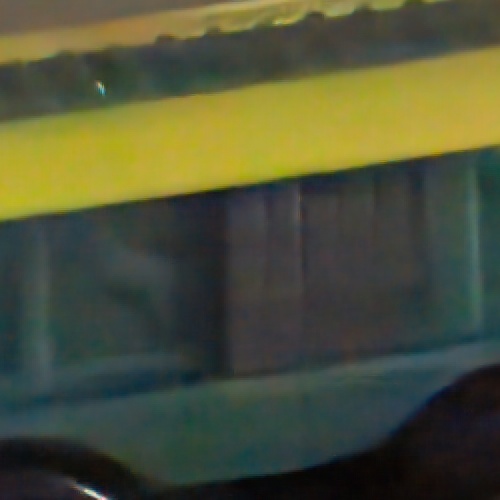} &
					\includegraphics[width=.1127\linewidth]{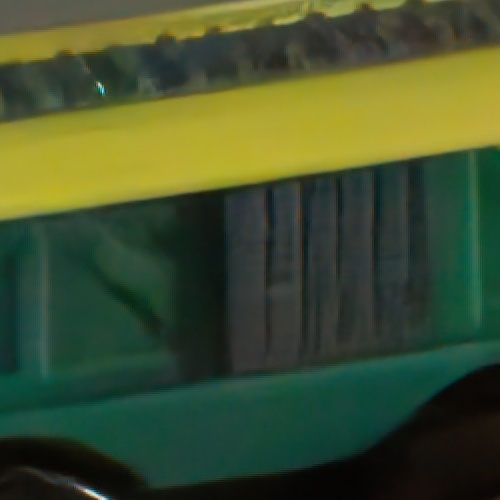}\\
					NoiseFlow\cite{NoiseFlow} & Starlight\cite{CVPR22/DUS} & LLD\cite{CVPR23/LLD} & LRD\cite{ICCV23/LRD} & PNNP (Ours) \\
					34.06/0.8417 & 33.38/0.8228 & 39.45/0.9445 & 39.22/0.9450 & {\color{red}40.51}/{\color{red}0.9533} \\
				\end{tabular}
			\end{tabular}
		\end{center}
		\caption{Raw image denoising results on images from the SID dataset. The {\color{red}red} color indicates the best results and the {\color{blue}blue} color indicates the second-best results. \textbf{(Best viewed with zoom-in)}}
		\label{fig:SID-compare}
	\end{figure*}
	
	% 受启发于前人的工作，我们记录了每种噪声在标定时的误差under the Gaussian error assumption。在合成噪声时，我们基于这些记录的误差扰动噪声参数，这可以让去噪模型更加鲁棒。
	Inspired by previous works~\cite{ICME21/RethinkNM,CVPR20/ELD,TPAMI21/ELD}, we record the errors associated with each type of noise (including linear dark shading) under the Gaussian error assumption during calibration. When synthesizing noise, we perturb the noise parameters based on these recorded errors, thereby endowing the denoising model with enhanced robustness.
	
	\vspace{0.5em}
	\noindent \textbf{Low-light Raw Image Denoising.}
	We employ the same UNet network structure~\cite{Unet} as ELD for denoising.
	% 我们选择了SID数据集中的SonyA7S2子数据集用于训练与验证，并使用ELD数据集中的SonyA7S2子数据集进行验证。
	For the SID dataset and ELD dataset, we utilize raw images from the SID Sony training set to synthesize training data. 
	The quantitative results are reported on the ELD Sony dataset and the whole SID Sony dataset, including validation and testing sets. % We tabulate the performance of denoising models at different exposure ratios in Table.~\ref{tab:compare}.
	For the LRID dataset, we synthesize training data based on the clean images in the training set. The quantitative results are reported on the official testing set.
	We adopt the dark shading correction strategy~\cite{TPAMI23/PMN}, thus we correct the frame-wise noise before inference. For training, we pack the raw images into four channels and crop each image into non-overlapping 512$\times$512 patches. These patches are then randomly rotated and flipped as a batch.
	We visualize the raw images as sRGB images for viewing through RawPy (a Python wrapper for LibRaw) based on the metadata of reference images. All of the quantitative results are computed on the raw images.
	
	We train denoising models using Adam and $L_1$ loss. Each epoch consists of data pairs from various scenarios and exposure ratios. The learning rate will vary with iterations similar to SGDR. 
	The entire training process comprises 1000 epochs, during which a single noisy image under each exposure ratio is traversed in each epoch. The training process includes two distinct stages: a coarse-tuning stage running for 600 epochs, initialized with a base learning rate of 2$\times$10$^{-4}$, and a subsequent fine-tuning stage extending over 400 epochs with a reduced base learning rate of 1$\times$10$^{-4}$. Both stages share the minimum learning rate of 1$\times$10$^{-5}$. The optimizer restarts every 200 epochs, accompanied by a halving of the learning rate upon each restart. 
	
	\vspace{0.5em}
	\noindent \textbf{Compared Methods.}
	In order to demonstrate the reliability of our PNNP, we compare our noise model with the following methods:
	\begin{itemize}[leftmargin=0.5cm]
		\item The denoising model trained with paired real data (\ie, Paired Data), serves as a baseline for evaluating the performance of noise modeling applied to denoising.
		
		\item Physics-based noise modeling methods including P-G (Poisson-Gaussian)~\cite{P-G}, ELD~\cite{TPAMI21/ELD} and SFRN~\cite{ICCV21/SFRN}.
		
		\item Learning-based noise modeling methods including NoiseFlow~\cite{NoiseFlow}, starlight~\cite{CVPR22/DUS}, LLD~\cite{CVPR23/LLD} and LRD~\cite{ICCV23/LRD}.
	\end{itemize}
	
	{\textbf{To ensure a fair comparison, we employ the same UNet-like network structure across all noise modeling methods.}
		The results of P-G, ELD, SFRN, and Paired Data are obtained using the code and weights released by the open-source project~\cite{TPAMI23/PMN}. 
		Due to the lack of implementation details, learning-based noise modeling methods often pose challenges for reproduction to match their claimed performance. Consequently, we only provide the complete results on NoiseFlow based on our implementation.
		The results of starlight and LLD are directly cited from the original authors of LLD~\cite{CVPR23/LLD}, while the results of LRD are obtained using the officially released weights.
		On the LRID dataset, PNNP is compared only with NoiseFlow among learning-based methods.}
	
	It is important to emphasize that precisely measuring noise distribution discrepancies in real-world low-light image denoising datasets is impractical.
	On the one hand, classic distribution metrics, such as Kullback-Leibler Divergence (KLD)~\cite{statistical}, are limited to characterize complex noise distributions under low-light conditions, as they are tailored for pixel-wise i.i.d. noise modeling~\cite{CVPR23/LLD}. On the other hand, inevitable data defects in the dataset, such as misalignments and residual noise, render the differences between clean data and noisy data inadequate for precisely representing the real noise distribution~\cite{TPAMI23/PMN}.
	In response to these measurement complexities, we pivot towards adopting the performance of denoising networks trained on synthetic data generated by different noise modeling methods as the sole criterion for evaluation.
	
	\subsection{Denoising Results}\label{subsec:public}
	\subsubsection{ELD dataset and SID dataset}
	Table~\ref{tab:compare} summarizes the denoising performances over different exposure ratios based on different noise models. Fig.~\ref{fig:ELD-compare} and Fig.~\ref{fig:SID-compare} shows the comparisons on ELD dataset and SID dataset, respectively. 
	
	The model trained with synthetic data based on physics-based noise models exhibits limited effectiveness in removing real-world noise. In low-light scenarios, the synthetic data generated by the P-G~\cite{P-G} lacks authenticity due to significant deviations from the real-world sensor noise model. As a result, denoising results exhibit noticeable color bias and stripe artifacts. While the ELD~\cite{TPAMI21/ELD} enhances the modeling of signal-independent noise upon the P-G, it still falls short compared to the real-world sensor noise model. The absence of fixed-pattern noise modeling in the ELD leads to prominent pattern noise in denoising results and a failure to restore fine textures. In the case of the SFRN~\cite{ICCV21/SFRN}, which synthesizes data by sampling real signal-independent noise, the complexity of real noise poses challenges for the network to learn accurate data mappings. Therefore, persistent fixed-pattern noise remains in denoising results, indicating the potential for further improvement in restoring fine textures. 
	
	Learning-based noise modeling methods face challenges in ensuring the accuracy of the learned noise model.
	NoiseFlow~\cite{NoiseFlow} is limited by the representation ability of flow-based modules, which hinders NoiseFlow to precisely approximate complex noise distributions, resulting in significant blur.
	Starlight~\cite{CVPR22/DUS} seems to have not converged well due to the instability of GAN, resulting in obvious stripes and color bias. 
	LLD~\cite{CVPR23/LLD} reduces the difficulty of noise approximation through noise decoupling. However, the flow-based backbone bound the representation ability of LLD, resulting in slight artifacts and color bias. 
	Based on the Fourier Transformers, LRD~\cite{ICCV23/LRD} stands out as the state-of-the-art GAN-based method. Nonetheless, approximating complex noise distributions with neural networks remains a challenge, as evidenced by obvious stripes artifacts and blur.
	
	In contrast, our method achieves the most exact color and clearest textures in the majority of scenarios in public datasets. Benefiting from accurate and precise noise modeling, the corresponding denoising results are free from obvious residual noise, encompassing complex fixed-pattern noise.
	
	\begin{table*}[t!]
		\caption{Quantitative results (PSNR/SSIM) of different methods on the LRID dataset. The {\color{red}red} color indicates the best results and the {\color{blue}blue} color indicates the second-best results.}
		%\vspace{-6pt}
		\label{tab:IMX686}
		\small
		\setlength{\tabcolsep}{5.5pt}
		\begin{center}
			{%
				%\vspace{-6pt}
				\begin{tabular}{cccccccc}
					\toprule
					{} & {} & 
					{\makebox[0.108\textwidth][c]{NoiseFlow~\cite{NoiseFlow}}} &
					{\makebox[0.108\textwidth][c]{P-G~\cite{P-G}}} &
					{\makebox[0.108\textwidth][c]{ELD~\cite{TPAMI21/ELD}}} &
					{\makebox[0.108\textwidth][c]{SFRN~\cite{ICCV21/SFRN}}} &
					{\makebox[0.108\textwidth][c]{Paired Data}} & 
					{\makebox[0.108\textwidth][c]{PNNP (Ours)}}\\ %\cline{3-8} 
					\multirow{-2}{*}{Dataset} & \multirow{-2}{*}{\makebox[0.06\textwidth][c]{Ratio}} & PSNR / SSIM & PSNR / SSIM & PSNR / SSIM & PSNR / SSIM & PSNR / SSIM & PSNR / SSIM \\ \midrule
					\multirow{6}{*}{LRID-Indoor}
					& $\times$64 & {48.16} / {0.9901} & {46.14} / {0.9872} & {48.19} / {0.9898} & {47.94} / {0.9899} & {\color{red}48.77} / {\color{blue}0.9906} & {\color{blue}48.50} / {\color{red}0.9908} \\
					& $\times$128 & {46.19} / {0.9828} & {44.98} / {0.9809} & {46.55} / {0.9836} & {46.52} / {0.9854} & {\color{red}47.00} / {\color{blue}0.9860} & {\color{blue}46.94} / {\color{red}0.9863} \\  
					& $\times$256 & {43.91} / {0.9698} & {43.31} / {0.9682} & {44.39} / {0.9730} & {\color{blue}44.74} / {\color{blue}0.9789} & {\color{blue}44.74} / {0.9786} & {\color{red}45.06} / {\color{red}0.9797} \\
					& $\times$512 & {41.09} / {0.9442} & {40.80} / {0.9429} & {41.56} / {0.9452} & {\color{blue}42.46} / {\color{blue}0.9652} & {42.40} / {0.9647} & {\color{red}42.64} / {\color{red}0.9662} \\  
					& $\times$1024 & {37.76} / {0.8906} & {37.74} / {0.8905} & {37.50} / {0.8915} & {\color{blue}40.10} / {\color{blue}0.9453} & {40.07} / {0.9437} & {\color{red}40.30} / {\color{red}0.9460} \\ 
					\rowcolor{mygray} % \cline{2-8}
					\cellcolor{white}
					& Average & {43.42} / {0.9555} & {42.59} / {0.9539} & {43.64} / {0.9566} & {44.35} / {\color{blue}0.9729} & {\color{blue}44.60} / {0.9727} & {\color{red}44.69} / {\color{red}0.9738} \\ 
					\midrule
					\multirow{4}{*}{LRID-Outdoor}
					& $\times$64 & {45.34} / {0.9856} & {42.16} / {0.9796} & {45.00} / {0.9841} & {45.05} / {0.9850} & {\color{red}45.84} / {\color{red}0.9876} & {\color{blue}45.62} / {\color{blue}0.9873} \\   
					& $\times$128 & {43.82} / {0.9757} & {41.48} / {0.9709} & {43.48} / {0.9734} & {43.67} / {\color{blue}0.9766} & {\color{red}44.50} / {\color{red}0.9821} & {\color{blue}44.27} / {\color{red}0.9821} \\
					& $\times$256 & {41.92} / {0.9570} & {40.36} / {0.9525} & {41.31} / {0.9450} & {41.89} / {0.9591} & {\color{red}42.66} / {\color{blue}0.9709} & {\color{blue}42.63} / {\color{red}0.9724} \\
					\rowcolor{mygray} % \cline{2-8}
					\cellcolor{white}
					& Average & {43.69} / {0.9728} & {41.33} / {0.9677} & {43.26} / {0.9675} & {43.54} / {0.9736} & {\color{red}44.33} / {\color{blue}0.9802} & {\color{blue}44.17} / {\color{red}0.9806} \\
					\bottomrule
				\end{tabular}%
			}
		\end{center}
	\end{table*}
	
	\begin{figure*}[t!]
		\small %此处写字体大小控制命令
		\setlength\tabcolsep{0.9pt}
		\renewcommand\arraystretch{0.8}
		\begin{center}
			{
				\begin{tabular}{cccccccc}
					\makebox[0.121\textwidth][c]{Input} &
					\makebox[0.121\textwidth][c]{NoiseFlow~\cite{NoiseFlow}}  & 
					\makebox[0.121\textwidth][c]{P-G~\cite{P-G}} &
					\makebox[0.121\textwidth][c]{ELD~\cite{TPAMI21/ELD}} & 
					\makebox[0.121\textwidth][c]{SFRN~\cite{ICCV21/SFRN}} &
					\makebox[0.121\textwidth][c]{Paired Data}  &  
					\makebox[0.121\textwidth][c]{PNNP (Ours)} & 
					\makebox[0.121\textwidth][c]{Reference} \\
					{\includegraphics[width=0.121\textwidth]{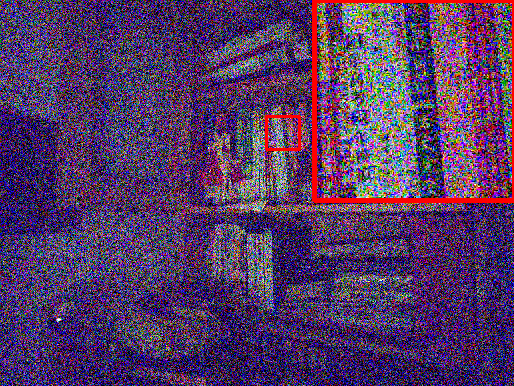}}&
					{\includegraphics[width=0.121\textwidth]{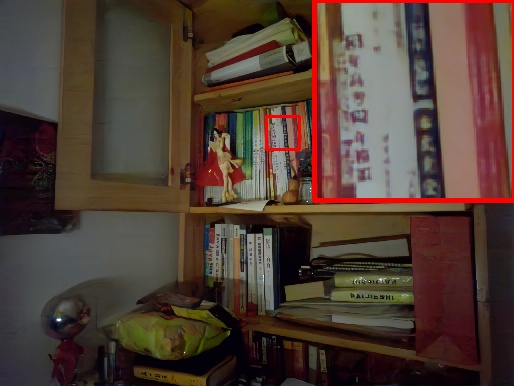}}&
					{\includegraphics[width=0.121\textwidth]{LRID04i/PG}} & 
					{\includegraphics[width=0.121\textwidth]{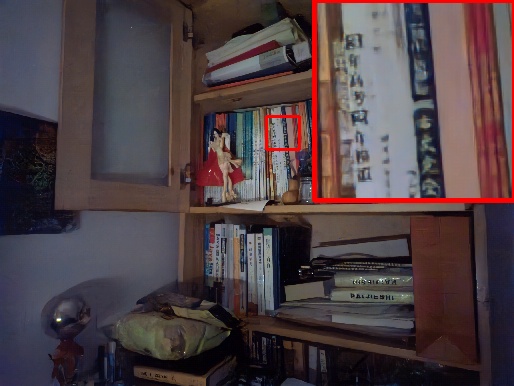}}&
					{\includegraphics[width=0.121\textwidth]{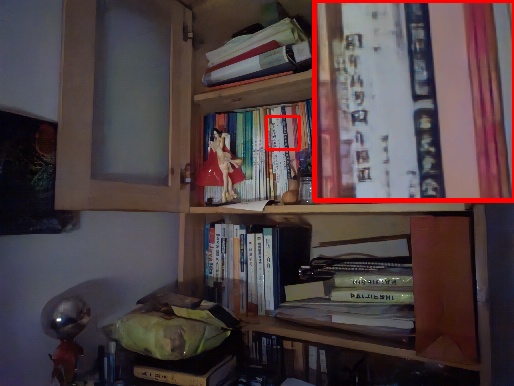}}&
					{\includegraphics[width=0.121\textwidth]{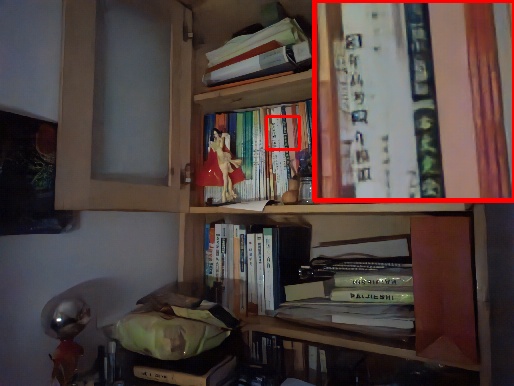}} &
					{\includegraphics[width=0.121\textwidth]{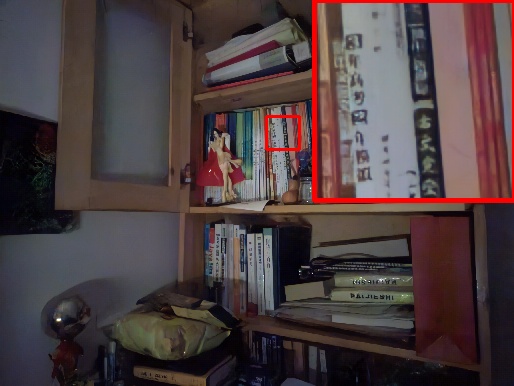}}&
					{\includegraphics[width=0.121\textwidth]{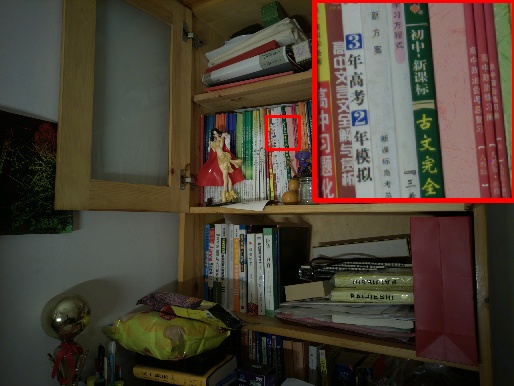}} \\
					{25.82} / {0.2597} & {45.14} / {0.9754} & {44.37} / {0.9721} & {45.27} / {0.9765} & {\color{blue}45.75} / {\color{blue}0.9810} & {45.41} / {0.9800} & {\color{red}46.10} / {\color{red}0.9825} & PSNR / SSIM \\
					\addlinespace[2pt]
					{\includegraphics[width=0.121\textwidth]{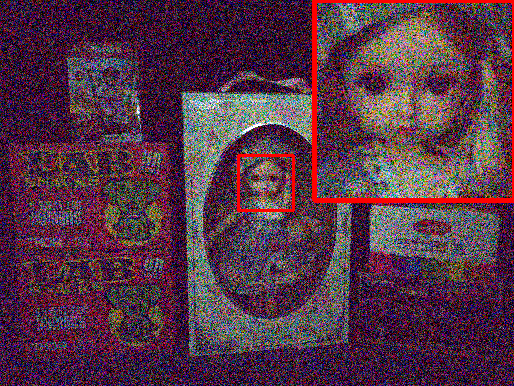}}&
					{\includegraphics[width=0.121\textwidth]{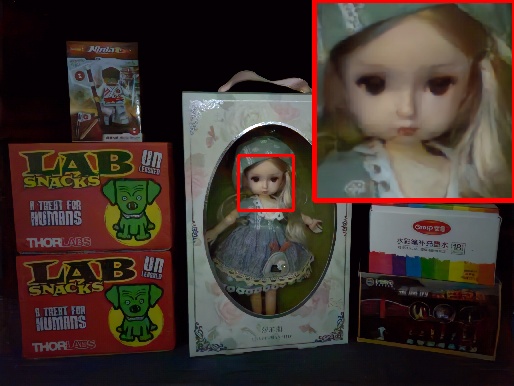}}&
					{\includegraphics[width=0.121\textwidth]{LRID53i/PG}} & 
					{\includegraphics[width=0.121\textwidth]{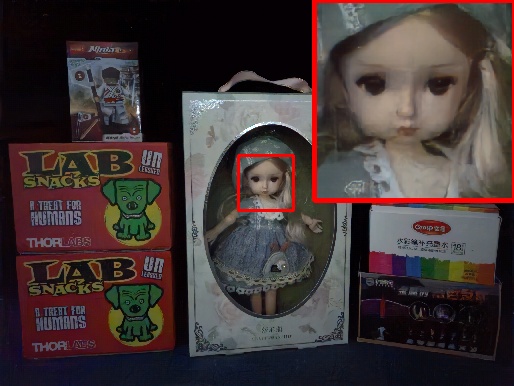}}&
					{\includegraphics[width=0.121\textwidth]{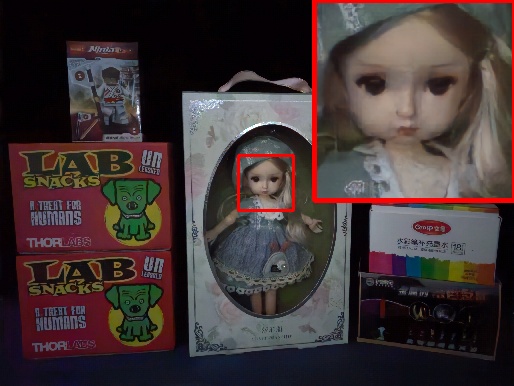}}&
					{\includegraphics[width=0.121\textwidth]{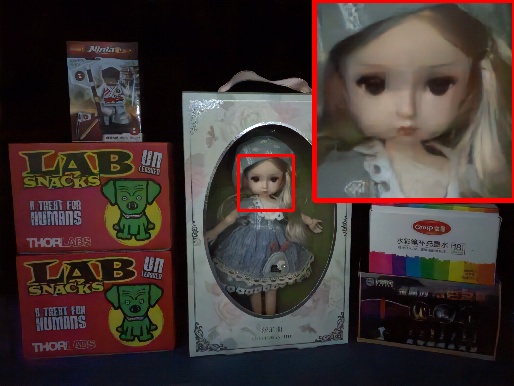}}&
					{\includegraphics[width=0.121\textwidth]{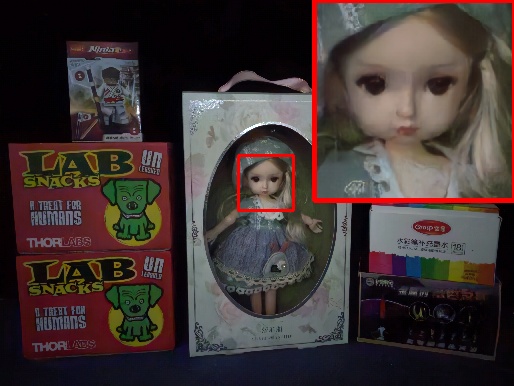}}&
					{\includegraphics[width=0.121\textwidth]{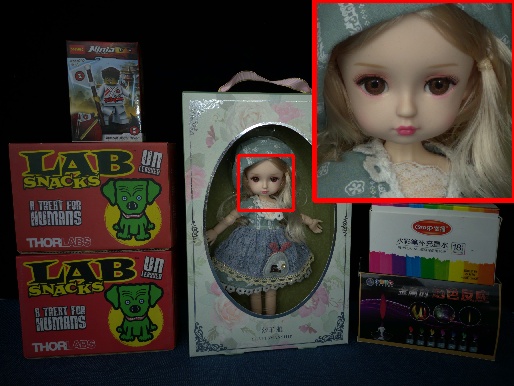}} \\
					{25.36} / {0.2236} & {41.94} / {0.9685} & {41.06} / {0.9653} & {41.92} / {0.9653} & {42.10} / {0.9735} & {\color{blue}42.40} / {\color{blue}0.9744} & {\color{red}42.50} / {\color{red}0.9751} & PSNR / SSIM \\
					% {23.56} / {0.2314} & 43.25 / 0.9815 & {41.42} / {0.9756} & {44.16} / {0.9848} & {44.32} / {0.9860} & {\color{blue}44.57} / {\color{blue}0.9862} & {\color{red}44.98} / {\color{red}0.9876} & PSNR / SSIM \\
					\addlinespace[2pt]
					{\includegraphics[width=0.121\textwidth]{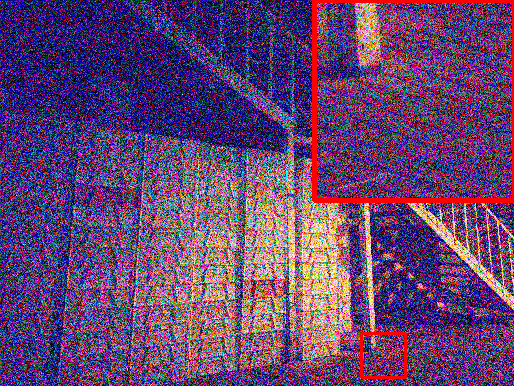}}&
					{\includegraphics[width=0.121\textwidth]{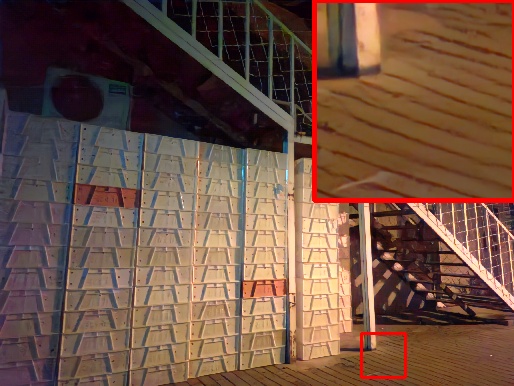}}&
					{\includegraphics[width=0.121\textwidth]{LRID21o/PG}} & 
					{\includegraphics[width=0.121\textwidth]{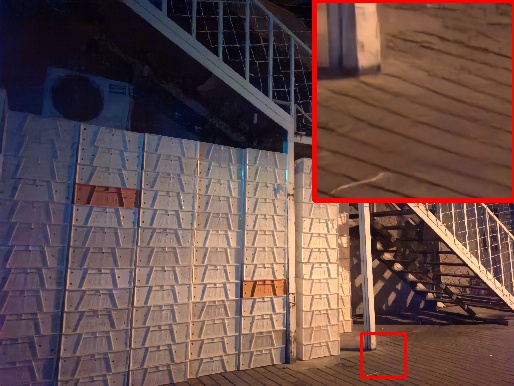}}&
					{\includegraphics[width=0.121\textwidth]{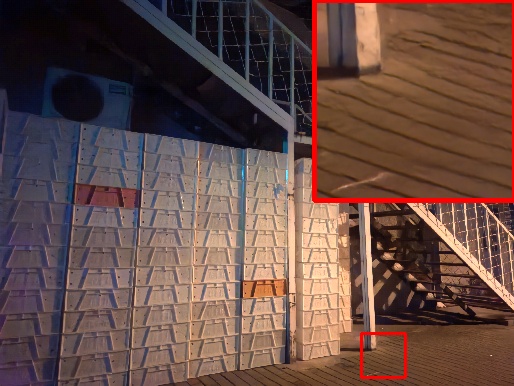}}&
					{\includegraphics[width=0.121\textwidth]{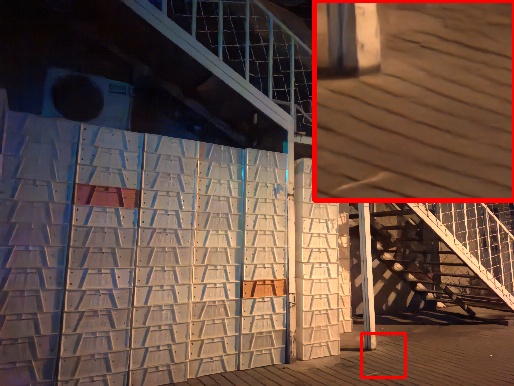}}&
					{\includegraphics[width=0.121\textwidth]{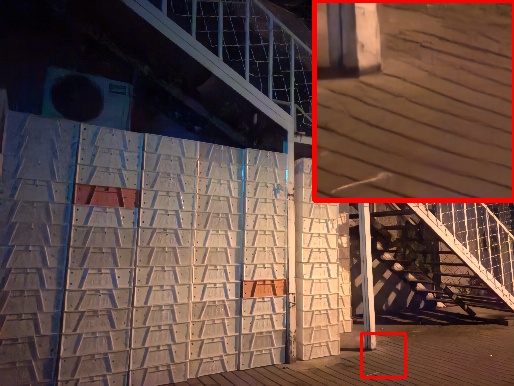}}&
					{\includegraphics[width=0.121\textwidth]{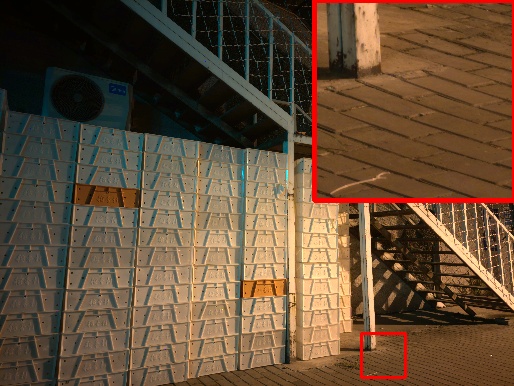}} \\
					{24.40} / {0.2593} & {40.70} / {0.9587} & {39.14} / {0.9506} & {40.49} / {0.9544} & {40.75} / {0.9625} & {\color{blue}40.95} / {\color{blue}0.9638} & {\color{red}41.00} / {\color{red}0.9650} & PSNR / SSIM \\
					\addlinespace[2pt]
					{\includegraphics[width=0.121\textwidth]{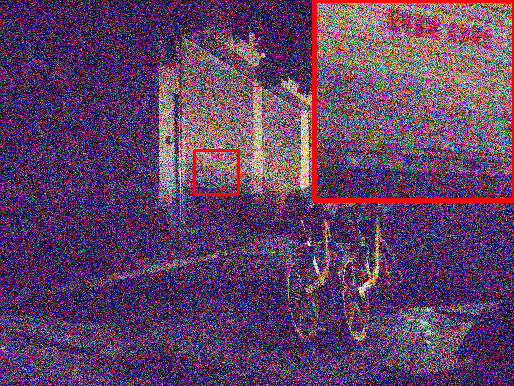}}&
					{\includegraphics[width=0.121\textwidth]{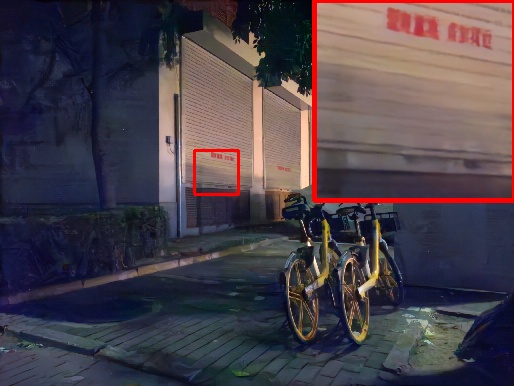}}&
					{\includegraphics[width=0.121\textwidth]{LRID51o/PG}} & 
					{\includegraphics[width=0.121\textwidth]{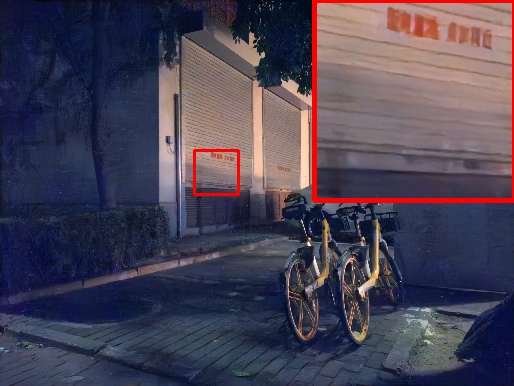}}&
					{\includegraphics[width=0.121\textwidth]{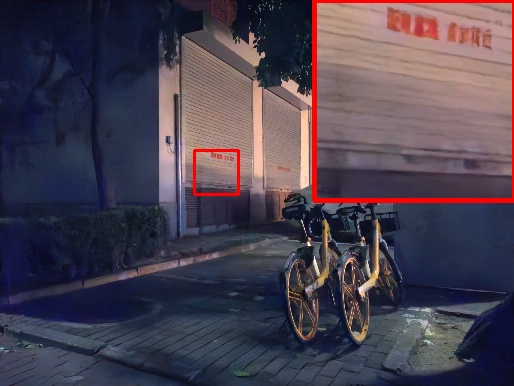}}&
					{\includegraphics[width=0.121\textwidth]{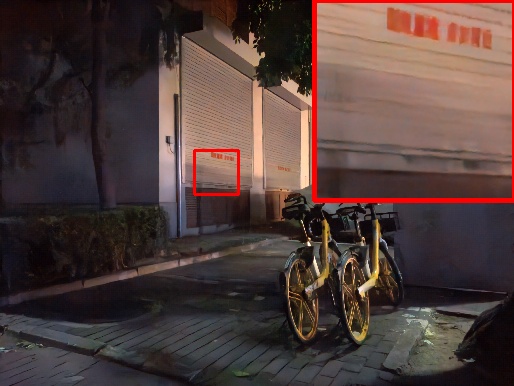}}&
					{\includegraphics[width=0.121\textwidth]{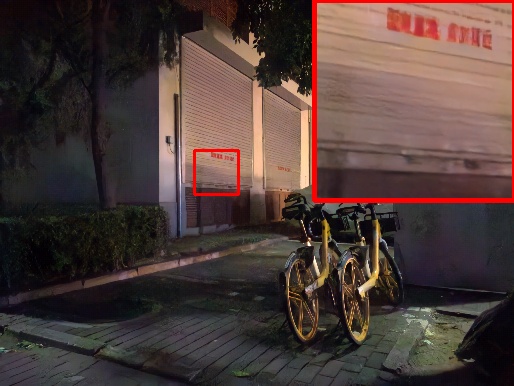}}&
					{\includegraphics[width=0.121\textwidth]{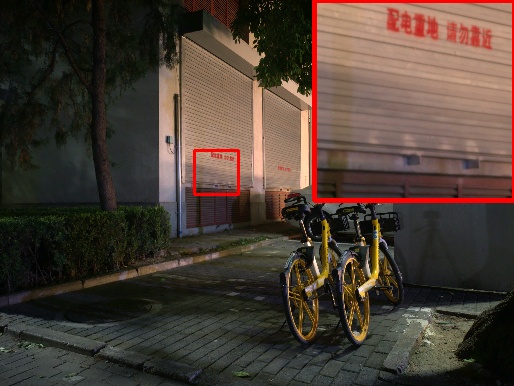}} \\
					{25.55} / {0.2505} & {43.03} / {0.9458} & {42.76} / {0.9468} & {41.19} / {0.9155} & {42.47} / {0.9367} & {\color{blue}44.81} / {\color{blue}0.9752} & {\color{red}44.89} / {\color{red}0.9760} & PSNR / SSIM
				\end{tabular}%
			}
		\end{center}
		\caption{Raw image denoising results on images from the LRID dataset. The {\color{red}red} color indicates the best results and the {\color{blue}blue} color indicates the second-best results. \textbf{(Best viewed with zoom-in)}}
		\label{fig:IMX686}
	\end{figure*}
	
	\subsubsection{LRID dataset}
	Table~\ref{tab:IMX686} summarizes the denoising performances over different exposure ratios based on different noise model. Fig.~\ref{fig:IMX686} shows the comparisons on some representative scenarios.
	The denoising performance rankings among noise modeling methods remain consistent with previous conclusions. Specifically, since the LRID dataset has fewer data defects compared to the SID dataset, the ``Paired Data" serves as an exceptionally high baseline on the LRID dataset. Due to the high-quality data, learning-based methods depending on paired real data also exhibit improved performance. 
	For instance, on the LRID dataset, NoiseFlow can be on par with or sometimes even outperform ELD. As mentioned in Table~\ref{tab:methods}, PNNP is independent of paired real data, therefore deriving minimal benefit from the improvement in data quality. Nonetheless, PNNP still demonstrates comparable denoising performance and further restore more details. The low data dependency and superior denoising performance demonstrate the practicality of our method.
	
	\subsection{Computational Cost}\label{subsec:complexity}
	\begin{table*}[t]
		\small
		\setlength\tabcolsep{8pt}
		\renewcommand\arraystretch{1.0}
		\caption{Comparison of computational costs among different noise modeling methods. ``Available" indicates that the process (network structure) for noise modeling is clearly known. The standard deviations of the runtimes are presented after ``$\pm$''. ``PSNR" corresponds to the denoising performance of each method on the ELD datasets. The {\textbf{bold}} metrics indicate the best results.}
		\label{tab:complexity}
		\centering
		\begin{tabular}{cccccccc}
			\toprule
			Approach & Method & Available & Parameters & Macs & Runtime (ms) & Device & PSNR \\
			\midrule
			\multirow{3}{*}{\makecell[c]{Physics-based\\Method}}
			&P-G~\cite{P-G} & \Checkmark & - & - & \makecell[c]{208 $\pm$ 31} & \makecell[c]{CPU} & 40.12 \\
			&ELD~\cite{TPAMI21/ELD} & \Checkmark & - & - & 1533 $\pm$ 181 & CPU & 44.44 \\
			&SFRN~\cite{ICCV21/SFRN} & \Checkmark & - & - & 3654 $\pm$ 159 & CPU & 45.38 \\
			\midrule
			\multirow{5.5}{*}{\makecell[c]{Learning-based\\Method}}
			&NoiseFlow~\cite{NoiseFlow} & \Checkmark & 2.82K & \textbf{13.57G} & \textbf{28.34 $\pm$ 1.63} & GPU & 41.90 \\
			&Starlight~\cite{CVPR22/DUS} & \Checkmark & 3.33M & 1095G & 111.96 $\pm$ 10.74 & GPU & 42.33 \\
			&LLD~\cite{CVPR23/LLD} & \XSolidBrush & - & - & - & GPU & 44.93 \\
			&LRD~\cite{ICCV23/LRD} & \XSolidBrush & - & - & - & GPU & 45.04 \\
			&PNNP (Ours) & \Checkmark & \textbf{2.30K} & 25.41G & 62.07 $\pm$ 0.18 & GPU & \textbf{46.39} \\
			\bottomrule
		\end{tabular}
		% 在表格下方添加注释
		%		\begin{flushleft}
			%			$^1$ ``Available" indicates that the process (network structure) for noise modeling is clearly known. \\
			%			$^2$ The standard deviations of the runtimes are presented after ``$\pm$''.\\
			%			$^3$ ``PSNR" corresponds to the denoising performance of each method on the ELD datasets.  
			%		\end{flushleft}
	\end{table*}
	
	\begin{figure*}[t!]
		\small
		\centering
		\renewcommand{\arraystretch}{0.5} % 调整行间距
		\setlength{\tabcolsep}{2pt}
		\begin{tabular}{c c c c c}
			& Dark Frame Noise$\qquad$ & Frame-wise Noise$\qquad$ & Band-wise Noise$\qquad$ & Pixel-wise Noise$\qquad$ \\
			\makecell[c]{\rotatebox[origin=c]{90}{Real}}
			& \makecell{
				\includegraphics[
				trim=5pt 12pt 5pt 8pt, 
				clip, 
				width=0.23\linewidth
				]{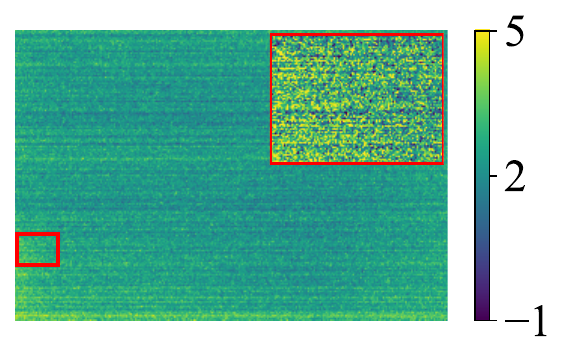}
			} 
			& \makecell{
				\includegraphics[
				trim=5pt 12pt 5pt 8pt, 
				clip, 
				width=0.23\linewidth
				]{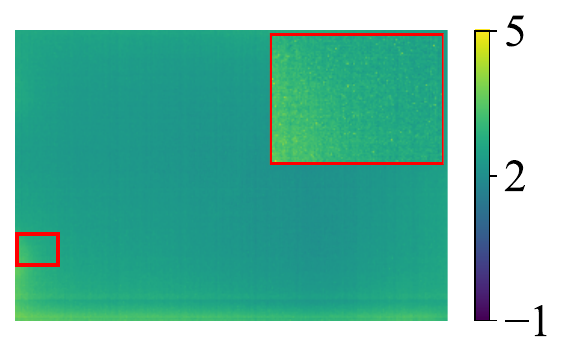}
			} 
			& \makecell{
				\includegraphics[
				trim=5pt 12pt 5pt 8pt, 
				clip, 
				width=0.23\linewidth
				]{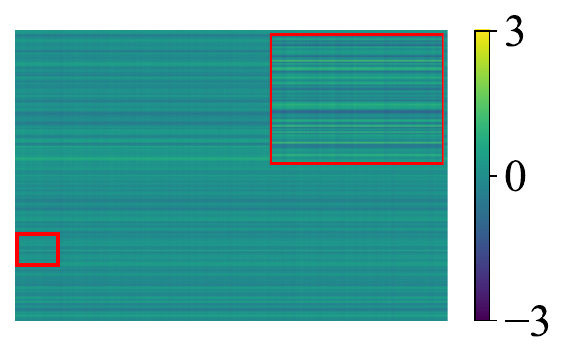}
			} 
			& \makecell{
				\includegraphics[
				trim=5pt 12pt 5pt 8pt, 
				clip, 
				width=0.23\linewidth
				]{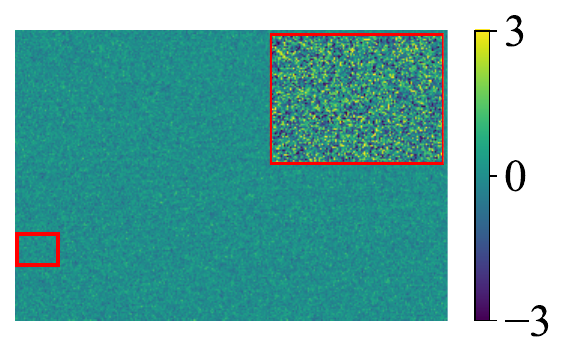}
			} \\
			%& KLD / R$^2\quad$ & KLD / R$^2\quad$ & KLD / R$^2\qquad$ & KLD / R$^2\qquad$ \\
			\addlinespace[2pt]
			\makecell[c]{\rotatebox[origin=c]{90}{PNNP (Ours)}}
			& \makecell{
				\includegraphics[
				trim=5pt 12pt 5pt 8pt, 
				clip, 
				width=0.23\linewidth
				]{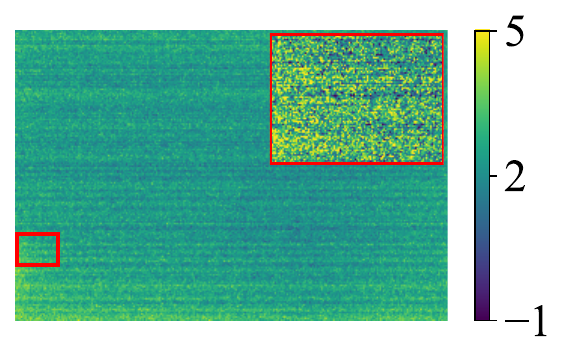}
			}
			& \makecell{
				\includegraphics[
				trim=5pt 12pt 5pt 8pt, 
				clip, 
				width=0.23\linewidth
				]{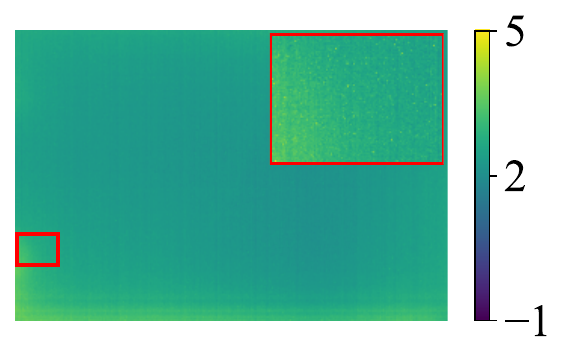}
			} 
			& \makecell{
				\includegraphics[
				trim=5pt 12pt 5pt 8pt, 
				clip, 
				width=0.23\linewidth
				]{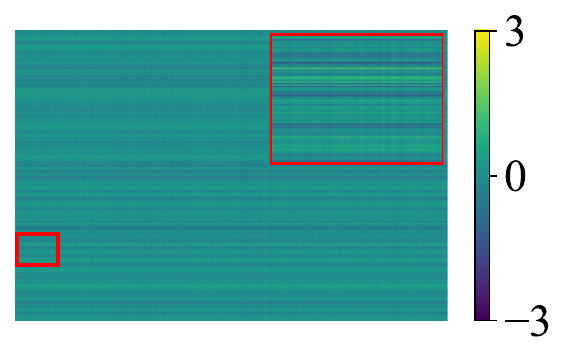}
			} 
			& \makecell{
				\includegraphics[
				trim=5pt 12pt 5pt 8pt, 
				clip, 
				width=0.23\linewidth
				]{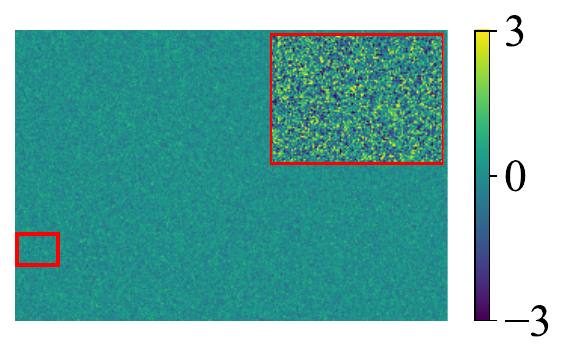}
			} \\
			& 0.00230 / 0.9985$\qquad$ & 0.00365 / 0.9664$\qquad$ & 0.00326 / 0.9975$\qquad$ & 0.00041 / 0.9993$\qquad$
			% & 0.00063 / 0.9996$\qquad$ & 0.00257 / 0.9799$\qquad$ & 0.00038 / 0.9750$\qquad$ & 0.00072 / 0.9997$\qquad$ \\3200
		\end{tabular}
		\caption{Comparison between real noise and PNNP noise across different levels of noise at ISO-1600. We provide corresponding KLD and R$^2$ at the bottom of the figures. {\color{red}Red} boxes highlight zoomed-in regions for detailed observation.}
		\label{fig:pnd_gen}
	\end{figure*}
	
	{Table~\ref{tab:complexity} presents a detailed comparison of computational costs among different noise modeling methods. The runtime reported in the table corresponds to the time required to generate a noisy raw image with 12 million pixels, which is also equivalent to the data volume processed in a single training iteration. Noise generation is performed 1000 times for each method, and the mean and standard deviation of the runtime are reported based on the statistical results. All runtimes are measured on a server equipped with an NVIDIA GeForce RTX 4090 GPU and an Intel Xeon Gold 6148 CPU. 
		We also present the denoising performance of each method as a reference. All methods use the same UNet-like network structure for denoising, ensuring a fair comparison with identical complexity. The denoising model consists of 7.76M parameters and requires 629 GMacs to process a noisy raw image with 12 million pixels, achieving a runtime of 40.02 $\pm$ 3.74 ms.
		
		P-G, ELD, and SFRN are physics-based noise modeling methods, whose parameters are not comparable to those of learning-based methods. Since the code and network structure for noise modeling in LLD and LRD are not publicly available, direct measurement of their computational costs is infeasible.
		The runtime bottleneck for physics-based methods primarily lies in the CPU. Although parallel acceleration can be applied in practice, the significant computational overhead of ELD and SFRN leads to relatively high runtimes. In contrast, learning-based methods benefit from GPU acceleration, achieving faster runtimes. Notably, PNNP delivers superior denoising performance with minimal parameters and the second-lowest runtime.
	}
	
	\subsection{Ablation Study}\label{subsec:ablation}
	
	We conduct ablation studies from both the perspectives of noise modeling and image denoising to evaluate the effectiveness of our proposed contributions: PND, PPM, and DDL. The quantitative results of the ablation studies are shown in Table~\ref{tab:ablation}. We visualize a representative comparison of ablation studies at ISO-1600 in Fig.~\ref{fig:pnd_gen}, Fig.\ref{fig:ab_score} and Fig.~\ref{fig:ablation}. All the ablation studies are performed on the SonyA7S2 camera.
	
	\setlength{\tabcolsep}{7pt}
	\begin{table*}[]
		\small
		\caption{Ablation studies of proposed methods under different conditions. The metrics listed in the table are all averages over the dataset. The {\textbf{bold}} metrics indicate the best results.}
		\label{tab:ablation}
		% \resizebox{\textwidth}{!}
		\centering
		{%
			\begin{tabular}{clcccc}
				\toprule
				\multirow{3.5}{*}{\makebox[0.07\linewidth]{Method}} & \multirow{3.5}{*}{\makebox[0.19\linewidth][l]{Condition}} & \multicolumn{2}{c}{Noise Modeling} & \multicolumn{2}{c}{Image Denoising} \\
				\cmidrule(lr){3-4} \cmidrule(lr){5-6}
				&  & Pixel-wise Noise & Dark Frame Noise & ELD~\cite{TPAMI21/ELD} & SID~\cite{CVPR18/SID} \\
				&  & KLD$\downarrow$ / R$^{2}\uparrow$ & KLD$\downarrow$ / R$^{2}\uparrow$ & PSNR$\uparrow$ / SSIM$\uparrow$ & PSNR$\uparrow$ / SSIM$\uparrow$ \\
				\midrule
				& Gaussian & 0.00730 / 0.9682 & 0.02252 / 0.9603 & 40.12 / 0.8274 & 37.05 / 0.8255 \\
				& ELD & 0.01342 / 0.9837 & 0.02464 / 0.9685 & 44.44 / 0.9649 & 39.05 / 0.9303 \\
				& w/o Noise Decoupling & 0.27335 / 0.9700 & 0.24726 / 0.9538 & 44.26 / 0.9591 & 38.45 / 0.9164 \\
				\rowcolor{mygray} \cellcolor{white}
				\multirow{-4}{*}{PND} & PNNP (Ours) & \textbf{0.00048} / \textbf{0.9995} & \textbf{0.01775} / \textbf{0.9789} & \textbf{46.39} / \textbf{0.9834} & \textbf{40.83} / \textbf{0.9479}\\
				\midrule
				& w/ Flow-based Module & 0.00186 / 0.9932 & 0.01947 / 0.9724 & 46.00 / 0.9797 & 40.53 / 0.9444 \\
				& w/ 3$\times$3 Conv & 0.00049 / 0.9994 & \textbf{0.01775} / \textbf{0.9789} & 44.69 / 0.9539 & 39.68 / 0.9213 \\
				& w/o Dual Branch & 0.00500 / 0.9961 & 0.01784 / \textbf{0.9789} & 45.89 / 0.9727 & 40.55 / 0.9388 \\
				\rowcolor{mygray} \cellcolor{white}
				\multirow{-4}{*}{PPM} & PNNP (Ours) & \textbf{0.00048} / \textbf{0.9995} & \textbf{0.01775} / \textbf{0.9789} & \textbf{46.39} / \textbf{0.9834} & \textbf{40.83} / \textbf{0.9479}\\
				\midrule
				% \cmidrule(lr){1-6} \morecmidrules \cmidrule(lr){1-6}
				& w/o Quantile Loss & 0.00203 / 0.9876 & 0.01986 / 0.9682 & 46.05 / 0.9806 & 40.67 / 0.9457 \\
				& w/o CDF Loss & 0.00057 / 0.9994 & \textbf{0.01775} / \textbf{0.9789} & 46.16 / 0.9818 & 40.71 / 0.9466 \\
				% & w/o Random Sampling & 0.0241 / {\color{black}0.9996} & 0.0205 / 0.9784 & 45.77 / 0.9726 & 40.59 / 0.9392 \\
				\rowcolor{mygray} \cellcolor{white}
				\multirow{-3}{*}{DDL} & PNNP (Ours) & \textbf{0.00048} / \textbf{0.9995} & \textbf{0.01775} / \textbf{0.9789} & \textbf{46.39} / \textbf{0.9834} & \textbf{40.83} / \textbf{0.9479}\\
				\bottomrule
			\end{tabular}%
		}
	\end{table*}

	\begin{figure*}[t!]
		\small %此处写字体大小控制命令
		\setlength\tabcolsep{4pt}
		\renewcommand\arraystretch{0.8}
		\begin{center}
			{%
				\begin{tabular}{ccc}
					\subfigure[Gaussian]
					{\includegraphics[width=0.315\linewidth]{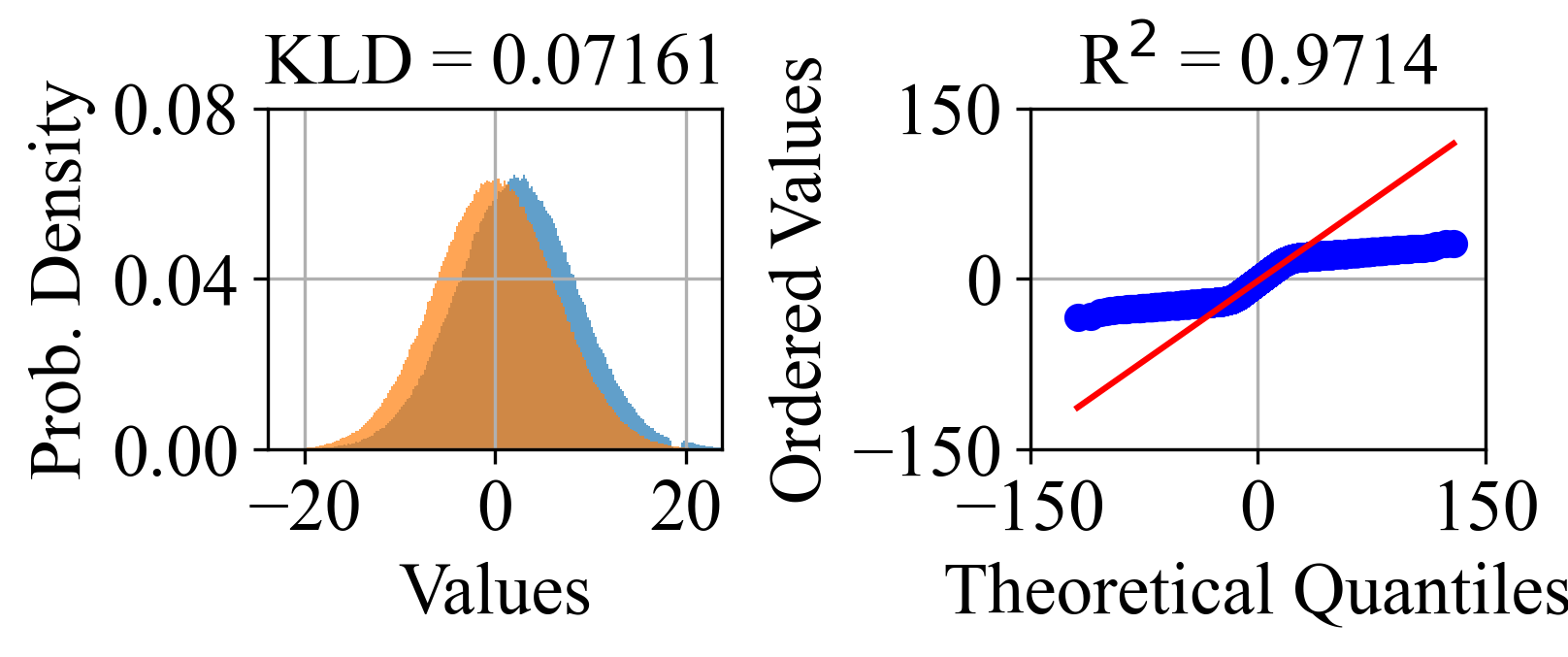}\label{fig:ab_a}} & 
					\subfigure[ELD]
					{\includegraphics[width=0.315\linewidth]{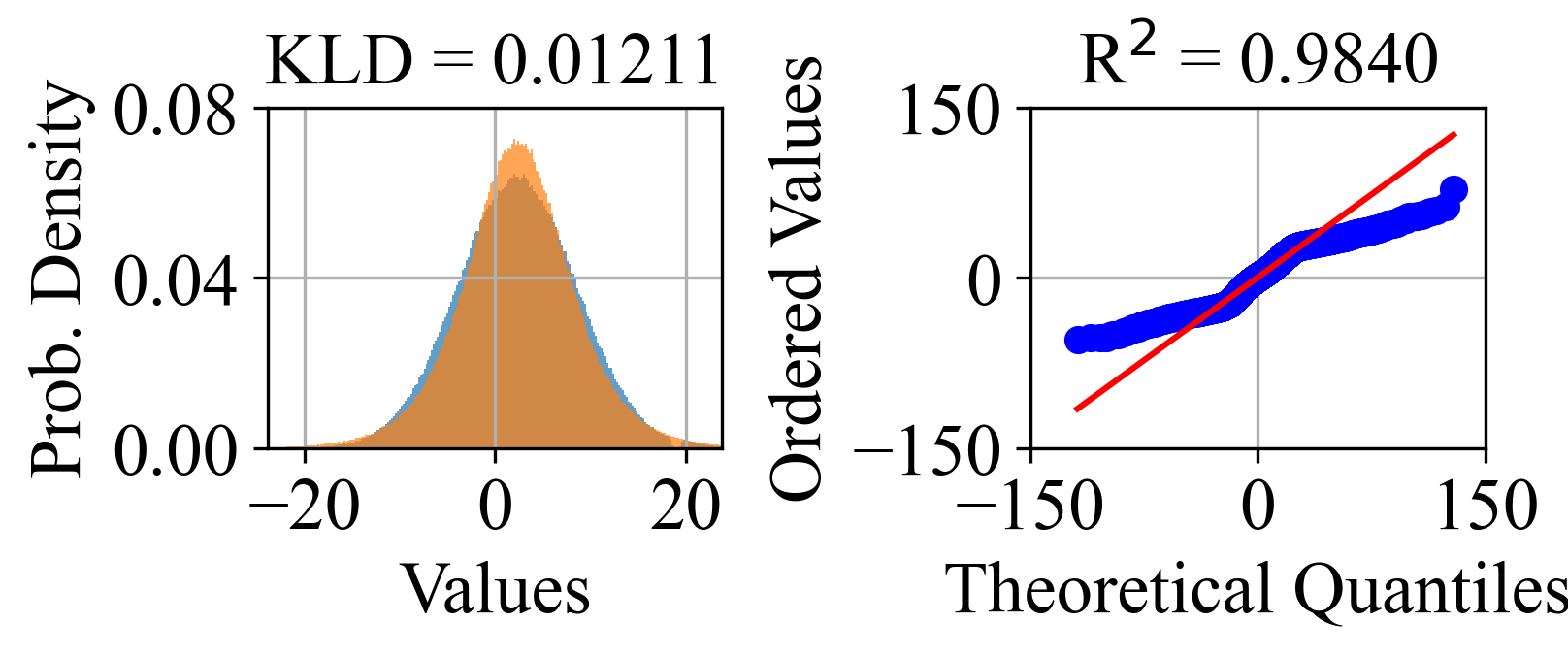}\label{fig:ab_b}} & 
					\subfigure[w/o Noise Decoupling]
					{\includegraphics[width=0.315\linewidth]{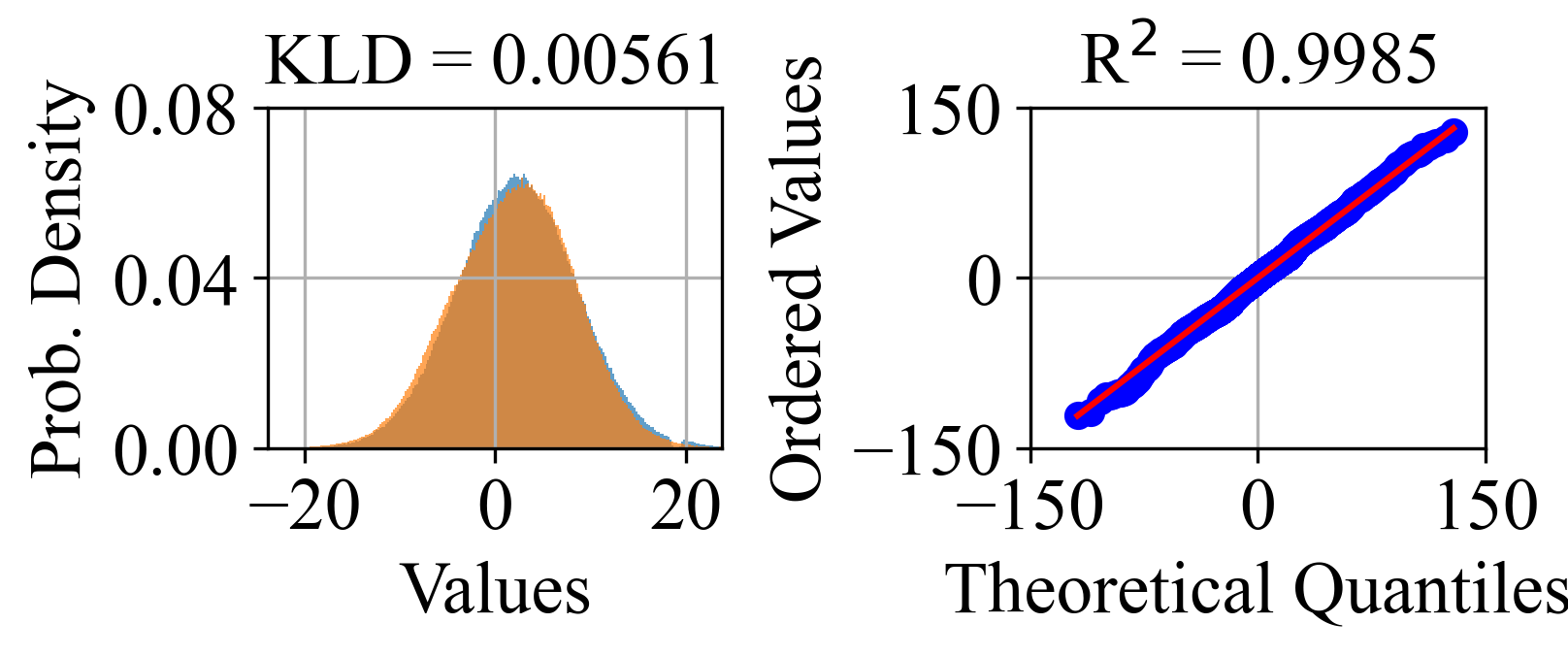}\label{fig:ab_c}} \\
					\subfigure[w/ Flow-based Module]
					{\includegraphics[width=0.315\linewidth]{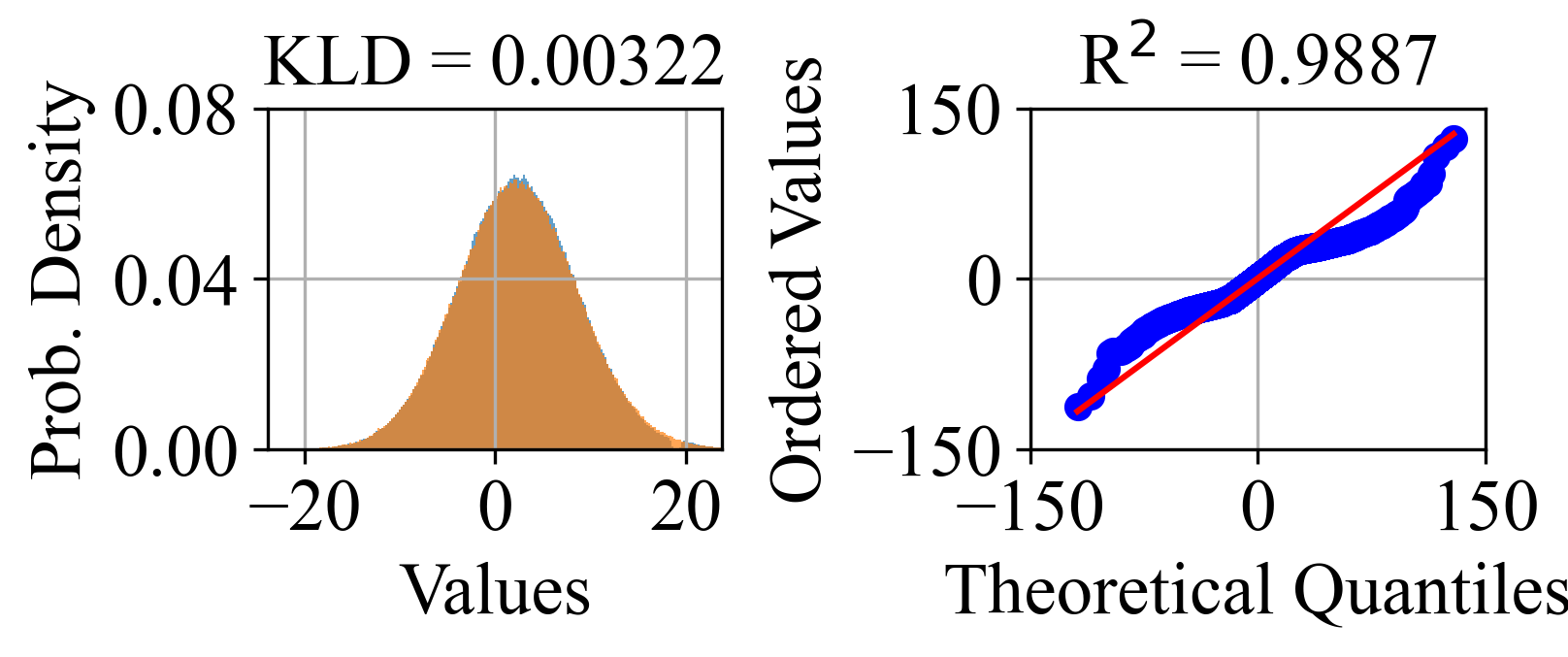}\label{fig:ab_d}} &
					\subfigure[w/ Conv3$\times$3]
					{\includegraphics[width=0.315\linewidth]{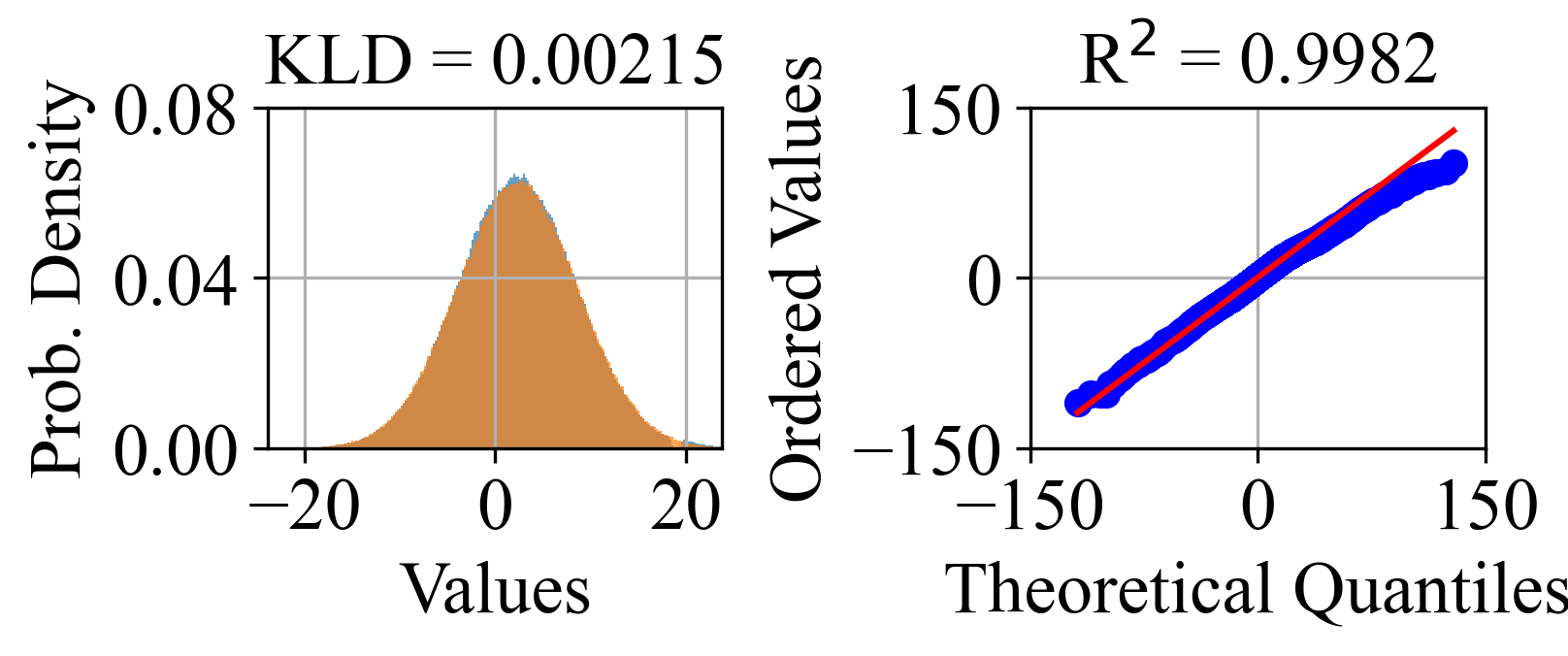}\label{fig:ab_e}} & 
					\subfigure[w/o Dual Branch]
					{\includegraphics[width=0.315\linewidth]{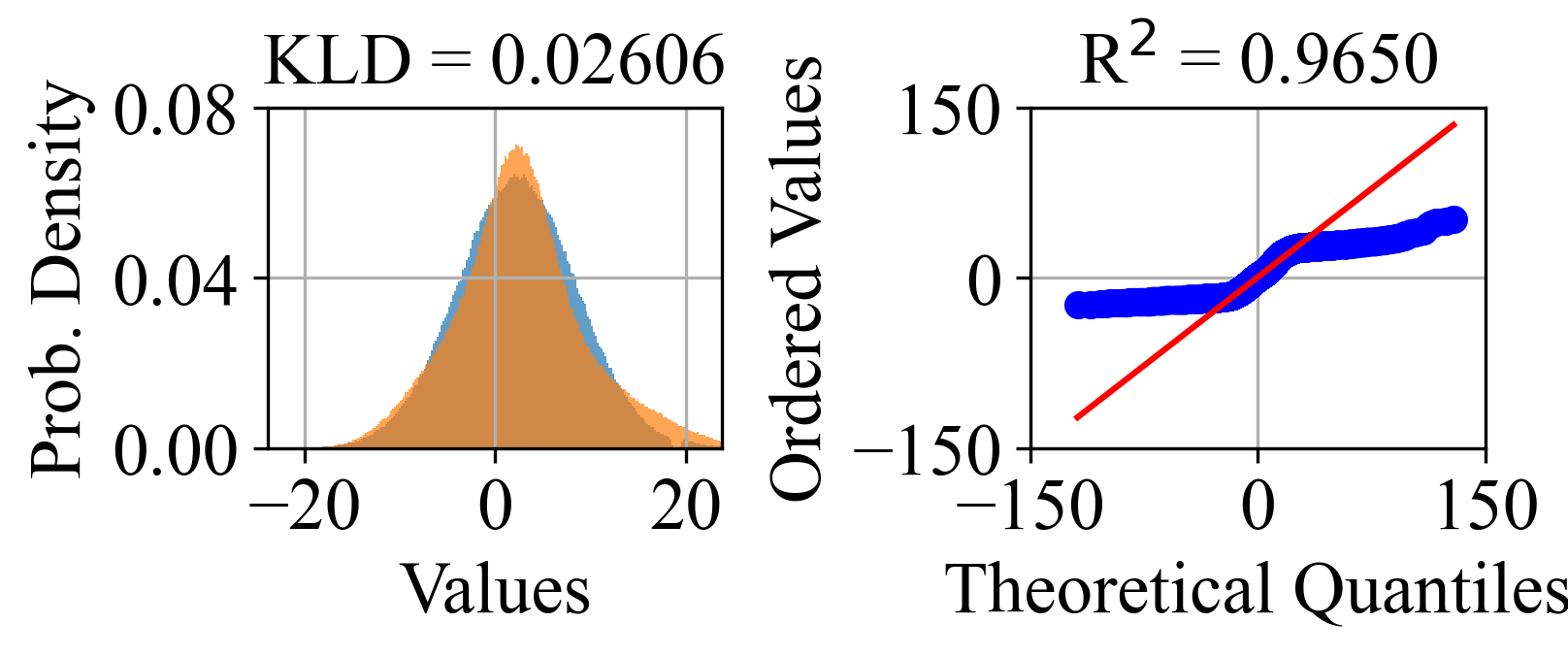}\label{fig:ab_f}} \\
					\subfigure[w/o Quantile Loss Branch]
					{\includegraphics[width=0.315\linewidth]{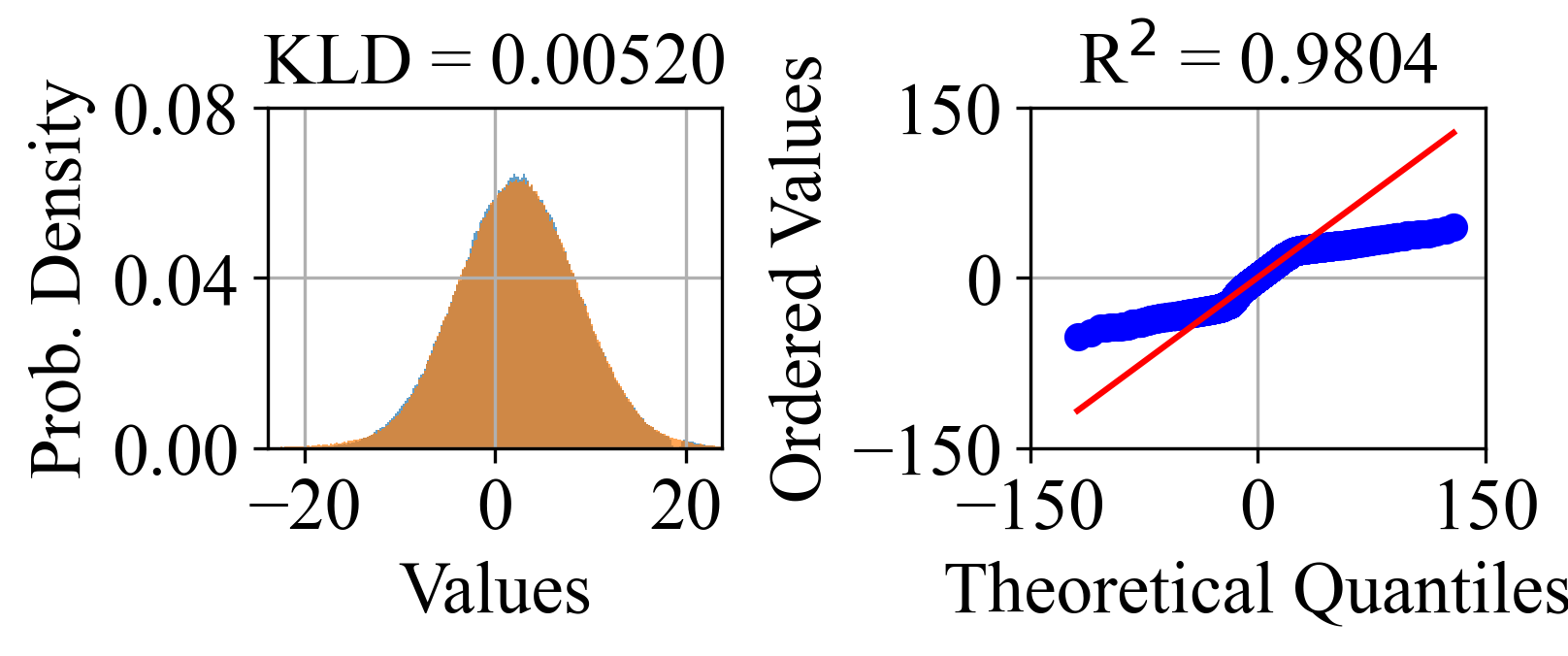}\label{fig:ab_g}} & 
					\subfigure[w/o CDF Loss Branch]
					{\includegraphics[width=0.315\linewidth]{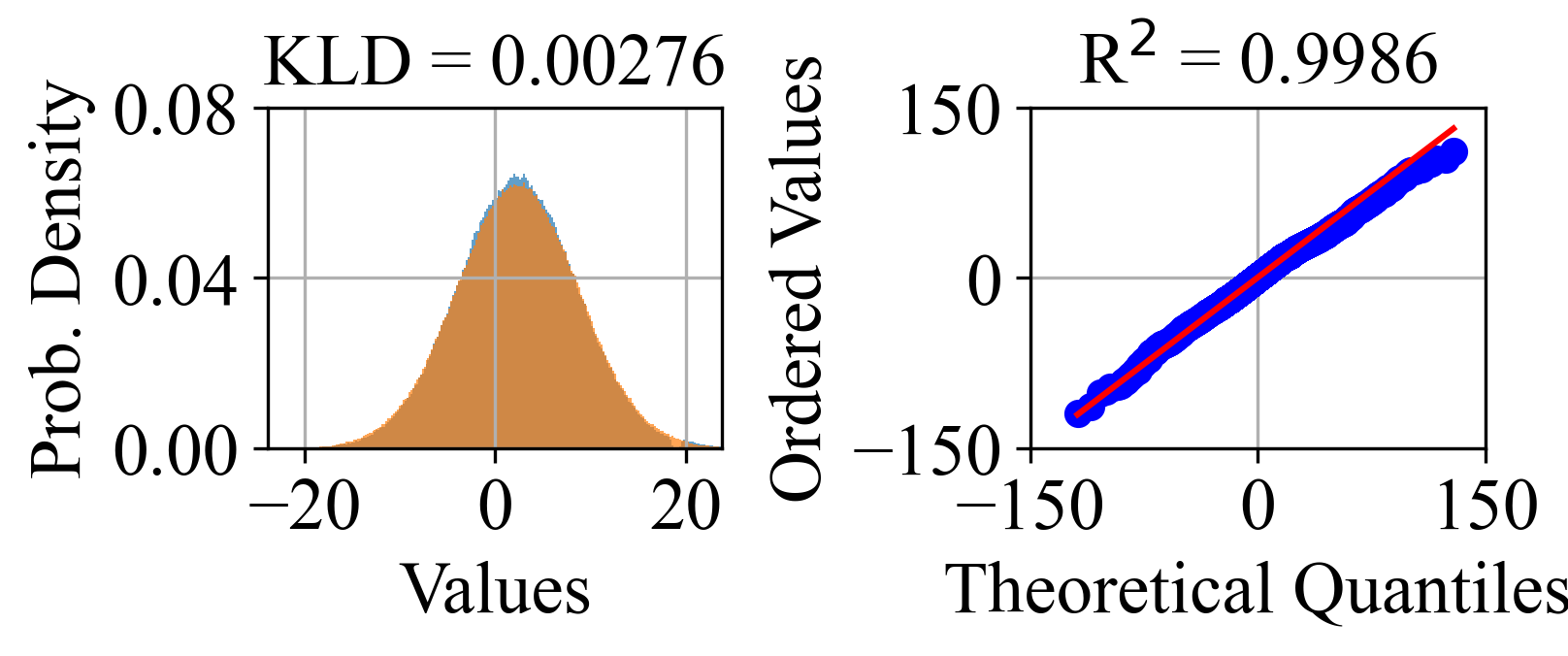}\label{fig:ab_h}} & 
					\subfigure[PNNP (Ours)]
					{\includegraphics[width=0.315\linewidth]{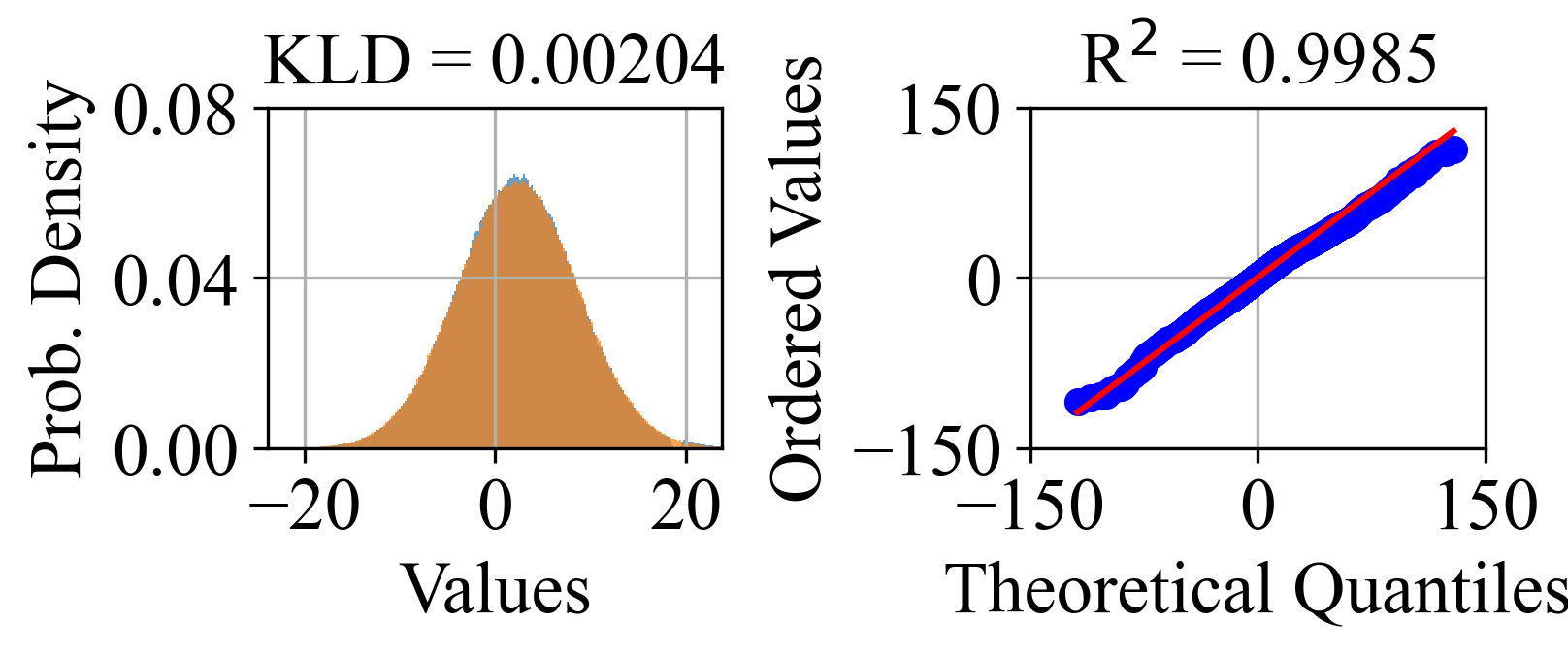}\label{fig:ab_i}}\\

				\end{tabular}%
			}
		\end{center}
		\caption{A distribution comparison of ablation studies on dark frame noise at ISO-1600. Each subfigure consists of two parts: the left part shows a comparison of the probability density functions, while the right part shows the probability plot~\cite{statistical}. The blue histogram represents the distribution of real noise, while the orange histogram represents the distribution of synthetic noise. The blue points represent the quantiles of the real noise distribution, and the red line represents the quantiles of the synthetic noise distribution.}
		\label{fig:ab_score}
	\end{figure*}
	
	To evaluate the performance of noise modeling, we employ KLD and coefficient of determination (R$^2$) as metrics on real noise~\cite{statistical}. Two types of real noise are used: pixel-wise noise and dark frame noise. Pixel-wise noise represents the high-bit pixel-wise noise decoupled by PND, which serves as the ground truth during training. Dark frame noise refers to the original dark frame without PND, which serves as the noise modeling target for our PNNP. The calculation process of KLD and R$^2$ is kept pace with the approaches described in NoiseFlow~\cite{NoiseFlow} and ELD~\cite{TPAMI21/ELD}, respectively.
	% 如图~\ref{fig:ab_score}所示，我们可视化了KLD and R$^2$在ISO-1600下对消融实验中各方法的评价。
	%For instance, we visualize a distribution comparison of ablation studies on pixel-wise noise in Figure~\ref{fig:ab_score}.
	% Such a measure ignores the ability of a model to capture correlations but focuses on a model's ability to capture the most basic characteristics of the distribution.
	% It is worth noting that KLD and R$^2$ do not capture spatial information comprehensively, which only capture the pixel-wise characteristics of the distribution. Therefore, the metrics on dark frame noise are provided for reference only. %The issue of noise distribution measurement will be discussed in detail in Section~\ref{discuss:measure}.
	
	To evaluate the performance of image denoising, we employ PSNR and SSIM as metrics on the public datasets. Since the metrics used for noise modeling cannot fully capture all the characteristics of the noise model, we consider the denoising performance corresponding to the noise model as the most important metric.
	
	\subsubsection{Ablation on PND}\label{ablation:PND}
	PND plays a vital role in bridging between physics-based noise modeling and learning-based noise modeling. 
	{The decoupled noise components are shown in Fig.~\ref{fig:pnd_gen}: frame-wise noise captures the spatial inconsistency of spatially variant noise, band-wise noise isolates row- and column-direction pattern noise, and pixel-wise noise exhibits no spatial correlation. Modeling global spatially variant noise is particularly challenging for neural networks with limited receptive fields. By significantly enhancing the learnability of the noise model, PND enables PNNP to accurately represent different levels of noise and generate realistic noise.}
	
\begin{figure*}[t!]
		\small %此处写字体大小控制命令
		\setlength\tabcolsep{1.2pt}
		\renewcommand\arraystretch{0.8}
		\centering
		{%
			\begin{tabular}{ccccc}
				Gaussian & ELD & w/o Noise Decoupling & w/ Flow-based Module & w/ Conv3$\times$3\\
				{\includegraphics[width=0.195\linewidth]{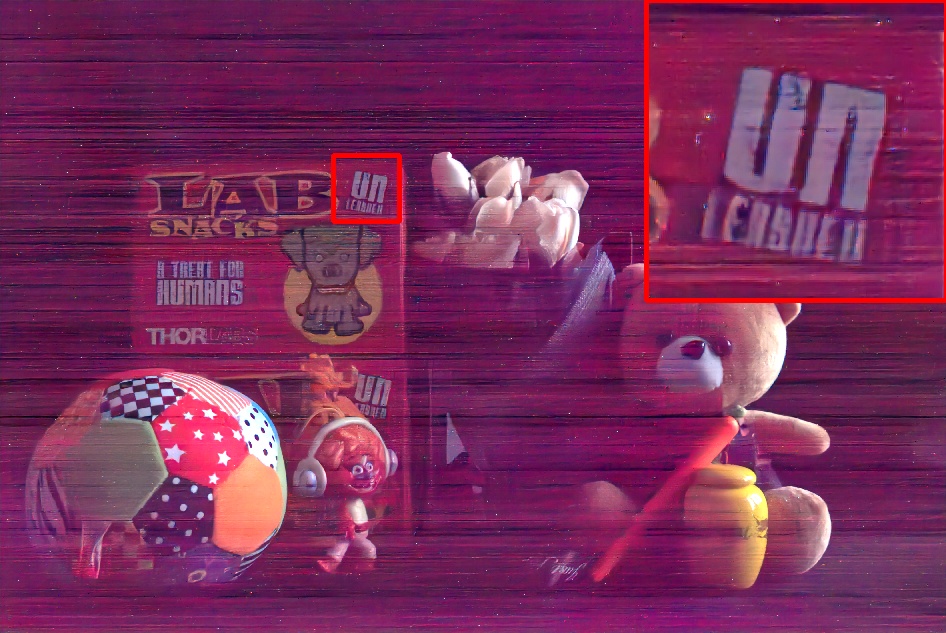}} & 
				{\includegraphics[width=0.195\linewidth]{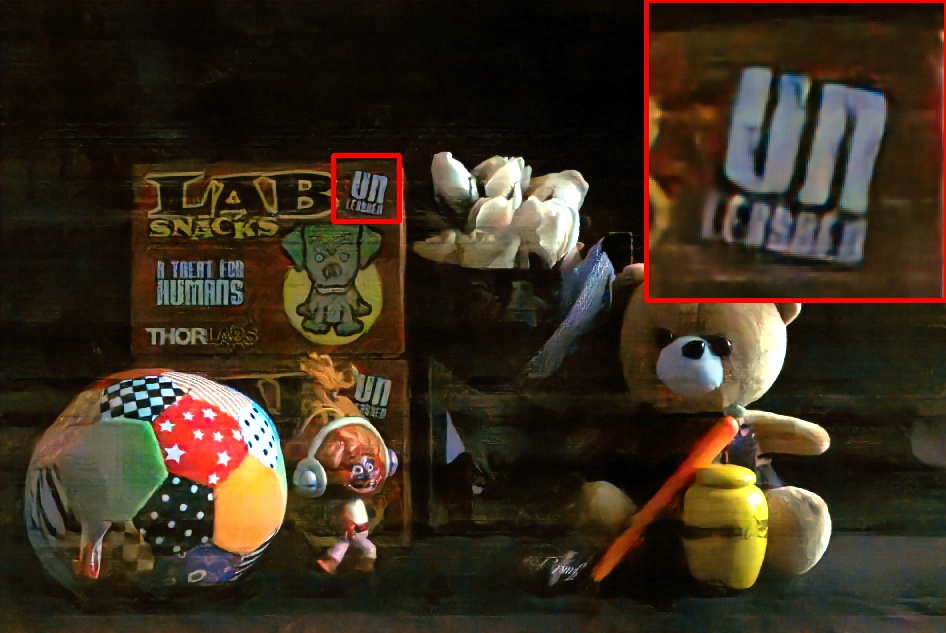}} & 
				{\includegraphics[width=0.195\linewidth]{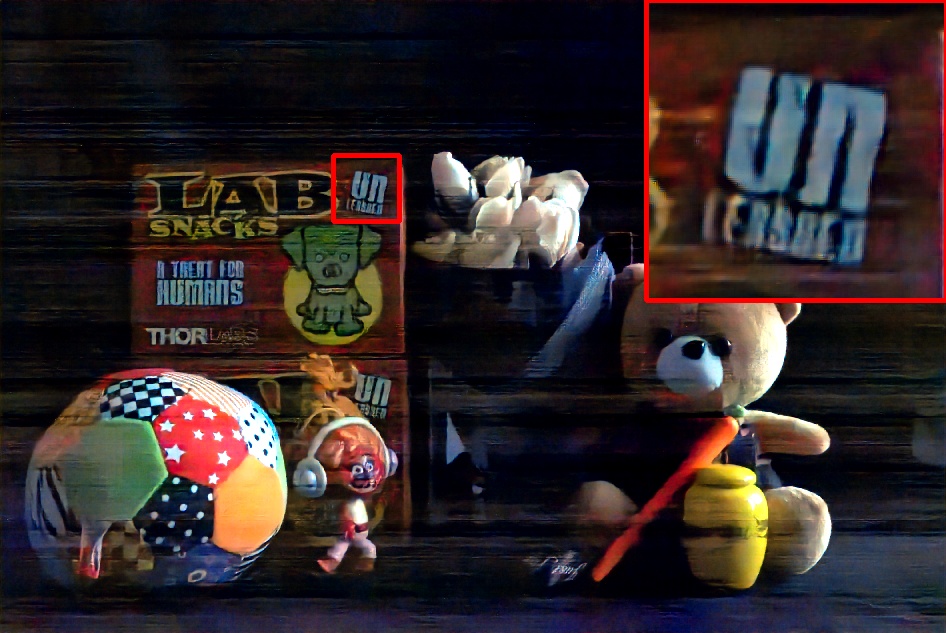}} & 
				{\includegraphics[width=0.195\linewidth]{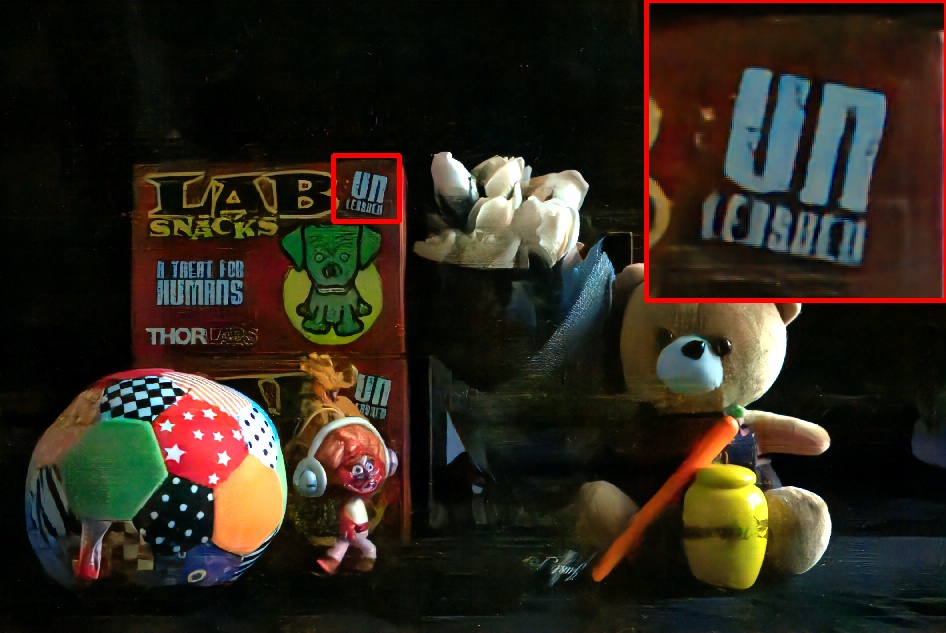}} & 
				{\includegraphics[width=0.195\linewidth]{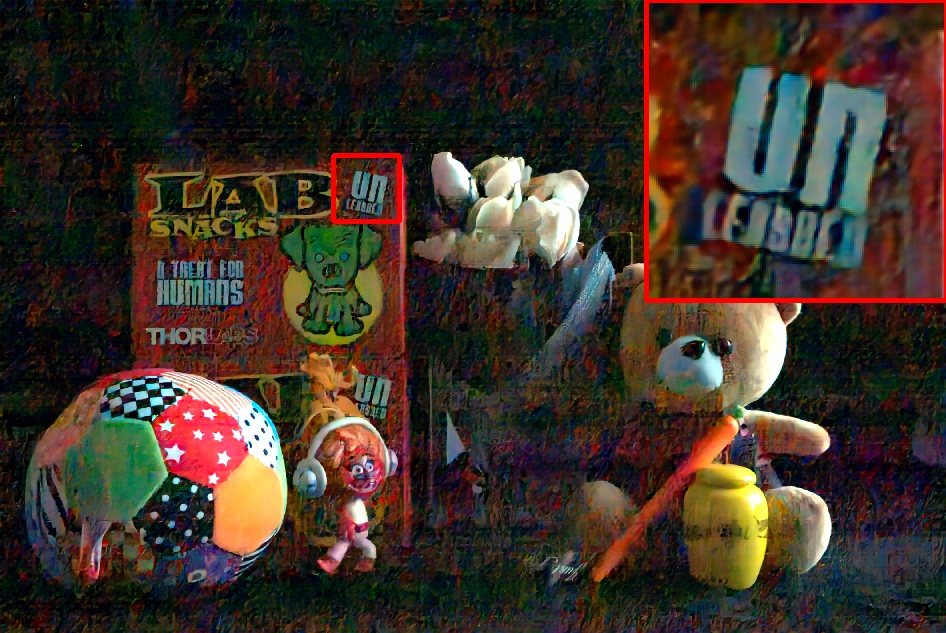}} \\
				{34.53} / {0.6103} & {42.75} / {0.9271} & {42.42} / {0.9174} & {44.17} / {0.9606} & {43.48} / {0.9469} \\
				\addlinespace[4pt]
				w/o Dual Branch & w/o Quantile Loss & w/o CDF Loss & PNNP (Ours) & Reference \\
				{\includegraphics[width=0.195\linewidth]{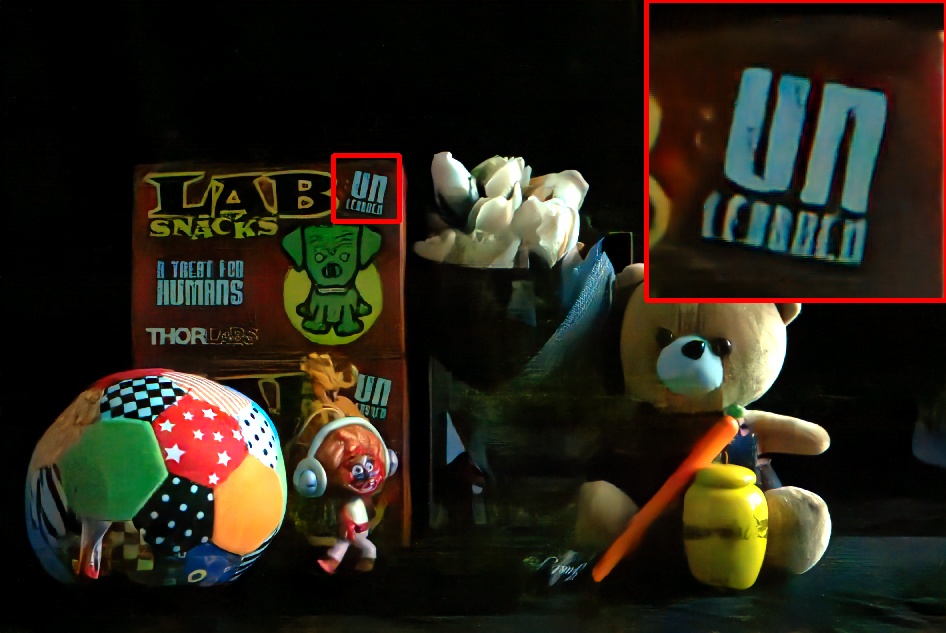}} &
				{\includegraphics[width=0.195\linewidth]{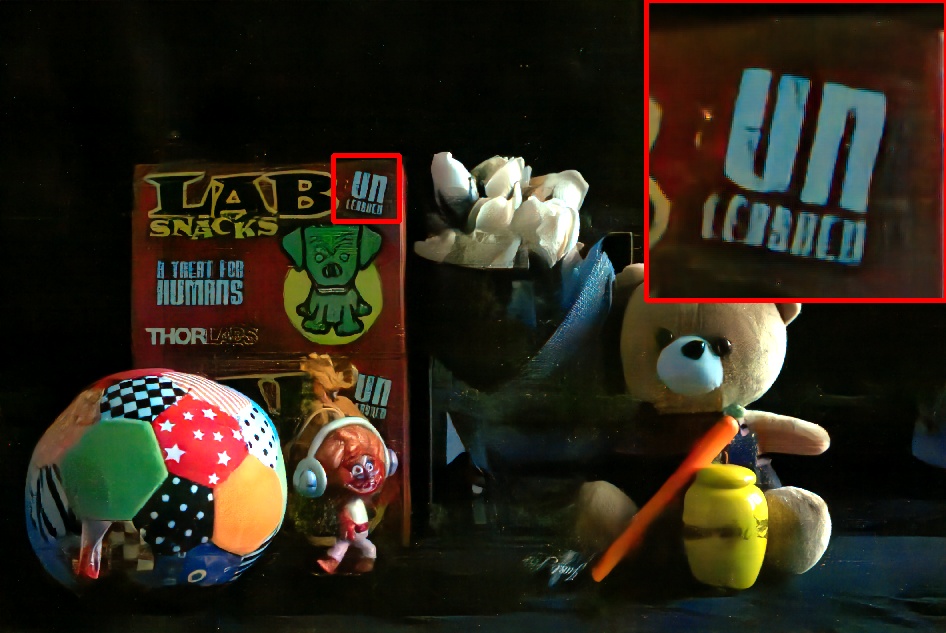}} & 
				{\includegraphics[width=0.195\linewidth]{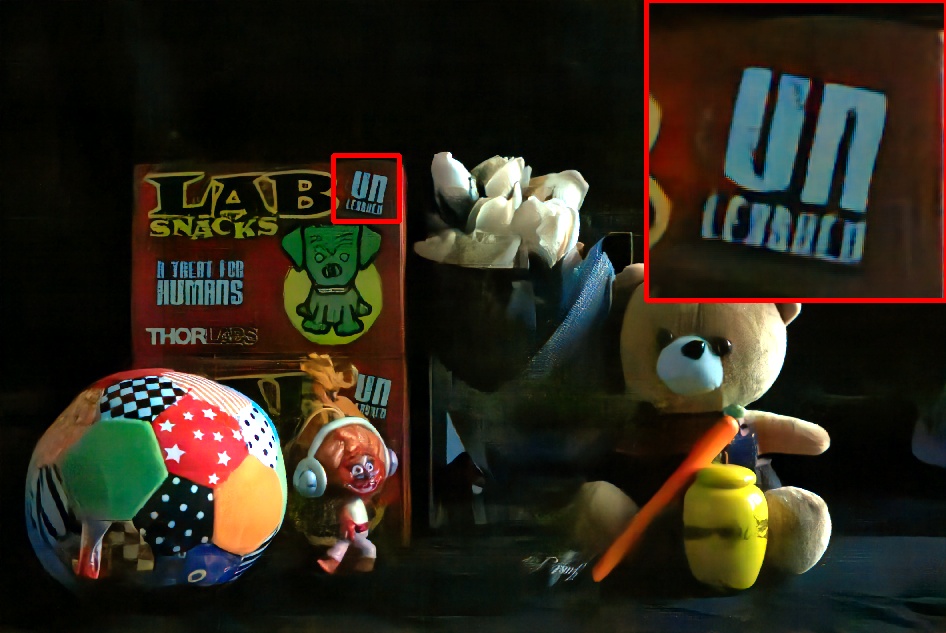}} &
				{\includegraphics[width=0.195\linewidth]{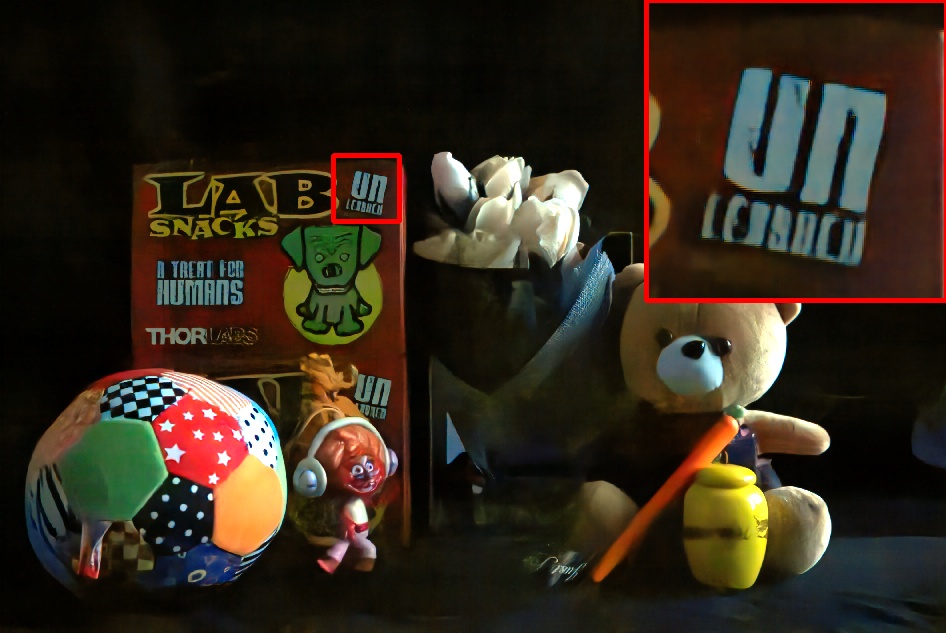}}&
				{\includegraphics[width=0.195\linewidth]{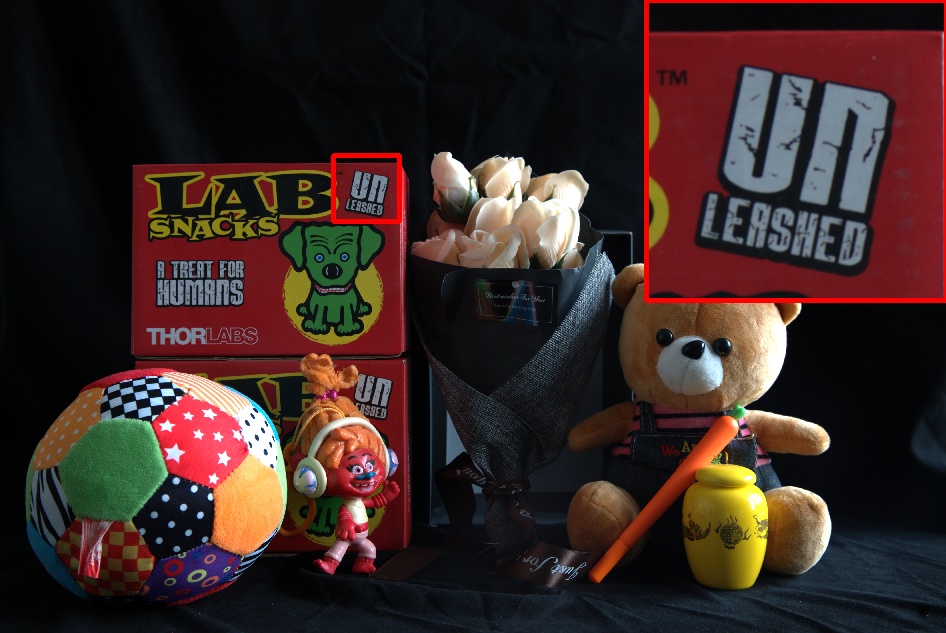}} \\
				{43.70} / {0.9332} & {44.20} / {0.9594} & {\color{blue}44.41} / {\color{blue}0.9636} & {\color{red}44.70} / {\color{red}0.9685} & PSNR / SSIM
			\end{tabular}%
		}
		\caption{A representative visual comparison of ablation studies on the ELD dataset. The {\color{red}red} color indicates the best results and the {\color{blue}blue} color indicates the second-best results. \textbf{(Best viewed with zoom-in)}}
		\label{fig:ablation}
	\end{figure*}

	To demonstrate the effectiveness of PND, we select two classical physics-based noise modeling methods (Gaussian and ELD) as baselines and then explore the case of direct learning without noise decoupling. As shown in Table~\ref{tab:ablation}, compared to the classical physics-based noise modeling, PNNP retains substantial superiority in noise modeling. As shown in Fig.~\ref{fig:ab_a} and Fig.~\ref{fig:ab_b}, both Gaussian and ELD exhibit low R$^2$, indicating their challenges in modeling long-tailed distributions. The comparison between Fig.~\ref{fig:ab_c} and Fig.~\ref{fig:ab_i} demonstrates the robust capability of our method in modeling long-tailed distributions. However, considering Table~\ref{tab:ablation} and Fig~\ref{fig:ablation}, PNNP without noise decoupling struggles to cover the complete noise distribution, leaving noticeable pattern noise residues, which indicates the indispensability of PND for the entire framework.
	
	\subsubsection{Ablation on PPM}\label{ablation:PPM}
	To demonstrate the effectiveness of PPM, we first substitute the PPM modules with flow-based modules (marked as modified NoiseFlow) as a baseline,
	As shown in Fig.~\ref{fig:ab_d}, the modified NoiseFlow performs well on KLD but falls short on R$^2$. Despite having more parameters than PNNP (2.8k vs. 2.3k), the modified NoiseFlow still struggles to approximate long-tailed distributions. Constrained by the limited representation ability of the flow model, DDL fails to bring significant improvements to the modified NoiseFlow.
	
	Next, we will evaluate our physics-aware module designs proposed in PPM.
	
	Our first design is using 1$\times$1 convolutions instead of 3$\times$3 convolutions to preserve the spatially independent prior of pixel-wise noise. In the ablation study, we replace all 1$\times$1 convolutions with 3$\times$3 convolutions. While the metrics for noise modeling may seem promising in Fig.~\ref{fig:ab_e}, the corresponding denoising results exhibit noticeable residual pattern noise in Fig.~\ref{fig:ablation}. In the absence of explicitly considering spatially correlated noise components, the network spontaneously learns to exploit neighborhood information to optimize the noise model, inadvertently introducing spatial correlation. The unacceptable pattern noise highlights the significance of our pixel-wise independent module design.
	
	Our second design is the network structure based on the sensor imaging mechanisms. In the ablation study, we change the ISO-aware dual-branch structure to a single branch. Without the physics-aware network structure design, the noise proxy model not only loses the physical interpretability but also exhibits a noticeable decrease in noise modeling, as shown in Fig.~\ref{fig:ab_f}, which highlights the importance of our ISO-aware dual-branch design.
	
	\subsubsection{Ablation on DDL}\label{ablation:DDL}
	DDL can be briefly divided into three components: quantile loss and CDF loss. We conducted an ablation study on each of these components, as shown in Table~\ref{tab:ablation}. 
	
	The quantile loss focuses on supervising the long-tail region of the noise distribution, which is consistent with the emphasis of R$^2$. Without quantile loss, the modeling ability on the long-tail distribution significantly decreases. It is reflected in the degradation of metrics in Fig~\ref{fig:ab_g} and the appearance of artifacts resembling defect pixels in denoising results, as shown in Fig.~\ref{fig:ablation}. The CDF loss focuses on supervising the central region of the noise distribution, which is the same as the emphasis of KLD. Without CDF loss, the modeling ability of PNNP slightly degrades, resulting in a slight decrease in KLD as shown in Fig~\ref{fig:ab_h}. The complete DDL consistently provides reliable and stable supervision for PPM, contributing to the best performance in both noise modeling and corresponding image denoising results.
	
	\section{Discussion}
	
	\begin{figure*}[t!]
		\small %此处写字体大小控制命令
		\setlength\tabcolsep{1.5pt}
		\renewcommand\arraystretch{0.8}
		\begin{center}
			\begin{tabular}{ccccccccc}
				{Input} & {Paired Data} & {SFRN~\cite{ICCV21/SFRN}} & {PMN~\cite{TPAMI23/PMN}} & {LLD~\cite{CVPR23/LLD}} & {LLD*~\cite{CVPR23/LLD}} & PNNP(Ours) & {PNNP*(Ours)} & {Reference} \\
				{\includegraphics[width=0.105\textwidth]{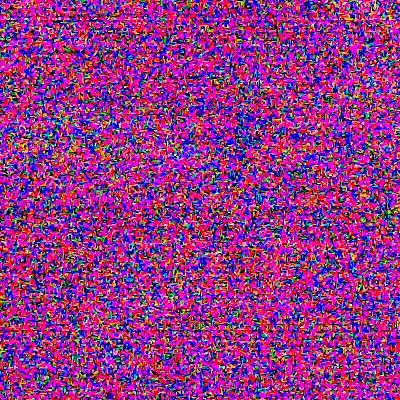}}& 
				{\includegraphics[width=0.105\textwidth]{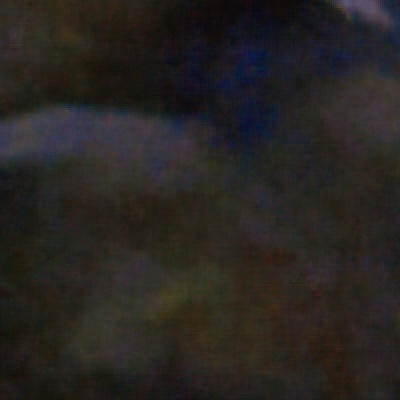}}&
				{\includegraphics[width=0.105\textwidth]{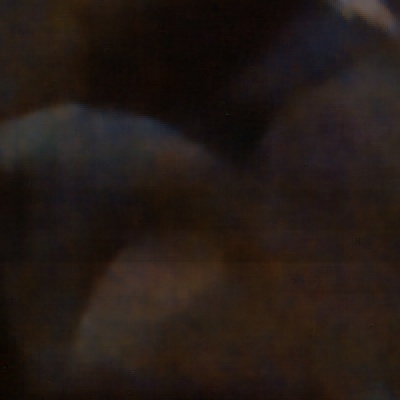}} & 
				{\includegraphics[width=0.105\textwidth]{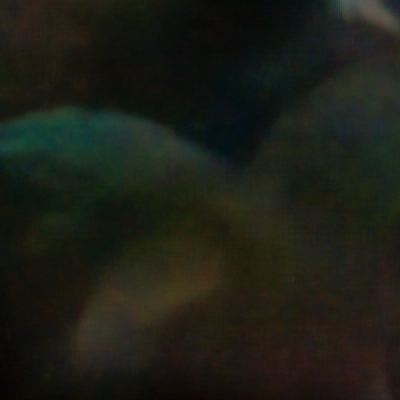}}&
				{\includegraphics[width=0.105\textwidth]{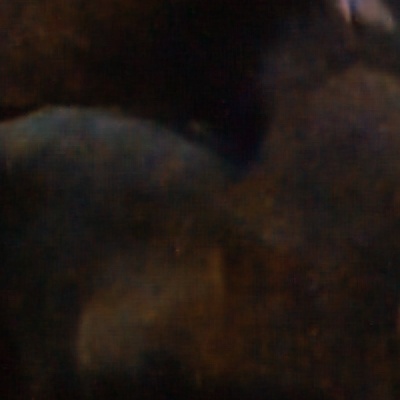}}&
				{\includegraphics[width=0.105\textwidth]{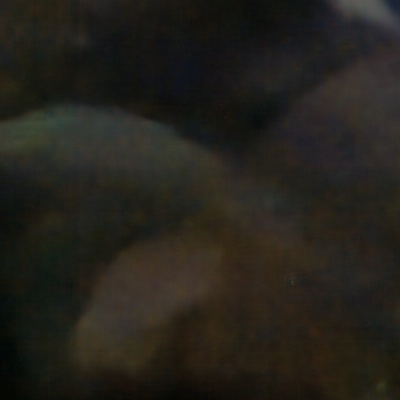}}&
				{\includegraphics[width=0.105\textwidth]{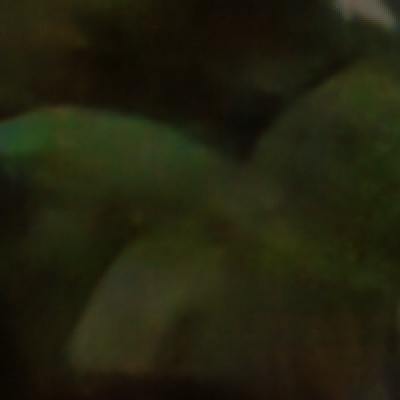}}&
				{\includegraphics[width=0.105\textwidth]{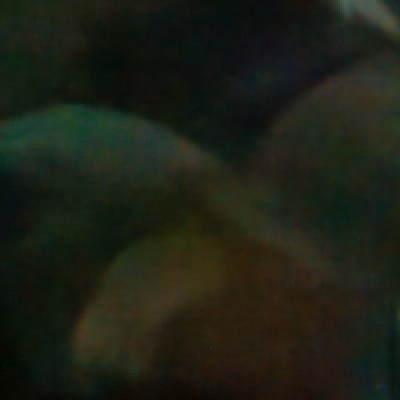}}&
				{\includegraphics[width=0.105\textwidth]{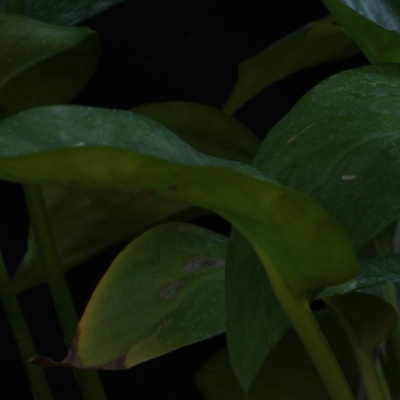}} \\
				{22.58}/{0.0565} & {41.04}/{0.8590} & {47.19}/{0.9531} & {48.82}/{0.9811} & {46.82}/{0.9544} & {49.24}/{0.9808} & {\color{red}49.83}/{\color{blue}0.9818} & {\color{blue}49.78}/{\color{red}0.9822} & PSNR/SSIM \\
				\addlinespace[2pt]
				{\includegraphics[width=0.105\textwidth]{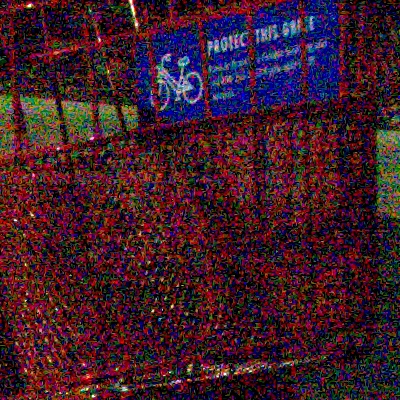}}& 
				{\includegraphics[width=0.105\textwidth]{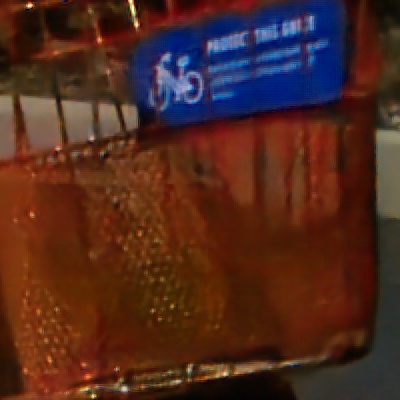}}&
				{\includegraphics[width=0.105\textwidth]{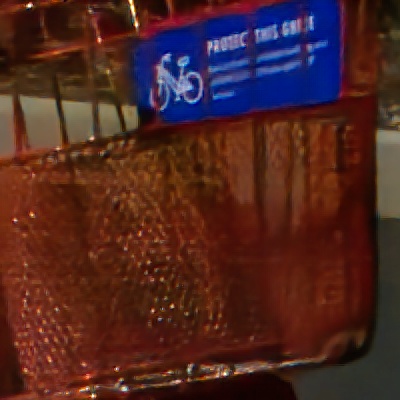}} & 
				{\includegraphics[width=0.105\textwidth]{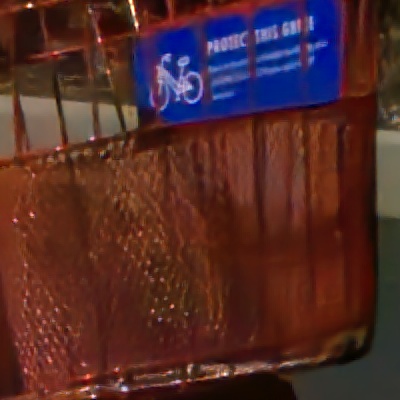}}&
				{\includegraphics[width=0.105\textwidth]{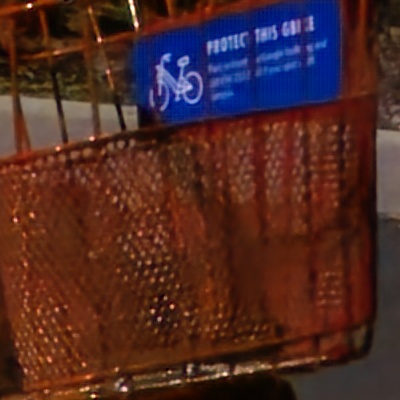}}&
				{\includegraphics[width=0.105\textwidth]{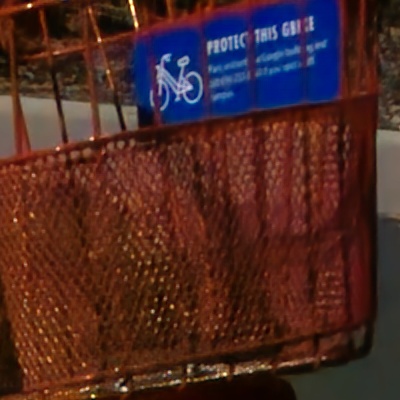}}&
				{\includegraphics[width=0.105\textwidth]{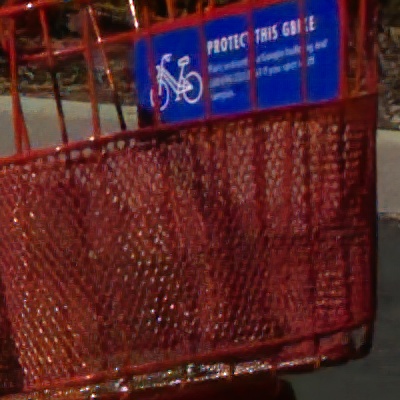}}&
				{\includegraphics[width=0.105\textwidth]{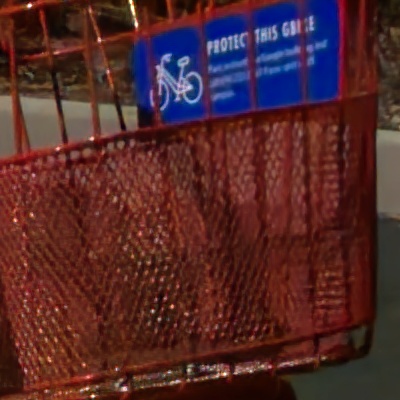}}&
				{\includegraphics[width=0.105\textwidth]{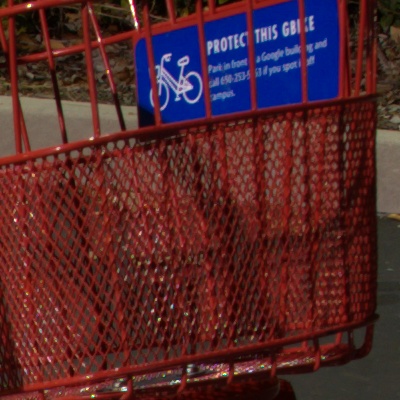}} \\
				{33.64}/{0.7154} & {43.15}/{0.9767} & {45.05}/{0.9801} & {44.76}/{\color{blue}0.9819} & {44.56}/{0.9788} & {44.91}/{0.9818} & {\color{red}45.90}/{\color{red}0.9840} & {\color{blue}45.10}/{0.9808} & PSNR/SSIM \\
				% \addlinespace[1pt]
				%{\includegraphics[width=0.105\textwidth]{SID20210/Input_croped}}& 
				%{\includegraphics[width=0.105\textwidth]{SID20210/Paired_croped}}&
				%{\includegraphics[width=0.105\textwidth]{SID20210/SFRN_croped}} & 
				%{\includegraphics[width=0.105\textwidth]{SID20210/PMN_croped}}&
				%{\includegraphics[width=0.105\textwidth]{SID20210/LLD_croped}}&
				%{\includegraphics[width=0.105\textwidth]{SID20210/LLD+_croped}}&
				%{\includegraphics[width=0.105\textwidth]{SID20210/PNNP_croped}}&
				%{\includegraphics[width=0.105\textwidth]{SID20210/PMNNP_croped}}&
				%{\includegraphics[width=0.105\textwidth]{SID20210/GT_croped}} \\
				%{15.14}/{0.0324} & {34.78}/{0.8655} & {35.61}/{0.8719} & {35.68}/{0.8729} & {35.39}/{0.8776} & {\color{blue}36.20}/{\color{blue}0.8866} & {\color{red}36.51}/{\color{red}0.8882} & {36.16}/{0.8806} & PSNR / SSIM \\
				\addlinespace[2pt]
				{\includegraphics[width=0.105\textwidth]{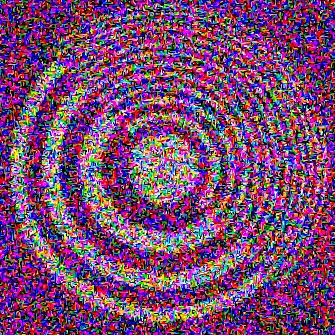}}& 
				{\includegraphics[width=0.105\textwidth]{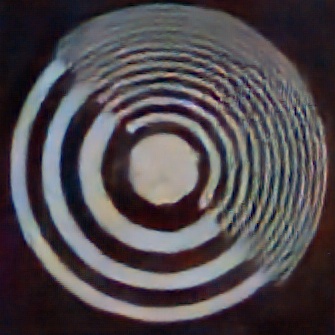}}&
				{\includegraphics[width=0.105\textwidth]{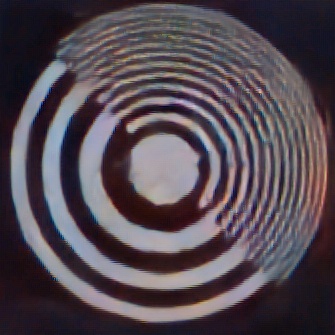}} & 
				{\includegraphics[width=0.105\textwidth]{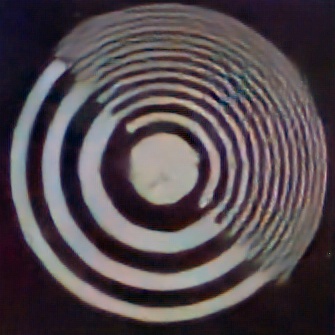}}&
				{\includegraphics[width=0.105\textwidth]{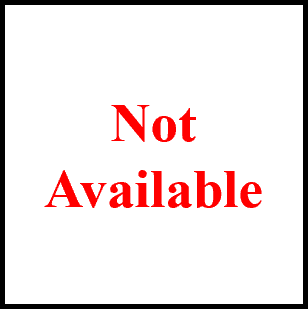}}&
				{\includegraphics[width=0.105\textwidth]{NotAvaliable}}&
				{\includegraphics[width=0.105\textwidth]{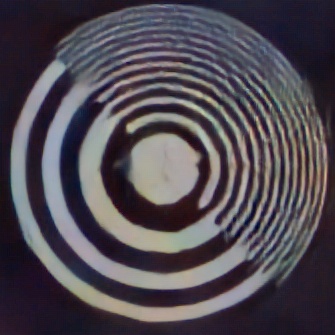}}&
				{\includegraphics[width=0.105\textwidth]{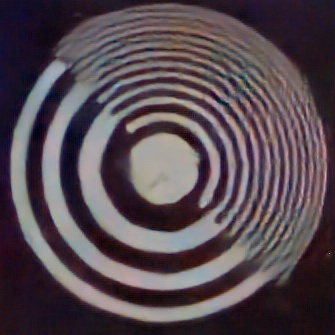}}&
				{\includegraphics[width=0.105\textwidth]{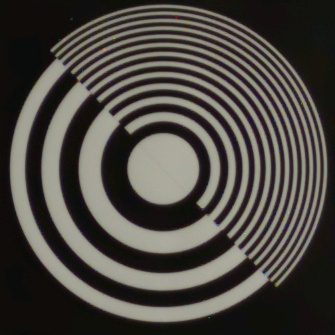}} \\
				% {22.58}/{0.0565} & {41.04}/{0.8590} & {47.19}/{0.9531} & {48.82}/{0.9811} & {46.82}/{0.9544} & {49.24}/{0.9808} & {\color{red}49.83}/{\color{blue}0.9818} & {\color{blue}49.78}/{\color{red}0.9822} & PSNR / SSIM \\
				{23.56}/{0.2314} & {44.57}/{0.9862} & {44.32}/{0.9860} & {\color{blue}45.44}/{\color{blue}0.9894} & - / -& - / - & {44.98}/{0.9876} & {\color{red}45.75}/{\color{red}0.9901} & PSNR/SSIM\\
			\end{tabular}%
		\end{center}
		%\vspace{-6pt}
		\caption{Comparison of denoising performance among noise modeling methods trained with and without paired real data. We select some representative results from ELD dataset~\cite{TPAMI21/ELD}, SID dataset~\cite{CVPR18/SID} and LRID datasets~\cite{TPAMI23/PMN}, respectively. The {\color{red}red} color indicates the best results and the {\color{blue}blue} color indicates the second-best results. \textbf{(Best viewed with zoom-in)}}
		\label{fig:pmn}
	\end{figure*}

	\subsection{Data for Noise Modeling}\label{discuss:data}
	% 噪声建模研究中，有一个重要的前提经常被忽略，即真实数据总是面向实用的噪声建模中不可或缺的成分。learning-based的噪声建模则需要真实数据来训练噪声模型，而physics-based噪声建模一般需要真实数据来标定噪声参数。尽管噪声建模方法一般是通用的，但是模型参数总是sensor–specific的。因此，实践中噪声建模总是需要传感器采集真实数据的。真实数据的质量决定了具体噪声模型的质量。从数据的角度出发，一个实用的噪声建模方法应当考虑获取真实数据源的获取难度，甚至适配并弥补数据的潜在缺陷。
	
	% 现有的learning-based噪声建模范式是learning the clean-to-noise mapping from paired real data，这从数据视角来看存在许多的问题。现有的范式依赖large-scale high-quality paired real data，然而这一前提很难被满足。一方面，欠发展的数据采集设置往往让paired real data中存在信号错位。另一方面，the coupling of excessive noise models within the paired real data往往让神经网络难以逼近。总而言之，数据质量的缺陷让现有的learning-based噪声建模范式在实战中表现不佳。
	
	%相较与learning-based噪声建模，physics-based噪声建模依赖的真实数据只是相机采集的标定数据，比如用于标定信号相关噪声的flat-field frames和用于标定信号无关噪声的dark frames。标定数据的采集方式在很长一段时间的研究下已经比较完善了，因此物理类噪声建模所依赖的数据的质量比学习类噪声建模的更高。learning the noise model from 标定数据（dark frames）而不是配对真实数据，这就是我们从数据视角提出的新的噪声建模范式。
	
	% 我们注意到最新的学习类噪声建模方法LLD和我们有相似的思路，这源于他们也重视了数据质量的问题。他们重新采集了配对真实数据集用于训练噪声建模并进行了噪声解耦。然而，不彻底的噪声解耦让他们的方法难以从过耦合的噪声中精确地代理到真实噪声模型。充分考虑真实数据质量是我们方法去噪表现优秀的根本原因之一。
	One important premise often overlooked in noise modeling research is that real data is always essential in practical noise modeling. 
	Learning-based approaches depend on real data to train noise models, while physics-based methods utilize real data for calibrating noise parameters. Although noise modeling methods are generally applicable, the model parameters exhibit sensor-specific characteristics. Hence, a robust and practical noise modeling approach should address the inherent challenges associated with real data quality.
	
	The quality of real data directly influences the quality of the resulting noise model. The overlook for data is particularly severe in learning-based noise modeling methods. 
	The existing strategy of learning-based noise modeling, which involves learning the clean-to-noise mapping from paired real data, has several problems from a data perspective. 
	This strategy heavily depends on large-scale high-quality paired real data, which is often challenging to obtain. 
	On one side, underdeveloped data acquisition protocol often results in signal misalignment within the paired real data. 
	On the other side, the coupling of excessive noise models within the paired real data makes it challenging for neural networks to accurately approximate the real-world sensor noise model. 
	In summary, data defects hinder the performance of the existing learning-based noise modeling strategy in practice.
	
	Compared to learning-based noise modeling, physics-based noise modeling relies on real data collected by cameras specifically for calibration purposes, such as flat-field frames for calibrating signal-dependent noise and dark frames for calibrating signal-independent noise. The calibration process is independent of paired real data, eliminating signal misalignment. From the perspective of data dependency, the data required for physics-based noise modeling is easier to obtain and of higher quality compared to learning-based noise modeling. 
	
	Based on these insights, we propose the strategy of learning the noise model from dark frames instead of paired real data. The data-centric perspective~\cite{data-centric} serves as the foundational principle guiding our analysis and problem-solving approach.
	
	% 2025.10
	% 实际上，在最近的两年间我们已经在多个项目的诸多的sensor上应用了PNNP及其改进版本PNNP*，包括但不限于监控相机系列（例如SC450AI、OS04J10）、手机sensor（例如OV48C、IMX766）、DSLR sensor（例如SonyA7R4、SonyA6700）。遗憾的是，尽管获取暗帧在实践中并不困难，但多数的项目不同时提供公开的暗帧与公开的数据。因此，本文仅提供了在两款传感器SonyA7S2与IMX686上的详细对比结果。特别的，我们提到的DSLR sensor来自一个ICCV 2025的workshop，AIM 2025 Real-World RAW Denoising Challenge。我们使用PNNP参与了比赛，在初赛获得了第二名，并在决赛获得了第三名，且在TOPIQ这一感知指标上排名第一。我们相信这个结果证明PNNP的泛化能力。
	{
		PNNP and its variant have been applied across a wide range of sensors in multiple projects, including but not limited to surveillance sensors (\eg, SC450AI, OS04J10), smartphone sensors (\eg, OV48C, IMX766), and DSLR sensors (\eg, SonyA7R4, SonyA6700). Unfortunately, although acquiring dark frames is often straightforward in practice, most of these projects lack simultaneously publicly available dark frames and datasets. Accordingly, this paper presents detailed comparisons only on the SonyA7S2 and IMX686 sensors, for which such data is available.
		Notably, the \textit{AIM 2025 Real-World RAW Denoising Challenge}~\cite{aim2025} provides a relevant benchmark, which includes several of the DSLR sensors we have applied PNNP to. Our PNNP-based solution achieves second place in the preliminary round, third place in the final round, and ranks first in TOPIQ, demonstrating strong generalization capability of PNNP.
	}
	
	\begin{figure*}[t!]
		\begin{center}
			\includegraphics[width=\linewidth]{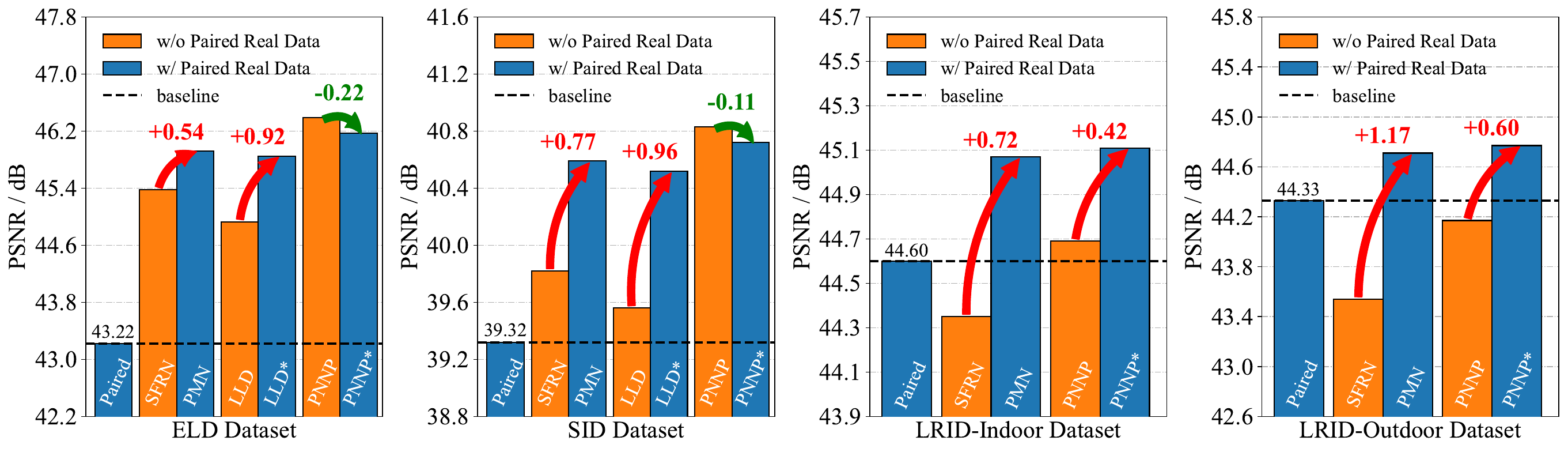}
		\end{center}
		%\vspace{-6pt}
		\caption{Comparison of denoising performance among noise modeling methods trained with and without paired real data. Arrows indicate changes in denoising performance after incorporating paired real data, with {\color{red}red} arrows indicating positive gains and {\color{darkgreen}green} arrows indicating negative gains.}
		\label{tab:pmn}
	\end{figure*}
	
	\subsection{Paired Real Data and Noise Modeling}\label{discuss:pmn}
	Recently, PMN~\cite{TPAMI23/PMN} introduces a novel strategy that integrates paired real data and noise modeling, forming a kind of augmented hybrid data. Moreover, LLD~\cite{CVPR23/LLD} proposes a hybrid training strategy based on PMN, comprising two strategies: the dark shading correction strategy and the zero-mean noise strategy. The dark shading correction strategy relies on paired real data, while the zero-mean noise strategy relies on data synthesized based on noise modeling. By integrating the signal-independent noise model of LLD via the zero-mean noise strategy, LLD* achieves competitive performance in low-light denoising.
	The above hybrid-based methods not only involve noise modeling but also depend on the quality of paired real data. For the sake of controlling variables, we exclude hybrid-based methods from comparisons in Section~\ref{subsec:public}.
	
	% To facilitate a fair comparison with hybrid-based methods, we develop the noise augmentation method of PMN based on PNNP. PMN argues that only signal-dependent shot noise can be accurately modeled, thus proposing the shot noise augmentation (SNA) method to generate hybrid data. Fortunately, experiments in Section~\ref{sec:exp} have demonstrated the high precision of our PNNP in approximating signal-independent noise. Hence, inspired by SNA, we introduce a read noise augmentation method, supporting arbitrary interpolation between paired real data and synthetic data while maintaining the reliability of the noise model. We name the new hybrid-based method as PNNP*.
	To ensure a fair comparison with hybrid-based methods, we extend PNNP by developing the noise augmentation method PMN. Since only signal-dependent shot noise can be reliably modeled, PMN introduces shot noise augmentation (SNA) to generate hybrid data. However, as the modeling of signal-independent noise remains uncertain, SNA augments only the signal-dependent noise in real noisy images, limiting the data diversity. Fortunately, PNNP provides a high-precision approximation of signal-independent noise as demonstrated in Section~\ref{sec:exp}. {To overcome the diversity limitation of SNA, we propose a bidirectional shot noise augmentation (BiSNA) technique.
		
		BiSNA relaxes the strict constraints of noise model consistency to enhance data diversity. Assuming that the noise model of a real noisy image $D$ satisfies Eq.~\eqref{eq:pg} and Eq.~\eqref{eq:possion}, with $N_{indep} \sim \text{PNNP}(iso)$, our target is to generate a realistic noisy image $D'$ with $\alpha I$ light intensity by BiSNA under the same ISO setting, where $\alpha$ denotes the light intensity scaling factor. 
		
		For $\alpha \geq 1$, BiSNA introduces a noisy signal increment $\Delta N \sim K \mathcal{P}(\frac{(\alpha - 1)I}{K})$, following the same procedure as SNA. 
		For $\alpha < 1$, BiSNA first scales the expectation of real noisy image $D$ by $\alpha$, then separately augments shot noise and signal-independent noise to maintain variance consistency, yielding $D' = \alpha D + \Delta N$. Specifically, the variance of the scaled noise is reduced by a factor of $\alpha^2$, while the variance of the shot noise should scale with $\alpha$ (as the mean and variance of a Poisson distribution are equal), and the variance of the signal-independent noise should remain unchanged due to the constant ISO setting. To ensure variance consistency, the noisy signal increment $\Delta N$ is defined as:
		
		\begin{equation}
			\begin{aligned}
				\Delta N \sim K\mathcal{P}\left(\frac{(\alpha - \alpha^2)I}{K}\right) &- (\alpha - \alpha^2)I \\
				&+ \sqrt{1 - \alpha^2}\text{PNNP}(iso).
			\end{aligned}
		\end{equation}
		
		BiSNA enables arbitrary interpolation between real data and synthetic data while preserving noise model similarity. We denote the new hybrid-based method as PNNP*.}
	
	As shown in Fig.~\ref{fig:pmn} and Fig.~\ref{tab:pmn}, we compare our method with existing hybrid-based methods and their corresponding noise modeling methods. Either PNNP or PNNP* consistently achieves state-of-the-art performance with exact color and clear details, thereby demonstrating the superiority of our method.
	
	The noteworthy differences between PNNP and PNNP* deserve an in-depth analysis.  Intuitively, differences between paired real data are typically considered a reliable representation of real noise. Therefore, the introduction of paired real data in hybrid data is generally believed to promote the denoising performance. While this assumption holds in most cases, it does not apply to PNNP and PNNP* on the ELD dataset and SID dataset. This counterintuitive phenomenon aligns with our emphasis on data quality in Section~\ref{discuss:data}. The performance of hybrid-based methods is influenced not only by the accuracy of noise modeling but also by the quality of paired real data. The SID dataset exhibits noticeable data defects, including residual noise, spatial misalignment, and intensity discrepancies. These defects reduce the reliability of data mapping between paired real data, resulting in PNNP* trained on the SID dataset performing less effectively than PNNP. The LRID dataset overcomes most of the data defects, making the model trained only on paired real data competitive with PNNP. The complementarity of high-quality paired real data and a large amount of realistic synthetic data (generated by PNNP) leads to superior performance for PNNP* trained on the LRID dataset.
	
	Finally, we note the efficient method proposed by LED~\cite{ICCV23/LED}, which requires only a small amount of paired real data to adapt a denoising method to new noise models. Although LED shares a similar motivation with hybrid-based methods, it lacks an explicit noise modeling process. Moreover, LED also faces challenges related to the quality of paired real data, resulting in suboptimal denoising performance (LED is notably inferior to SFRN). Hence, we do not include a direct comparison with LED in this study.
	
	%	\setlength{\tabcolsep}{10pt}
	%	\begin{table}[t!]
		%		\small
		%		\caption{Quantitative results (PSNR/SSIM) of different methods on SonyA7S2 Camera (ELD dataset and SID dataset). ``*" donate the method trained with paired real data. The {\color{red}red} color indicates the best results and the {\color{blue}blue} color indicates the second-best results.}
		%		\vspace{-1pt}
		%		\label{tab:pmn}
		%		\setlength{\tabcolsep}{3pt}
		%		\begin{center}
			%			{%
				%				% \vspace{-6pt}
				%				\begin{tabular}{lcccc}
					%					\toprule
					%					{} & {} & ELD~\cite{TPAMI21/ELD} & SID~\cite{CVPR18/SID} & LRID~\cite{TPAMI23/PMN}\\
					%					\cmidrule(lr){3-3}\cmidrule(lr){4-4}\cmidrule(lr){5-5}
					%					\multirow{-2.5}{*}{Method} & \multirow{-2.5}{*}{Data} & {PSNR/SSIM} & {PSNR/SSIM} & {PSNR/SSIM} \\ 
					%					\midrule
					%					Paired & R & 43.22/0.9479  & 39.32/0.9374 & 44.52/0.9748\\
					%					ELD~\cite{TPAMI21/ELD} & S & 44.44/0.9649 & 39.05/0.9303 & 43.53/0.9597\\
					%					PMN~\cite{TPAMI23/PMN} & R+S & {45.92}/0.9763 & 40.59/0.9465 & {\color{blue}44.97}/{\color{red}0.9761}\\
					%					\midrule
					%					LLD~\cite{CVPR23/LLD} & S & 44.93/0.9712 & 39.56/0.9348 & - / -\\
					%					LLD*~\cite{CVPR23/LLD} & R+S & {45.85}/{0.9814} & 40.52/{0.9472} & - / -\\
					%					\midrule
					%					\textbf{PNNP} & S & {\color{red}46.39}/{\color{red}0.9834} & {\color{red}40.83}/{\color{red}0.9479} & {44.54}/{\color{blue}0.9757} \\
					%					\textbf{PMNNP} & R+S & {\color{blue}46.13}/{\color{blue}0.9820} & {\color{blue}40.77}/{\color{blue}0.9480} & {\color{red}45.01}/{\color{blue}0.9757}\\
					%					\bottomrule
					%				\end{tabular}%
				%			}
			%		\end{center}
		%	\end{table}
	
	\begin{figure}[t!]
		\footnotesize %此处写字体大小控制命令
		\setlength\tabcolsep{1.2pt}
		\centering
		\begin{tabular}{c}
			\begin{tabular}{ccccc}
				Reference & & Real~\cite{CVPR18/SIDD} & P-G\cite{TPAMI21/ELD} & CA-GAN~\cite{ECCV20/CAGAN}\\
				\makecell{\includegraphics[width=0.23\linewidth,trim=0pt 8pt 0pt 0pt,clip]{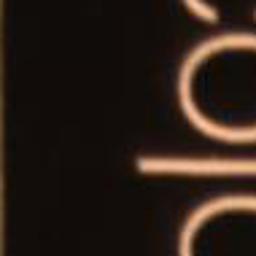}}&
				\rotatebox[origin=c]{90}{Noisy}&
				\makecell{\includegraphics[width=0.23\linewidth,trim=0pt 8pt 0pt 0pt,clip]{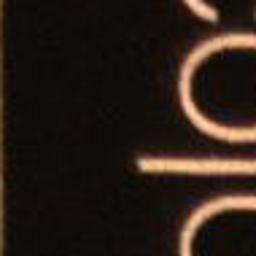}}&
				\makecell{\includegraphics[width=0.23\linewidth,trim=0pt 8pt 0pt 0pt,clip]{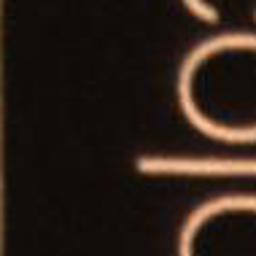}}&
				\makecell{\includegraphics[width=0.23\linewidth,trim=0pt 8pt 0pt 0pt,clip]{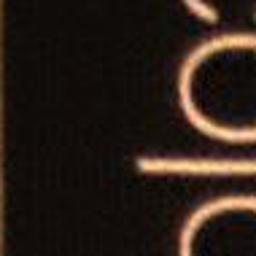}} \\
				&
				\rotatebox[origin=c]{90}{Noise}&
				\makecell{\includegraphics[width=0.23\linewidth,trim=0pt 8pt 0pt 0pt,clip]{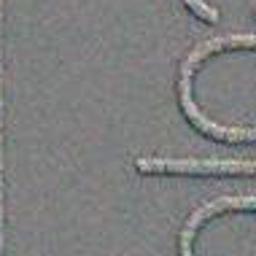}}&
				\makecell{\includegraphics[width=0.23\linewidth,trim=0pt 8pt 0pt 0pt,clip]{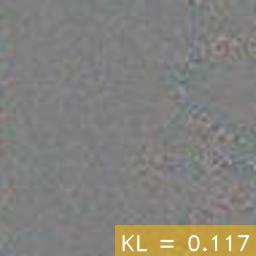}}& 
				\makecell{\includegraphics[width=0.23\linewidth,trim=0pt 8pt 0pt 0pt,clip]{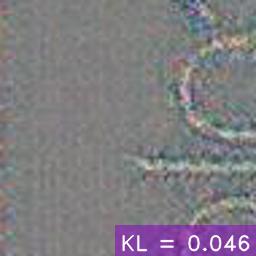}} \\
				&& KLD$\downarrow$ & 0.112 & 0.046
			\end{tabular}\\
			\addlinespace[4pt]
			(a) Spatial Misalginment{{\label{fig:defects1}}}\\ \\
			\begin{tabular}{ccccc}%\rotatebox[origin=c]{125}
				Reference & & Real~\cite{CVPR18/SID} & P-G\cite{TPAMI21/ELD} & LLD\cite{CVPR23/LLD}\\
				\makecell{\includegraphics[width=0.23\linewidth,trim=0pt 0pt 14pt 0pt,clip]{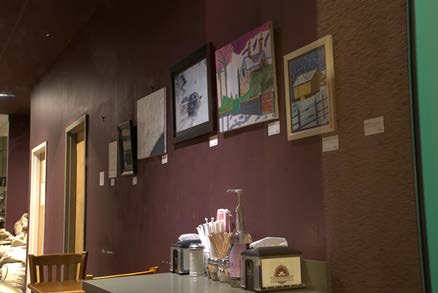}}&
				\rotatebox[origin=c]{90}{Noisy}&
				\makecell{\includegraphics[width=0.23\linewidth,trim=0pt 0pt 14pt 0pt,clip]{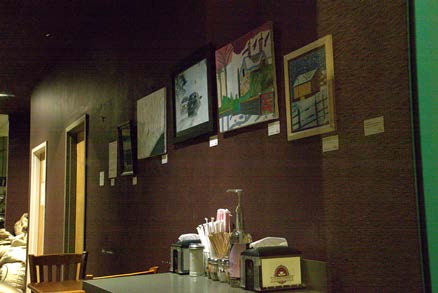}}&
				\makecell{\includegraphics[width=0.23\linewidth,trim=0pt 0pt 40pt 0pt,clip]{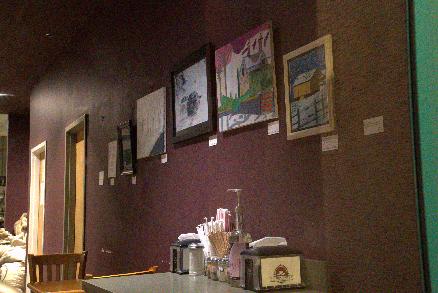}}&
				\makecell{\includegraphics[width=0.23\linewidth,trim=0pt 0pt 14pt 0pt,clip]{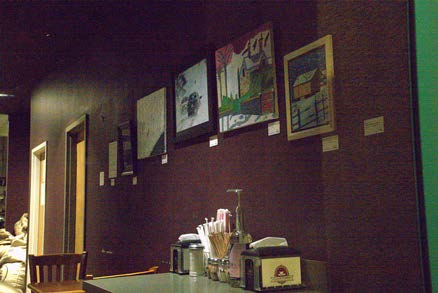}} \\
				&
				\rotatebox[origin=c]{90}{Noise}&
				\makecell{\includegraphics[width=0.23\linewidth,trim=0pt 0pt 40pt 0pt,clip]{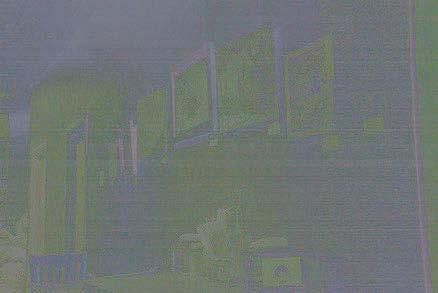}}&
				\makecell{\includegraphics[width=0.23\linewidth,trim=0pt 0pt 40pt 0pt,clip]{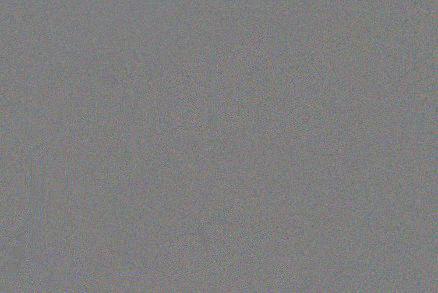}}& 
				\makecell{\includegraphics[width=0.23\linewidth,trim=0pt 0pt 40pt 0pt,clip]{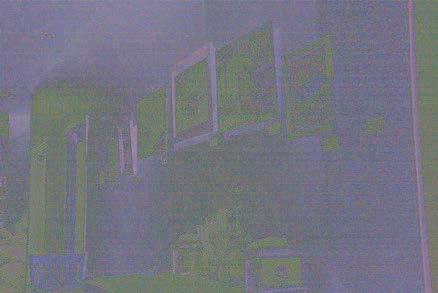}} \\
				&& KLD$\downarrow$ & 0.648 & 0.020
			\end{tabular}\\
			\addlinespace[4pt]
			(b) Brightness Misalginment{\label{fig:defects2}}
		\end{tabular}
		\caption{A representative comparison of physics-based method and learning-based method in response to data defects. (a) ``Real Noise" includes spatial misalignment in addition to noise, and CA-GAN overfits the misaligned details. (b) ``Real Noise" includes brightness misalignment in addition to noise, and LLD overfits the color bias.}
		\label{fig:defects}
	\end{figure}

	{
		\subsection{Practical Limitations of Noise Metrics}\label{discuss:measure}
		Measuring real-world noise distributions is challenging. Under low-light conditions, the spatial correlation cannot be ignored in real-world noise distribution. Existing noise distribution metrics, whether KLD or R$^2$, rely on the i.i.d. assumption and thus are unable to accurately characterize non-i.i.d. noise distributions.
		
		For instance, ``w/ 3×3 Conv" in Section~\ref{ablation:PPM} generates spatially correlated noise, resulting in denoised images filled with spatially correlated artifacts, as shown in Fig.~\ref{fig:ablation}. However, the KLD and R$^2$ of ``w/ 3×3 Conv" are almost identical to those of PNNP in Table~\ref{tab:ablation} and Fig.~\ref{fig:ab_e}. This counterexample demonstrates that existing metrics cannot accurately measure real-world noise distributions.
		
		Moreover, due to underdeveloped data acquisition settings and inevitable environmental disturbances, existing datasets generally suffer from data defects such as residual noise, spatial misalignment, and brightness misalignment~\cite{TPAMI23/PMN}. These data defects render noise distribution metrics based on paired real data unconvincing.
		
		Firstly, residual noise confounds the statistical results of noise distribution, introducing bias into the noise distribution metrics. Learning-based methods are trained on the real-world noise with such confounded distribution. Consequently, learning-based methods tend to yield biased noise distributions and overestimated metrics.
		
		Secondly, spatial misalignment directly disrupts the noise distribution, creating spurious spatial correlations that undermine the metric reliability. For example, CA-GAN claims to have achieved metrics superior to those of physics-based noise modeling methods. However, as shown in Fig.~\ref{fig:defects}(a), its excellent metrics demonstrably result from overfitting to spatial misalignment rather than accurately approximating the noise distribution.
		
		Lastly, brightness misalignment causes a shift in the noise distribution, leading to an unfair assessment between learning-based and physics-based noise modeling methods. As shown in Fig.~\ref{fig:defects}(b), a significant brightness misalignment exists between the reference and real noisy image. Consequently, P-G fails to match the biased real noise distribution, whereas LLD overfits the shifted distribution, which significantly impacts the metrics. Discrimination in metrics occurs in all methods using physics-based shot noise modeling, \ie, Eq.~\eqref{eq:possion}, to synthesize signal-dependent noise. 
		Notably, during noise synthesis, the brightness of the reference image is always augmented to increase the diversity of synthetic noise. Therefore, such brightness misalignment rarely affects the training of denoising networks using physics-based noise modeling methods in practice, which highlights the particular unfairness of such metric discrimination.
		
		Based on the above analysis, we abandon measuring noise distributions on real datasets. In Section~\ref{subsec:ablation}, we provide KLD and R$^2$ for dark frame noise and pixel-wise noise as reference. We prioritize the performance of denoising networks trained on synthetic data from different noise modeling methods as the main evaluation criterion. 
	}
	
	\section{Conclusion}
	In this paper, we propose a novel strategy for noise modeling: learning the noise model from dark frames instead of paired real data.
	Based on the new strategy, we introduce a physics-informed noise neural proxy framework. Our framework leverages the physical priors of the sensor to decouple the complex noise, constrain the optimization process, and provide reliable supervision, thereby further improving the performance of noise modeling.
	Firstly, we propose a physics-guided noise decoupling strategy to handle different levels of noise in a flexible manner. Secondly, we propose a physics-aware proxy model incorporating physical priors to constrain the synthetic noise. Finally, we propose a differentiable distribution loss function to efficiently supervise the random noise variables. 
	Benefiting from physics-informed designs, PNNP not only exhibits low data dependency but also facilitates easy training. These user-friendly features emphasize the practicality of PNNP.
	Experimental results on extensive low-light raw image denoising datasets demonstrate the superiority of our PNNP, highlighting its effectiveness and practicality in noise modeling.
	
	%	Appendix one text goes here.
	\ifCLASSOPTIONcompsoc
	\ifCLASSOPTIONcaptionsoff
	\newpage
	\fi
	
	\bibliographystyle{IEEEtran}
	\bibliography{Reference}
	
	\begin{IEEEbiography}[{\includegraphics[width=1in,height=1.25in,clip,keepaspectratio]{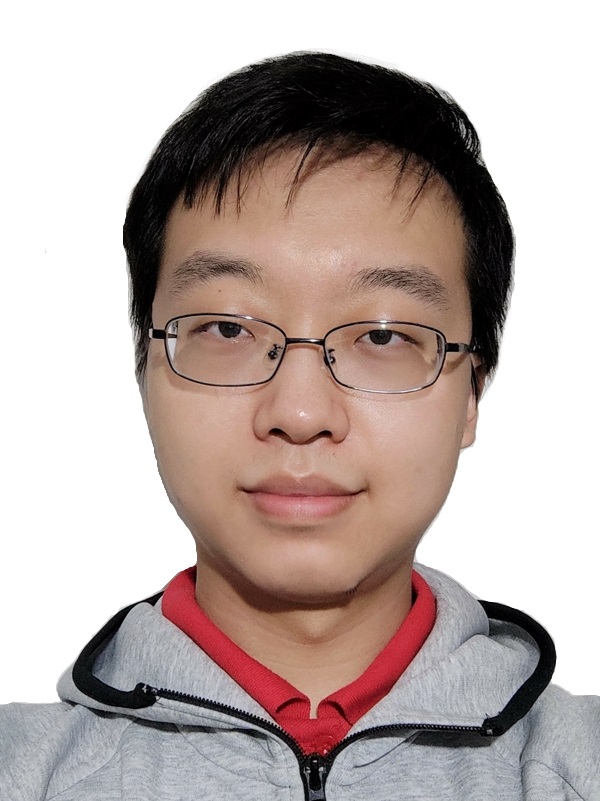}}]{Hansen Feng}
		received the BS degree from the University of Science and Technology Beijing, China, in 2020. He is currently a Ph.D. student with the School of Computer Science and Technology, Beijing Institute of Technology. His research interests include computational photography and image processing. He received the Best Paper Runner-Up Award of ACM MM 2022.
	\end{IEEEbiography}
	
	\begin{IEEEbiography}[{\includegraphics[width=1in,height=1.25in,clip,keepaspectratio]{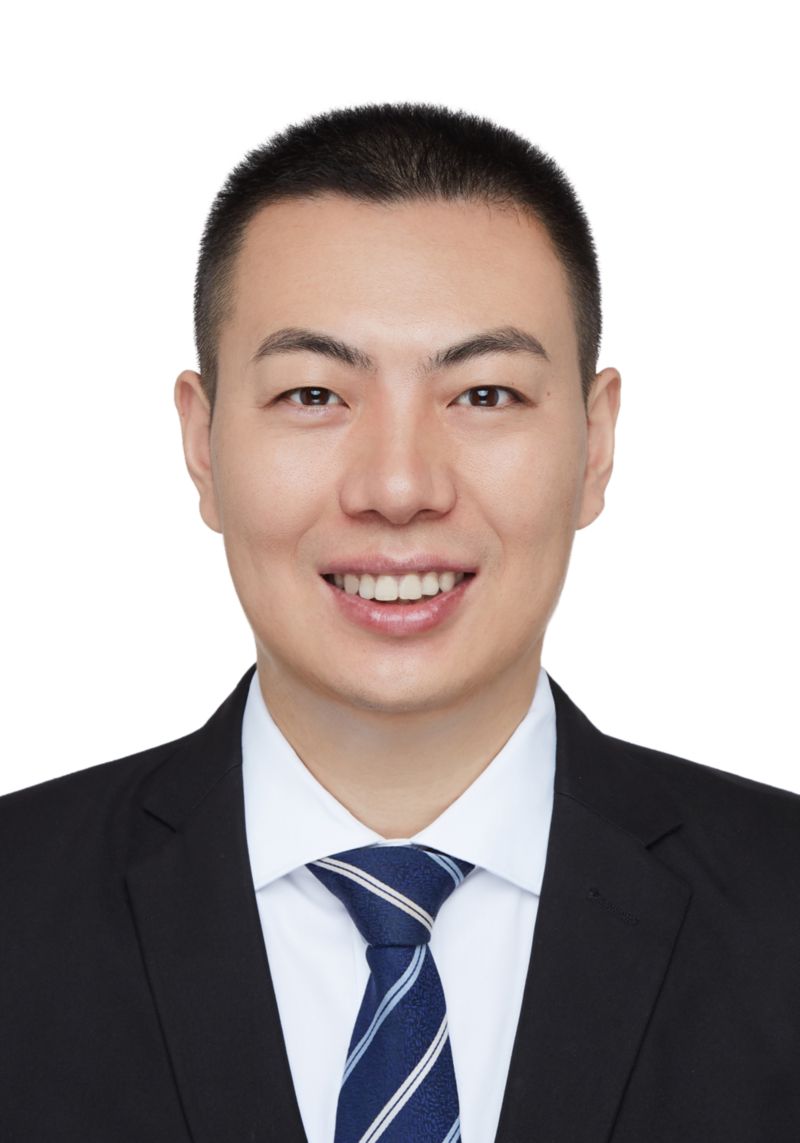}}]{Lizhi Wang}
		(Member, IEEE) received the B.S. and Ph.D. degrees from Xidian University, Xi'an, China, in 2011 and 2016, respectively. He is currently a professor in the School of Artificial Intelligence, Beijing Normal University. His research interests include computational photography and image processing. He is serving as an associate editor of IEEE Transactions on Image Processing. He received the Best Paper Runner-up Award of ACM MM 2022 and Best Paper Award of IEEE VCIP 2016.
	\end{IEEEbiography}
	
	\begin{IEEEbiography}[{\includegraphics[width=1in,height=1.25in,clip,keepaspectratio]{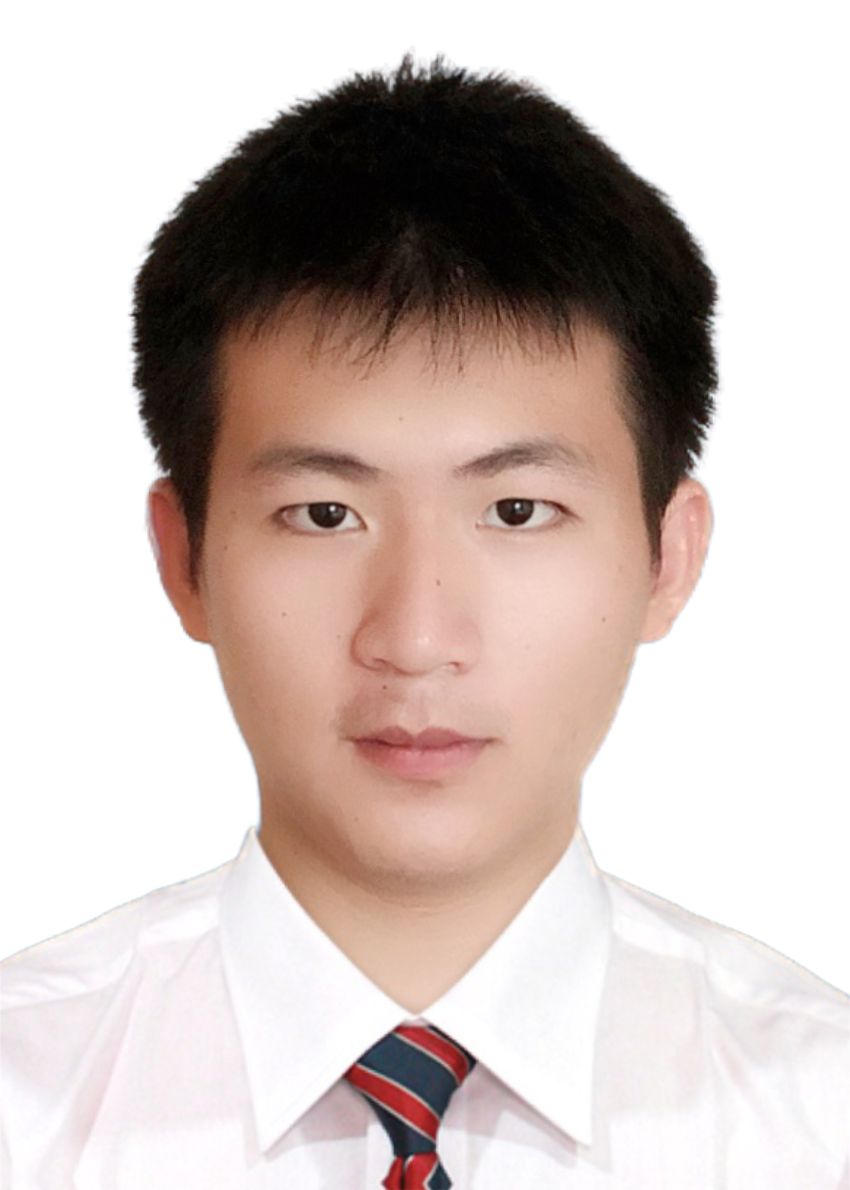}}]{Yiqi Huang}
		received the B.S. degree from the school of Information Science and Engineering, Lanzhou University, Lanzhou, China, in 2022. He is currently a M.D. student with the School of Computer Science and Technology, Beijing Institute of Technology. His research interests include computational photography and image processing.
	\end{IEEEbiography}
	
	%\newpage
	
	\begin{IEEEbiography}[{\includegraphics[width=1in,height=1.25in,clip,keepaspectratio]{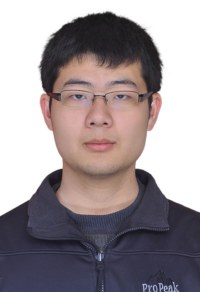}}]{Yuzhi Wang}
		received the B.S. degree from the School of Telecommunication Engineering, Xidian University, Xi’an, China, in 2012. He received the Ph.D. degree with the Department of Electronic Engineering, Tsinghua University, Beijing, China, under the supervision of Prof. H. Yang. His research interests include wireless sensor networks, computational photography, machine learning, and deep neural networks
	\end{IEEEbiography}
	
	\begin{IEEEbiography}[{\includegraphics[width=1in,height=1.25in,clip,keepaspectratio]{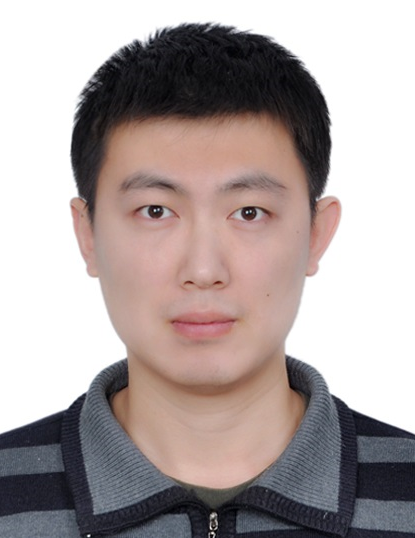}}]{Lin Zhu}
		(Member, IEEE) received the B.S. degree in computer science from the Northwestern Polytechnical University, China, in 2014, the M.S. degree in computer science from the North Automatic Control Technology Institute, China, in 2018, and the Ph.D. degree with the School of Electronics Engineering and Computer Science, Peking University, China, in 2022. He is currently an assistant professor with the School of Computer Science, Beijing Institute of Technology, China. His current research interests include image processing, computer vision, neuromorphic computing, and spiking neural network.
	\end{IEEEbiography}
	
	\begin{IEEEbiography}[{\includegraphics[width=1in,height=1.25in,clip,keepaspectratio]{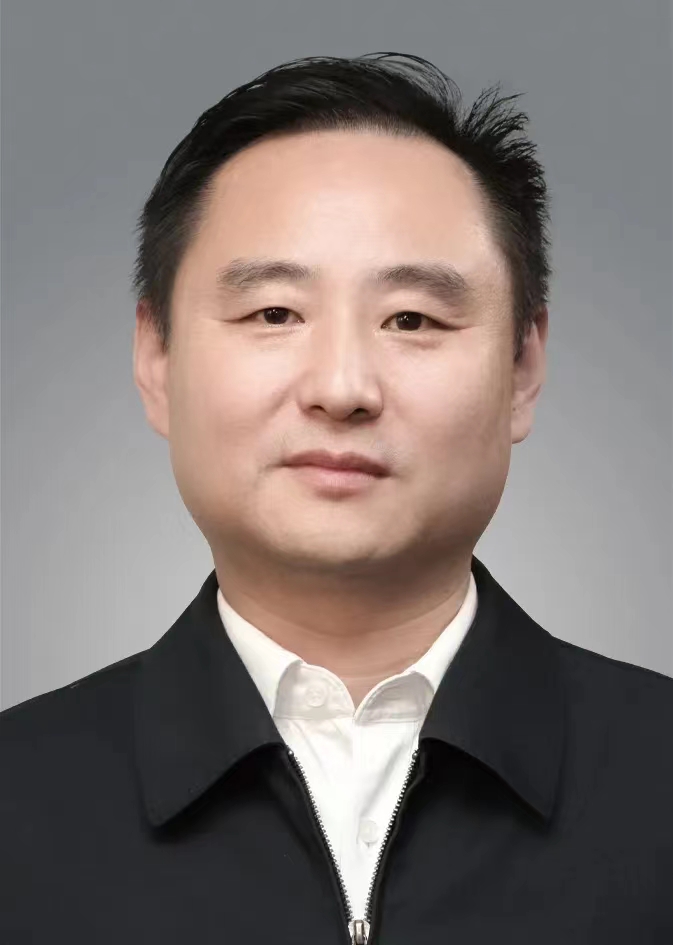}}]{Hua Huang}
		(Senior Member, IEEE) received the B.S. and Ph.D.	degrees from Xi’an Jiaotong University, in 1996	and 2006, respectively. He is currently a professor in the School of Artificial Intelligence, Beijing Normal University. He is also an adjunct professor at Xi’an Jiaotong University and Beijing Institute of Technology. His main research interests include image and video processing, computational photography, and computer graphics. He received the Best Paper Award of ICML2020 / EURASIP2020 / PRCV2019 / ChinaMM2017.
	\end{IEEEbiography}
	
	
	%\vfill
	

\end{document}

%% file: LizhiPackage.tex
\usepackage{ifpdf}
\ifpdf
  
\else
\fi
\newcommand{\eg}{\textit{e}.\textit{g}.}
\newcommand{\ie}{\textit{i}.\textit{e}.}

\usepackage{color}
\definecolor{cyan}{cmyk}{1,0,0,0}
\definecolor{darkgreen}{rgb}{0,0.5,0}
\definecolor{orange}{rgb}{1,0.5,0}
\definecolor{magenta}{cmyk}{0,1,0,0}
\definecolor{darkyellow}{cmyk}{0,0,0.75,0}
\definecolor{gray}{rgb}{0.8,0.8,0.8}